\documentclass[usenatbib]{mn2e}
\usepackage{graphicx,natbib,amssymb,amsmath}
\usepackage{txfonts, stfloats}
\usepackage{xcolor}
\usepackage{textcomp}
\usepackage{hyperref}
\usepackage{soul}

\newcommand{\expf}[1]{{{\rm e}^{#1}}}

\newcommand{\id}{{\,\rm d}}

\newcommand{\beq}{\begin{equation}}   %

\newcommand{\eeq}{\end{equation}}   %

\newcommand{\beqa}{\begin{eqnarray}}   %

\newcommand{\eeqa}{\end{eqnarray}}   %

\newcommand{\beal}{\begin{align}}

\newcommand{\enal}{\end{align}}

\newcommand{\bspl}{\begin{split}}

\newcommand{\espl}{\end{split}}

\newcommand{\bsub}{\begin{subequations}}

\newcommand{\esub}{\end{subequations}}

\newcommand{\bmulti}{\begin{multline}}   %

\newcommand{\beqm}{\begin{mathletters}}   %

\newcommand{\eeqm}{\end{mathletters}}   %

\newcommand{\me}{m_{\rm e}}

\newcommand{\Ne}{N_{\rm e}}

\newcommand{\Te}{T_{\rm e}}

\newcommand{\The}{\theta_{\rm e}}

\newcommand{\sigT}{\sigma_{\rm T}}

\newcommand{\vek} [1]{\mbox{\boldmath${#1}$\unboldmath}}

\newcommand{\pot}[2]{#1 \times 10^{#2}}

\newcommand{\lit}[1]{{\color{red}[{\sc REF}]}}

\title[Bremsstrahlung Gaunt factors]{Improved calculations of electron-ion Bremsstrahlung Gaunt factors for astrophysical applications}
\author{Jens Chluba, Andrea Ravenni and Boris Bolliet}

\begin{document}
\setstcolor{red}
\author[Chluba, Ravenni \& Bolliet]{
Jens Chluba$^1$\thanks{E-mail:jens.chluba@manchester.ac.uk},
Andrea Ravenni$^1$\thanks{E-mail:andrea.ravenni@manchester.ac.uk}
and 
Boris Bolliet$^1$\thanks{E-mail:boris.bolliet@manchester.ac.uk}
\\
$^1$Jodrell Bank Centre for Astrophysics, School of Physics and Astronomy, The University of Manchester, Manchester M13 9PL, U.K.
}

\date{\vspace{-7mm}{Accepted 2019 --. Received 2019 --}}

\maketitle

\begin{abstract}
Electron-ion Bremsstrahlung (free-free) emission and absorption occur in many astrophysical plasmas for a wide range of physical conditions. This classical problem has been studied multiple times, and many analytical and numerical approximations exist. However, accurate calculations of the transition from the non-relativistic to the relativistic regime remain sparse. Here we provide a comprehensive study of the free-free Gaunt factors for ions with low charge ($Z\leq 10$). We compute the Gaunt factor using the expressions for the differential cross section given by Elwert \& Haug (EH) and compare to various limiting cases.
We develop a new software package, {\tt BRpack}, for direct numerical applications. This package uses a combination of pre-computed tables and analytical approximations to efficiently cover a wide range of electron and photon energies, providing a representation of the EH Gaunt factor to better than $0.03\%$ precision for $Z\leq 2$. Our results are compared to those of previous studies highlighting the improvements achieved here. {\tt BRpack} should be useful in computations of spectral distortions of the cosmic microwave background, radiative transfer problems during reionization or inside galaxy clusters, and the modeling of galactic free-free foregrounds. The developed computational methods can furthermore be extended to higher energies and ion charge.
\end{abstract}


\section{Introduction}
The Bremsstrahlung (BR) or free-free emission process is highly relevant in many astrophysical plasmas \citep[e.g.,][]{Blumenthal1970, Rybicki1979}. As such it has been studied extensively in the literature \citep[e.g.,][]{Menzel1935, Karzas1961, Brussaard1962, Johnson1972, Kellogg1975, Hummer1988}, with early theoretical works reaching all the way back to the pioneering stages of quantum mechanics \citep{Kramers1923, Gaunt1930, Sommerfeld1931, BetheHeitler1934, Sommerfeld1935, Elwert1939}. 

Bremsstrahlung is the main process responsible for the X-ray radiation of galaxy clusters \citep[e.g.,][]{Gursky1972,Cavaliere1976, Sarazin1986}; it provides a source of soft photons relevant to the thermalization of spectral distortions of the cosmic microwave background \citep{Sunyaev1970mu, Sunyaev1970SPEC, Hu1993, Chluba2011therm}; and is a very important radiation mechanism close to compact objects \citep{Shakura1973, Narayan1995, McKinney2017}. In addition it is one of the main galactic foregrounds for cosmic microwave background (CMB) temperature anisotropy studies \citep{PlanckSM2015}. 
It is thus important to have an accurate representation of this process, a problem that can be cast into computations of the {\it free-free Gaunt factor}, which extend the classical Kramers formula \citep{Kramers1923} by quantum and relativistic corrections.

Here we are interested in typical electron energies corresponding to temperatures of $\simeq 10^{-7}{\rm keV}$ (a few K) up to a few tens of keV ($\lesssim 10^9$ K). This broad range of conditions is present in astrophysical plasmas of the early and late Universe (redshift $z\simeq 1-10^9$), covering both non-relativistic and mildly relativistic thermal electron populations. 
Two main approaches have featured in the literature: at non-relativistic energies, the analytic expressions summarized by \citet[][henceforth KL]{Karzas1961} can be applied, while at higher energies the Bethe-Heitler formula \citep[][henceforth BH]{BetheHeitler1934} is valid. The KL formulae provide a non-perturbative description of the BR emissivity assuming non-relativistic electron velocities\footnote{The ion is assumed to rest before and after the interaction} (i.e., electron speeds $|\varv|/c \ll 1$), while the BH expression utilizes the first order Born approximation ($\alpha Z \ll 1$) for relativistic electrons. Although it is well-known that higher order Coulomb corrections and shielding effects become important for high ion charge $Z$ and extreme electron energies \citep[e.g.,][]{Tseng1971, Roche1972, Haug2008}, the KL and BH formulae are accurate in their respective regimes. 
For intermediate energies, no simple expressions exist that allow describing the Gaunt factor in the transition between the KL and BH limits.

The computation of the KL and BH Gaunt factors and their thermal averages is fairly straightforward, and various approximations and computational schemes have been developed \citep{Karzas1961, Brussaard1962, Itoh1985, Hummer1988, Nozawa1998, Itoh2002}. 
To bridge the gap between these two limits, \citet{vanHoof2015} combined the non-relativistic KL expressions and BH formula to mimic the transition.
It is, however, possible to directly model the transition using the differential BR cross section of \citet[][EH hereafter]{ElwertHaug1969}. This cross section is based on Sommerfeld-Maue eigenfunction \citep{Sommerfeld1935} and is valid for low ion charge over a wide range of electron and photon energies. It was shown that the cross section naturally approaches the non-relativistic and relativistic limits \citep{ElwertHaug1969}, thus joining the two regimes.
However, it still has to be integrated over the particle momenta and thermally-averaged, a task that will be studied here.

In this paper we investigate the EH expression computing the total BR Gaunt factor and thermal averages for ionic charge $Z\leq 10$, having applications to the evolution of CMB spectral distortions and the reionization process in mind, where hydrogen and helium (i.e., $Z\leq 2$) dominate.
We numerically integrate the differential EH cross section and compare the obtained results to various limiting cases. The differential cross section is simplified and several new approximations are presented (Sect.~\ref{sec:general_sigma}). The main numerical challenge is the demanding evaluation of hypergeometric functions, which we reduce to the evaluation of one real function (see Appendix~\ref{sec:EH_rewrite}). We in detail discuss the domains of validity of the various expressions (Sect.~\ref{sec:GF_numerical} and Sect.~\ref{sec:therm_results}) and directly compare with previous calculations (Sect.~\ref{sec:Itoh_and_vH}). 
All our results can be reproduced with {\tt BRpack}\footnote{{\tt BRpack} will be made available at \url{www.chluba.de/BRpack}.}, which uses a combination of pre-computed tables and analytic approximations to efficiently represent the EH, BH and KL Gaunt factors over a wide range of electron and photon energies. A compression of the required data at low and high photon energies is achieved by analytic considerations.

\vspace{-4mm}
\section{Bremsstrahlung cross sections}
\label{sec:BR_definitions}
In this section, we provide a comprehensive summary of existing analytic expressions for the BR emission cross section\footnote{The absorption cross section can be deduced by interchanging the roles of the initial/final electron, denoted by momenta $p_1$ and $p_2$, respectively. All momenta and energies are expressed in units of $\me c$ and $\me c^2$, respectively.}. An improved expression for the differential cross section was given by EH. One crucial feature is that at high energies the EH formula naturally reduces to the BH formula, while at low energies the non-relativistic expression of KL is recovered. Thus, the EH formalism allows computing  the total BR cross section for the intermediate case. However, the evaluation of the cross section is cumbersome and it is therefore crucial to understand its limiting cases.

\vspace{-3mm}
\subsection{Classical Kramers BR formula}
\label{sec:Kramers}
In the classical limit, the BR cross section for the emission of a photon at energy $\omega=h\nu/\me c^2$ by an electron with momentum $p_1$ reads \citep{Kramers1923, Karzas1961}
\begin{align}
\label{eq:dsig_domega_K}
\frac{\id \sigma_{\rm K}(\omega, p_1)}{\id\omega}  
&=\frac{2\alpha Z^2}{\sqrt{3}} \frac{\sigT}{p_1^2 \omega}.
\end{align}
Here, $\alpha$ is the fine structure constant, $Z$ the ion charge and $\sigma_\mathrm{T}$ the Thomson cross-section. Due to energy conservation, only photons with energy $\omega\leq \omega_{\rm max}=\gamma_1-1$ can be emitted. Here $\gamma_1=(1+p_1^2)^{1/2}$ is the Lorentz factor of the initial electron. The ratio of the BR emission cross sections discussed in the following sections and the Kramers approximation then defines the related Gaunt factor. 

\vspace{-3mm}
\subsection{Exact non-relativistic BR cross section}
The exact non-relativistic (NR) BR emission cross section can be cast into the form\footnote{We modified the definitions of $\eta_i$ to match the relativistic form of EH. The effects of this modification will be illustrated below and is found to slightly improve the agreement with the EH result.} \citep{Karzas1961, Hummer1988}
\bsub
\label{eq:dsig_domega_K}
\begin{align}
\frac{\id \sigma_{\rm NR}(\omega, p_1)}{\id\omega}  
&=\frac{\id \sigma_{\rm K}(\omega, p_1)}{\id\omega}\,g_{\rm NR}(\omega, p_1)
\nonumber \\[1mm]
g_{\rm NR}(\omega, p_1)
&=\frac{\sqrt{3}}{4\pi}
\,\mathcal{F}(\eta_1,\eta_2)\,G_0
\Bigg\{
\left[\eta_1 \, \eta_2+\frac{1}{2}\left(\frac{\eta_1}{\eta_2}+\frac{\eta_2}{\eta_1}\right)\,\right] G_0
\nonumber\\
&\qquad\qquad\qquad\qquad\quad - \frac{(1+\eta_1^2)(1+\eta_2^2)}{6} \,G_1
\Bigg\}
\\
\label{eq:def_Gell}
G_\ell(\eta_1, \eta_2, x)&=
\left(-x\right)^{\ell +1}
(1-x)^{\frac{i(\eta_1+\eta_2)}{2}}\,\expf{-\pi\eta_1}
\\
&\qquad\quad  _2F_1\left(1+\ell+i\eta_1, 1+\ell+i\eta_2, 2\ell+2, x \right)
\nonumber\\[1mm]
\eta_i=\frac{\alpha Z \gamma_i}{p_i},& \quad 
x=-\frac{4\eta_1\eta_2}{(\eta_1-\eta_2)^2},\quad 1-x=\frac{(\eta_1+\eta_2)^2}{(\eta_1-\eta_2)^2}
\nonumber\\
\label{eq:dsig_domega_K_F}
\mathcal{F}(\eta_1,\eta_2) &=\frac{4\pi^2\eta_1\eta_2}{(1-\expf{-2\pi\eta_1})(1-\expf{-2\pi\eta_2})},
\end{align}
\esub
with $p_2=\sqrt{p_1^2+\omega(\omega-2\gamma_1)}$. The functions $G_\ell$ are all real functions (see Appendix~\ref{app:G_ell_Real}). Since the scattered electron momentum obeys $p_2\leq p_1$, one also has $\eta_1\leq \eta_2$. 
As shown in Appendix~\ref{app:G_ell_Real_rewrite}, the NR Gaunt factor can be further simplified to
\begin{align}
\label{eq:gaunt_NR_mod}
g_{\rm NR}(\omega, p_1)
&\equiv -\frac{\sqrt{3}}{2\pi} \,\mathcal{F}(\eta_1,\eta_2) \, \frac{(\eta_1+\eta_2)^2}{(\eta_1-\eta_2)^2}
\,G_0\,G'_0
\end{align}
with $G'_0=\partial_x G_0$ evaluated at $x=-4\eta_1\eta_2/(\eta_1-\eta_2)^2$. 
This eases the numerical computation of $g_{\rm NR}$ greatly because only $G_0$ has to be computed. In a similar manner we will reduce the EH expression to a function of $G_0$ and $G'_0$ (Appendix~\ref{sec:EH_rewrite}).

The functions, $G_\ell(\eta_1, \eta_2)$, are rather hard to evaluate for the range of momenta we require. In particular for $\omega<10^{-6}\omega_{\rm max}$ and at $p_1<10^{-3}$ the computations become difficult due to catastrophic cancellations of large numbers.
At $p_1>10^{-3}$, we use simple recursion relations similar to \citep{Karzas1961, Hummer1988} outlined in Appendix~\ref{sec:recursion}. At $\omega<10^{-20}\omega_{\rm max}$ we use the soft-photon limit of Eq.~\eqref{eq:dsig_domega_K} derived in Appendix~\ref{sec:NR_low_approx} (where we also kept higher order terms). It can be cast into the simple form
\begin{align}
\label{eq:NR_approx_more}
g^{\rm soft}_{\rm NR}(\omega, p_1)
&\approx \frac{\sqrt{3}}{2\pi} \,\mathcal{F}_{\rm E}(\eta_1,\eta_2)\,
\left\{\ln\left(\frac{4\eta_1\eta_2}{(\eta_1-\eta_2)^2}\right)-{\rm Re}\left[H(i \eta_1)\right]\right\}
\nonumber\\
\mathcal{F}_{\rm E}(\eta_1,\eta_2)&=\frac{\eta_2}{\eta_1}\frac{1-\expf{-2\pi\eta_1}}{1-\expf{-2\pi\eta_2}},
\end{align}
which closely matches the NR calculation even at higher frequencies (up to $\omega \simeq 10^{-3} \omega_{\rm max}$). Here, $H(z)$ denotes the harmonic number (see Appendix~\ref{sec:NR_low_approx}).
The remaining cases can be evaluated using arbitrary number precision (e.g., with {\tt Mathematica}). Alternatively, the differential equation for $G_0$ can be solved, which also directly gives $G_0'$ without further effort (see Appendix~\ref{app:G_0_ODE}).

To quickly compute the non-relativistic Gaunt factor we tabulate it for charge $Z=1$ as a function of $p_1$ and $w=\omega/\omega_{\rm max}$ at $p_1\in [\pot{5}{-8}, 10^{-3}]$ and\footnote{This is one of the benefits of using the emission Gaunt factor as it has a finite upper limit at $\omega_{\rm max}=\gamma_1-1$.} $w \in [10^{-20}, 1]$, which in turn allows us to obtain the thermally-averaged Gaunt factor down to temperatures comparable to $\Te \simeq 1\,{\rm K}$. Tables for the non-relativistic absorption Gaunt factor were also given by \citet{vanHoof2014} and can be reproduced using the emission Gaunt factor.
The results for ionic charge $Z>1$ can be obtained by interpolating those for $Z=1$ using the simple mapping $g_{\rm NR}(p_1,\omega)\rightarrow g_{\rm NR}(p_1^*,\omega^*)$ with
\begin{equation}
\label{eq:prime}
p^*_i=\frac{p_i/Z}{\sqrt{1+(p_i/Z)^2[Z^2-1]}}, \qquad \omega^*=\gamma_1^*-\gamma_2^*.
\end{equation}
Overall our procedure gives better than 0.01\% numerical precision for the non-relativistic cross section at all $\omega$ and $p_1$. To further improve the non-relativistic Gaunt factor one can multiply it by $\gamma_1^2$ to capture the leading order relativistic correction
\begin{equation}
\label{eq:g_NR_rel}
g^{\rm corr}_{\rm NR}(p_1,\omega)= \gamma_1^2 g_{\rm NR}(p_1,\omega).
\end{equation}
As we will show this indeed improves the range of applicability of the KL formula (see Sect.~\ref{sec:BH_sigma} for discussion). 

\vspace{-3mm}
\subsection{Bethe-Heitler cross section}
\label{sec:BH_sigma}
At high energies ($p_1\gtrsim 10^{-2} \,Z$), the Bethe-Heitler cross section \citep{BetheHeitler1934}, derived using the first order Born approximation, becomes valid. It can be cast into the form\footnote{Note a missing factor of 2 in the $L$-term of \cite{Jauch1976}.} \citep[e.g.,][]{BetheHeitler1934, Jauch1976}
\begin{align}
\label{eq:dsig_domega_BH}
\frac{\id \sigma_{\rm BH}(\omega, p_1)}{\id\omega}  
&=\frac{\id \sigma_{\rm K}(\omega, p_1)}{\id\omega}\,g_{\rm BH}(\omega, p_1)
\nonumber \\[1mm]
g_{\rm BH}(\omega, p_1)
&=
%
\frac{\sqrt{3}}{\pi}\Bigg[
\frac{p_1  p_2}{4}-\frac{3}{8} \gamma_1\gamma_2\left(\frac{p_1}{p_2}+\frac{p_2}{p_1}\right)
+\gamma_1\gamma_2\,L
\nonumber\\
&\!\!\!\!\!\!\!\!\!\!\!\!\!\!\!\!\!\!\!\!\!\!\!\!\!\!\!\!\!\!\!\!\!
+\frac{3}{8}\omega L
\Bigg\{
\left(1+\frac{\gamma_1\gamma_2}{p^2_1}\right) \frac{\lambda_1}{p_1}
-\left(1+\frac{\gamma_1\gamma_2}{p^2_2}\right) \frac{\lambda_2}{p_2}
+\omega
\left[1\!+\!\frac{\gamma_1\gamma_2}{p_1^2p_2^2}\!+\!\frac{\gamma^2_1\gamma^2_2}{p_1^2p_2^2} \right]
\Bigg\}
\nonumber\\
&\quad+\frac{3}{8}\left(\frac{\gamma_2 p_2}{p^2_1}\lambda_1 +\frac{\gamma_1 p_1}{p^2_2}\lambda_2 -2\lambda_1\lambda_2\right)
\Bigg],
\\[1mm] \nonumber
\lambda_i&=\ln(\gamma_i+p_i), \qquad L=\ln\left[\frac{\gamma_1\gamma_2+p_1p_2-1}{\omega}\right].
\end{align}
Since this expression only involves elementary functions it can be evaluated very efficiently.  It is equivalent to the one used in \citet{Itoh1985, Nozawa1998} and \citet{vanHoof2015} after transforming to their variables.

At low frequencies, $p_1\simeq p_2$ and $\gamma_1\simeq \gamma_2$, such that
\begin{align}
\label{eq:Gaunt_BH}
g_{\rm BH}
&\approx
\frac{\sqrt{3}}{\pi}\left\{\gamma^2_1\left[\ln\left(\frac{2 p_1^2}{\omega}\right)-\frac{1}{2}-\frac{1}{4\gamma_1^2}\right]
+ \frac{3}{4} \left(\frac{\gamma_1}{p_1} -\lambda_1\right)\lambda_1\right\}.
\end{align}
For increasing $p_1$, this expression scales like $\simeq \gamma_1^2$, which causes a large boost of the BR emissivity. For convenience it is therefore good to absorb this extra factor into the Kramers approximation and define the BR Gaunt factor with respect to this modified Kramers approximation, i.e., $\mathrm{d}\sigma_\mathrm{K}^\mathrm{corr}/\mathrm{d}\omega=\gamma_1^2\mathrm{d}\sigma_\mathrm{K}/\mathrm{d}\omega$. The modified Kramers cross section can still be thermally-averaged analytically [see Eq.~\eqref{eq:dN_dt_K_rel}] such that this modification does not cause any additional complications. 

It is also well-known that the BH approximation can be improved by adding the so-called Elwert factor \citep{Elwert1939}, which already appeared in Eq.~\eqref{eq:NR_approx_more}. This then yields
\begin{align}
\label{eq:Gaunt_BH_EW}
g^*_{\rm BH}(p_1, \omega)
&\approx  \mathcal{F}_{\rm E}(\eta_1,\eta_2) \, g_{\rm BH}(p_1, \omega),
\end{align}
which improves the agreement with the EH Gaunt factor in particular in the short-wavelength limit ($\omega \simeq \omega_{\rm max}$). In our computations we shall always use $g^*_{\rm BH}(p_1, \omega)$ for the Bethe-Heiter limit.

\vspace{-0mm}
\subsection{Elwert-Haug cross section}
\label{sec:general_sigma}
Considering BR in the EH case is a lot more challenging. No analytic expression for the total cross section, $\id \sigma/\id \omega$, has been given. However, EH provide an expression for the differential cross section that allows us to describe the transition between the non-relativistic and relativistic regimes.

Starting from EH, but significantly rewriting the differential cross section (see Appendix~\ref{sec:EH_rewrite}), we find
\bsub
\begin{align}
\label{eq:EH_final}
\frac{\id^3 \sigma_{\rm EH}}{\id\mu_1\!\id\mu_2\!\id\phi_2}
&=\frac{\id \sigma_{\rm K}(\omega, p_1)}{\id\omega}\,\frac{\id^3 g_{\rm EH}(\omega, p_1)}{\id\mu_1\!\id\mu_2\!\id\phi_2}
\\[0mm]
\frac{\id^3 g_{\rm EH}}{\id\mu_1\!\id\mu_2\!\id\phi_2}
&=\frac{3\sqrt{3}}{8\pi^2}\,p_1 p_2\,\mathcal{F}(\eta_1,\eta_2)\,\mathcal{M}^2(\omega, p_1, \mu_1, \mu_2, \phi_2),
\end{align}
\esub
where $\id^3 g_{\rm EH}/\id \mu_1\!\id\mu_2\!\id\phi_2$ defines the EH Gaunt factor that is differential in three angles, characterized by the direction cosines, $\mu_i=\vek{p}_i\cdot\vek{k}/p_1\omega$, and the polar angle $\phi_2$ between the incoming photon and outgoing electron. 
After introducing the auxiliary variables:
\begin{align}
\eta_\infty&=\alpha Z, \quad\eta_\pm=\eta_1\pm\eta_2
\nonumber \\
\mu_i&=\frac{\vek{p}_i\cdot\vek{k}}{p_1\omega},
\quad
\mu_{12}=\frac{\vek{p}_1\cdot\vek{p}_2}{p_1 p_2}=\mu_1\mu_2+\cos(\phi_2)\sqrt{1-\mu_1^2}\sqrt{1-\mu_2^2}
\nonumber \\
\pi_{i}&=p_i\mu_1, \quad \pi_{12}=p_1p_2\mu_{12}, 
\quad
\kappa_i=2(\gamma_i-p_i\mu_i)=2(\gamma_i-\pi_i)
\nonumber\\
\chi_i&=p_i\sqrt{1-\mu_i^2}, \quad \chi_{12}=\chi_1 \chi_2 \cos(\phi_2)
\nonumber\\
\tau_i&=4\gamma_i^2-q^2, \quad \tau_{12}=4\gamma_1\gamma_2-q^2
\nonumber\\[1mm]
q^2&=|\vek{p}_1-\vek{p}_2-\vek{k}|^2
=p_1^2+p_2^2+\omega^2+2\left[\omega(\pi_2-\pi_1)-\pi_{12}\right]
\nonumber\\[1mm]
\xi&=\left[\left(\frac{p_1+p_2}{\omega}\right)^2-1\right] \frac{q^2}{\kappa_1\kappa_2}
\equiv\frac{\tilde{\mu} q^2}{\kappa_1\kappa_2}
\nonumber\\[0.5mm]
\kappa&=\frac{\eta_+}{\eta_\infty}=\frac{\gamma_1}{p_1}+\frac{\gamma_2}{p_2},
\; \rho=\frac{1}{p_1}+\frac{1}{p_2},
\nonumber
\end{align}
the required matrix element can be cast into the compact form
\begin{align}
\label{eq:EH_G_final_G0}
\mathcal{M}^2&
=\frac{1}{q^4}\Bigg\{ 
\left[J_{\rm BH}
-2\frac{\eta_-}{\eta_+}\xi \, D_1
+\frac{\eta^2_-}{\eta^2_+} \xi^2\, D_2
\right] \frac{\eta_+^2 G_0^2}{4(1-\xi)^2} 
+ J_{\rm BH} \, [\xi G'_0]^2 
\Bigg\}
\nonumber
\\[1.5mm]
J_{\rm BH}&= \tau_1 \frac{\chi_2^2}{\kappa_2^2}+\tau_2 \frac{\chi_1^2}{\kappa_1^2} 
-\tau_{12} \frac{2\chi_{12}}{\kappa_1\kappa_2}
+ \left(\chi_1^2+\chi_2^2-2\chi_{12}\right)\,\frac{2\omega^2}{\kappa_1\kappa_2}
\nonumber
\\
D_1&=
\tau_1 \frac{\chi_2^2}{\kappa_2^2}-\tau_2 \frac{\chi_1^2}{\kappa_1^2}
+(\chi^2_{1}-\chi^2_{2})\frac{2 \omega^2}{\kappa_1\kappa_2}
+ \left(\frac{L_1}{\kappa_1}+\frac{L_2}{\kappa_2}\right)\frac{\omega}{\rho}
\nonumber\\[1.5mm]
D_2&=
\tau_1 \frac{\chi_2^2}{\kappa_2^2}+\tau_2 \frac{\chi_1^2}{\kappa_1^2}+\tau_{12} \frac{2\chi_{12}}{\kappa_1\kappa_2}
+\left(\chi_1^2+\chi_2^2+2\chi_{12}\right)\frac{2 \omega^2}{\kappa_1\kappa_2} 
\nonumber\\
&\qquad\qquad
+\frac{8 \omega^2}{\kappa_1\kappa_2} -\left(\frac{L_1}{\kappa_1}-\frac{L_2}{\kappa_2}\right)\frac{2\omega}{\rho}
+L_3\frac{\omega^2}{\rho^2}
\nonumber\\[1.5mm]
L_1&=\kappa\left[\pi_1(\pi_{12}+p_2^2)-\left(\pi_1+\pi_2-\omega\right)p_1p_2+(2-\pi_1\pi_2)\omega\right]
\nonumber\\
&\qquad\qquad+2\frac{\omega}{p_1}\left(\pi_1+\pi_2-\omega\right)
\nonumber\\
L_2&=\kappa\left[\pi_2(\pi_{12}+p_1^2)-\left(\pi_1+\pi_2+\omega\right)p_1p_2-(2-\pi_1\pi_2)\omega\right]
\nonumber\\
&\qquad\qquad-2\frac{\omega}{p_2}\left(\pi_1+\pi_2+\omega\right)
\nonumber\\
L_3&=\tilde{\mu}\,\omega^2\left[1-\frac{\pi_1\pi_2}{p_1p_2}+\frac{\gamma_1+\gamma_2}{p_1p_2}\,\frac{\gamma_1+\gamma_2+\pi_1+\pi_2}{p_1p_2}\right]-2\rho^2.
\end{align}
where $G_0$ and $G'_0$ are both evaluated at $x=1-\xi$ in Eq.~\eqref{eq:dsig_domega_K}. 
Expressed in this way indeed simplifies the computation of the cross section significantly and also allows one to more directly read off limiting cases. 
For instance, in the BH limit, one has $\mathcal{M}^2=J_{\rm BH}/q^4$ \citep{ElwertHaug1969}.
Alternatively, the cross section in the form Appendix~\eqref{eq:EH_G_final_M2} can be applied. Both approaches give excellent results when using the numerical method described next.

\vspace{-1.0mm}
\subsubsection{Numerical evaluation of the EH cross section}
\label{sec:general_sigma_int}
To evaluate the total EH cross section, we have to integrate Eq.~\eqref{eq:EH_final} over $\mu_1$, $\mu_2$ and $\phi_2$. This is a non-trivial task even for modern computers. At low frequencies, large cancelation issues arise which can be cured using suitable variables. At both large and small values of $p_1$, the evaluation of hypergeometric functions furthermore becomes cumbersome even when applying suitable transformations for the argument. Luckily, in many of the extreme cases we can resort to the non-relativistic and Bethe-Heitler formulae. Nevertheless, intermediate cases have to be explicitly computed.

Firstly, it is helpful to convert the integral over $\phi_2$ into an integral over $\xi$. This also reduces the number of evaluations for the hypergeometric functions, which significantly improves the computational efficiency. The symmetry of the integrand in $\phi_2$ implies $\int_0^{2\pi} \id \phi_2=2\int_0^{\pi} \id \phi_2=2\int_{\xi_{\rm min}}^{\xi_{\rm max}} \frac{\id \phi_2}{\id \xi} \id \xi=2\int_{0}^{\Delta \xi_{\rm tot}} \frac{\id \phi_2}{\id \xi} \id \Delta \xi$ with 
\bsub
\begin{align}
\Delta \xi &=\xi - \xi_{\rm min}, 
\qquad \cos(\phi_2)=1-2\frac{\Delta \xi}{\Delta \xi_{\rm tot}}
\\
\Delta \xi_{\rm tot} &=\frac{4\,\tilde{\mu}\,\chi_1\chi_2}{\kappa_1\kappa_2}, 
\qquad \frac{\id \phi_2}{\id \xi}=\frac{1}{\sqrt{\Delta \xi(\Delta \xi_{\rm tot}-\Delta \xi)}}
\\[1mm]
\xi_{\rm min} &=\frac{\tilde{\mu} (p_1^2+p_2^2+\omega^2+2\left[\omega(\pi_2-\pi_1)-(\pi_1\pi_2+\chi_1\chi_2)\right])}{\kappa_1\kappa_2}.
\end{align}
This transformation is crucial for improving the stability of the code near the maxima of $1/q^4$; however, to further improve matters one also has to use $\Delta \mu_{21}=\mu_2-\mu_1$ instead of $\mu_2$. At low frequencies, the integrand picks up most of its contributions from around $\mu_1\simeq \mu_2$. Hence this variable more naturally allows us to focus evaluations around the poles. After these transformation, we also regroup contributions and analytically cancel leading order terms $\propto \mu_1$ and $\propto p_1$. As an example, for $\xi_{\rm min}$ we find
\begin{align}
\xi_{\rm min} &=\frac{\tilde{\mu}}{\kappa_1\kappa_2} \Bigg\{\Delta p_{21}^2
-2 p_1 p_2(\mathcal{S}_1 \Delta \mathcal{S}_{21}+\mu_1\Delta \mu_{21})
\nonumber\\
&\qquad\qquad\quad +2[p_1\Delta \mu_{21}+\Delta p_{21} (\mu_1+\Delta \mu_{21})]\omega+\omega^2 
\Bigg\}.
\end{align}
\esub
with $\Delta p_{21}=p_2-p_1$, $\mathcal{S}_i=\sqrt{1-\mu^2_i}$ and $\Delta \mathcal{S}_{21}=\mathcal{S}_2-\mathcal{S}_1$. It is also important to treat the differences $\Delta p_{21}$ and $\Delta \mathcal{S}_{21}$ analytically for small $\omega$ and $\Delta \mu_{21}$. 

The contributions $\propto D_i$ in Eq.~\eqref{eq:EH_G_final_G0} become small at low frequencies and do not cause any serious numerical issues. However, we have to regroup the terms in $J_{\rm BH}$. We found
\begin{align}
J_{\rm BH}&= 4\left(\gamma_2\frac{\chi_1}{\kappa_1}-\gamma_1\frac{\chi_2}{\kappa_2}\right)^2 
\!- \left(\frac{\chi_1}{\kappa_1}-\frac{\chi_2}{\kappa_2}\right)^2q^2
\nonumber\\
&\qquad\qquad
+(\tau_{12}+2\omega^2)\frac{\Delta \xi}{\tilde{\mu}} 
+ \left(\chi_1-\chi_2\right)^2 \frac{2\omega^2}{\kappa_1\kappa_2}
\label{eq:J_rewrite}
\end{align}
to provide numerically stable results. Here we used the identity $\chi_{1}\chi_{2}-\chi_{12}=\kappa_1\kappa_2 \Delta \xi/[2\tilde{\mu}]$. This procedure allows us to compute the Bethe-Heitler limit by numerical integration of the differential cross section even at extremely low frequencies $(w=\omega/[\gamma_1-1]\simeq 10^{-14})$, highlighting the numerical precision of our method.

The computations for the EH case over a wide range of energies requires a few additional steps. The biggest remaining problem is the evaluation of terms related to the hypergeometric functions. These can be either treated by using the real functions $G_0$ and $G_0'$ like for the NR case or by expressing matters in terms of $|\mathcal{A}|^2$ and $|\mathcal{W}|^2$ (see Appendix~\ref{eq:relations_AWIM} for definitions). We studied both approaches but eventually used the former in our final calculations, finding it to be more efficient as evaluation of $G_0$ using the differential equation approach simultaneously yield $G'_0$ without further effort. Both approaches gave consistent results and we also validated the various versions of writing the EH cross section given the significant steps involved in the derivation (see Appendix~\ref{sec:EH_rewrite}).

At $w>10^{-6}$ and also $\xi\lesssim 10^{-8}$, we used the recursion relations for $G_0$ and the hypergeometric function series to compute $G_0$ and $G'_0$. At $w\leq 10^{-6}$ we tabulated $G_0$ and $G'_0$ for $\xi\in [ 10^{-8}, 10^7]$ every time $p_1$ and $w$ changed to accelerate the evaluations. To extend the evaluation to $\xi\gtrsim 10^7$ in this regime, we used the following asymptotic expansions for $G_0$ and $G'_0$ 
\begin{align}
G_0^2
&\approx \left(\frac{1-\expf{-2\pi \eta_1}}{2\pi \eta_1}\right)^2
\left[
\phi^2+2\eta_1^2\,\frac{\phi(2+\phi)}{\xi}
\right]
\nonumber\\[1mm]
[\xi G_0']^2
&\approx \left(\frac{1-\expf{-2\pi \eta_1}}{2\pi \eta_1}\right)^2\left[1-2\eta_1^2\frac{(1+\phi)}{\xi}\right]
\nonumber\\[1mm]
\phi&=\ln\xi - 2\,{\rm Re}[H(i\eta_1)].
\label{eq:G0_rewrite}
\end{align}
Finally, instead of computing the total cross section we numerically integrate the difference with respect to the BH case (modified by the Elwert factor). This improves the numerical precision and rate of convergence. To obtain the Gaunt factors at all energies of the photon we use the standard variables (i.e., $\phi_2$ and $\mu_2$) at $w\gtrsim 10^{-2}$, which we found to perform better in this regime. 

At very low photon energies ($w\lesssim 10^{-14}$), even the procedure described above was no longer sufficient. However, just like for the non-relativitistic Gaunt factor, at those energies the asymptotic behavior is reached. Motivated by Eq.~\eqref{eq:NR_approx_more}, we thus used
\begin{align}
\label{eq:EH_soft_template}
g^{\rm soft}_{\rm EH}(\omega, p_1)
&\approx \frac{\sqrt{3}}{2\pi} \,\mathcal{F}_{\rm E}(\eta_1,\eta_2)\,
A\left\{\ln\left(\frac{4\eta_1\eta_2}{(\eta_1-\eta_2)^2}\right)-B\right\},
\end{align}
with the free parameters $A$ and $B$ to extrapolate the Gaunt factor towards low energies. In practice we use $w=10^{-8}$ and $10^{-6}$ to determine the free parameters for any of the cases at $w\lesssim 10^{-10}$. Since we are able to directly compute cases down to $w\simeq 10^{-14}$ we could validate these extrapolations explicitly. 

Given the numerical challenges of multi-dimensional integration we expect errors to become noticeable at $\simeq 10^{-4}$ relative precision. To carry out the numerical integrals we used nested Patterson quadrature rules and the {\tt CUBA} library\footnote{\url{http://www.feynarts.de/cuba/}}. Both procedures yield consistent results. We also checked many of our results using {\tt Mathematica}, however, {\tt BRpack} was found to be faster.

\begin{figure}
\includegraphics[width=\columnwidth ]{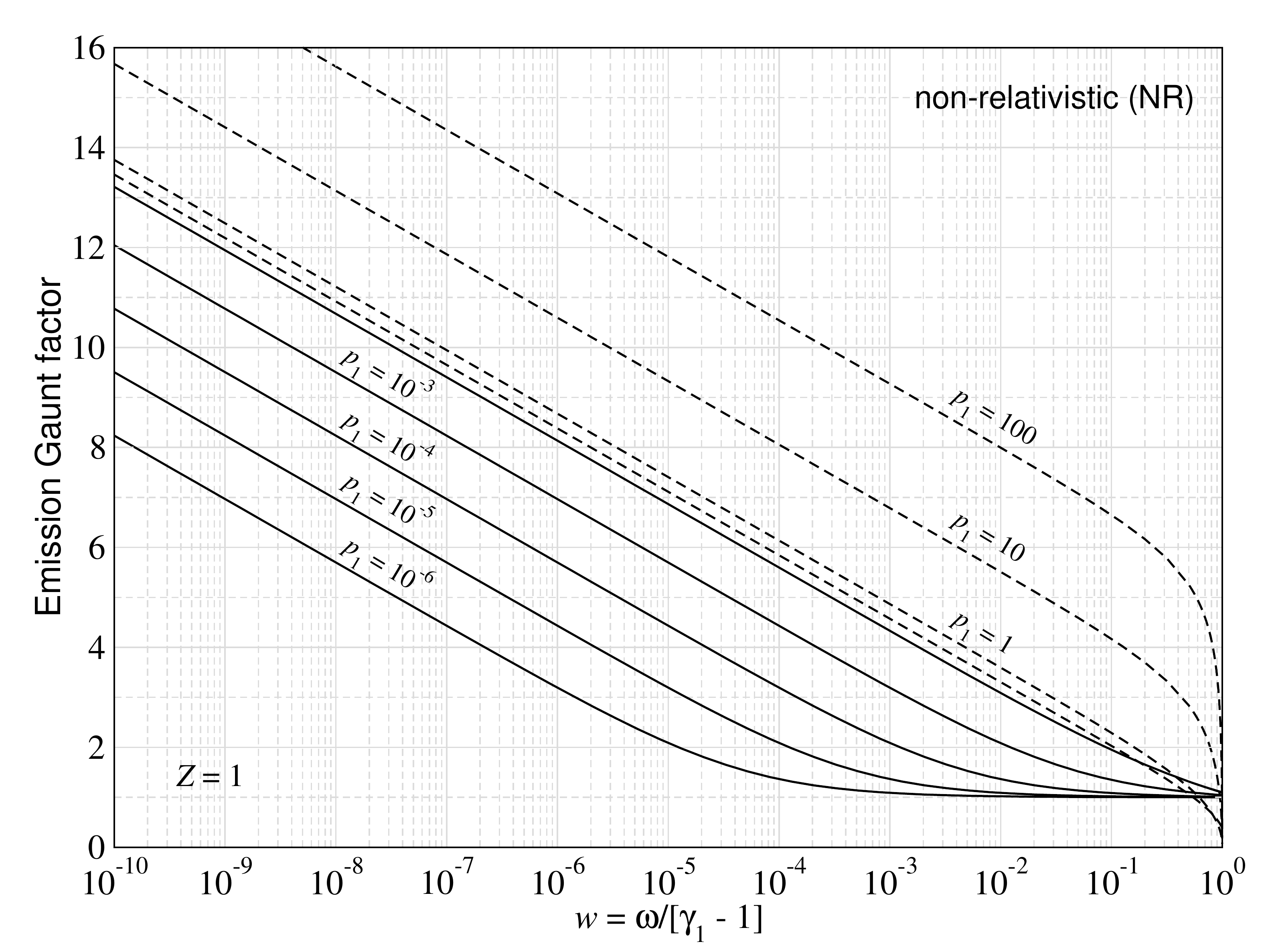}
\caption{Non-relativistic Gaunt factor [Eq.~\eqref{eq:dsig_domega_K}] for $Z=1$. The usual definition relative to the Kramers formula ($\!\id \sigma/\id \omega \propto 1/[p_1^2 \omega]$) is applied. The lines are for different values of $p_1$, varied by factors of 10. The non-relativistic formula becomes inaccurate at $p_1\gtrsim 10^{-1}$ (dashed lines).}
\label{fig:gaunt_Z_1}
\end{figure}

\begin{figure}
\includegraphics[width=\columnwidth]{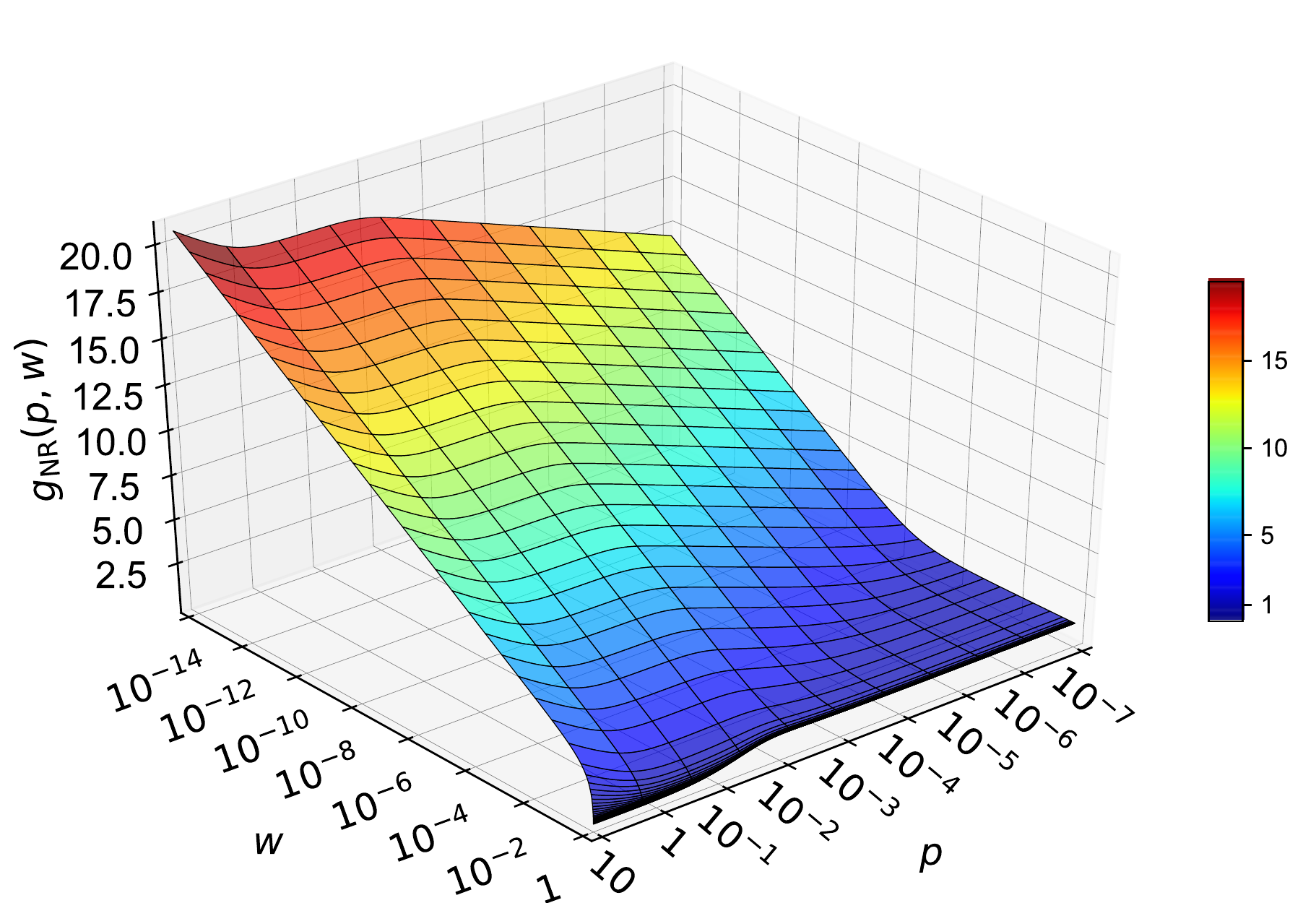}
\caption{Numerically evaluated non-relativistic Gaunt factor [Eq.~\eqref{eq:dsig_domega_K}] for $Z=1$. The results for $Z>1$ can be constructed using simple variable mapping, as given in Eq.~\eqref{eq:prime}.}
\label{fig:gaunt_table}
\end{figure}

\vspace{-2mm}
\section{Results for the Gaunt factor}
\label{sec:GF_numerical}
In this section, we present our results for the BR emission Gaunt factor, illustrating its main bahavior. We also determine the range of applicability of the various approximations, focusing of low-charge ions ($Z\leq 10$). In particular, the cases $Z=1$ and $2$ (hydrogen and doubly-ionized helium) are of relevance to us, as these ions are the most common in the early Universe. To present and store the results it is convenient to use $w=\omega/[\gamma_1-1]$ as the frequency variable, implying $w\leq 1$. Indeed, this is one of the benefits of working with the emission Gaunt factor instead of the absorption Gaunt factor, as the $w$ is bounded from above.

\subsection{Non-relativistic approximation}
In Fig.~\ref{fig:gaunt_Z_1}, we illustrate the non-relativistic Gaunt factor for ionic charge $Z=1$. At low photon energies, the simple asymptotic dependence given by Eq.~\eqref{eq:NR_approx_more} is observed. At $p_1\simeq 10^{-1}$, the non-relativistic approximation becomes inaccurate as we will see more quantitatively below (cases with dashed lines in the figure). 
The Gaunt factors for $Z>1$ and $p_1\leq 10^{-2}\,Z$ show similar characteristics as those for $Z=1$. They can be obtained by simple mapping of variables [Eq.~\eqref{eq:prime}], which essentially leads to $p_1\rightarrow p_1/Z$ and $\omega\rightarrow \omega/Z^2$ to leading order. For larger values of $Z$, additional corrections become important.
A wider parameter range is covered in Fig.~\eqref{fig:gaunt_table}, for further illustration.

In Eq.~\eqref{eq:dsig_domega_K} we used $\eta_i=\alpha Z \gamma_i/p_i$ instead of $\eta_i^{\rm KL}=\alpha Z /p_i$ given in \citet{Karzas1961}. This choice is motivated by the expression of EH, which also depend this modified variable. The main difference is that at $p_1\gtrsim 1$ (i.e., $\eta_i\rightarrow 0$) the KL expression yields a Gaunt factor that asymptotes to a constant shape. In the thermally-averaged Gaunt factor this leads to a significant drop at high photon energies as we will see below (cf., Fig.~\ref{fig:Gaunt_NR_KL_therm_Z1}). This drop is not seen for the EH result and the change of variables indeed reduces the departures. From our precomputed tables, the KL case can be obtained by simply replacing $p_i\rightarrow p_i/\gamma_i$ in the evaluation. However, in the following discussion we shall use the modified version of the NR expression, as the main conclusions do not change.

\begin{figure}
\includegraphics[width=\columnwidth ]{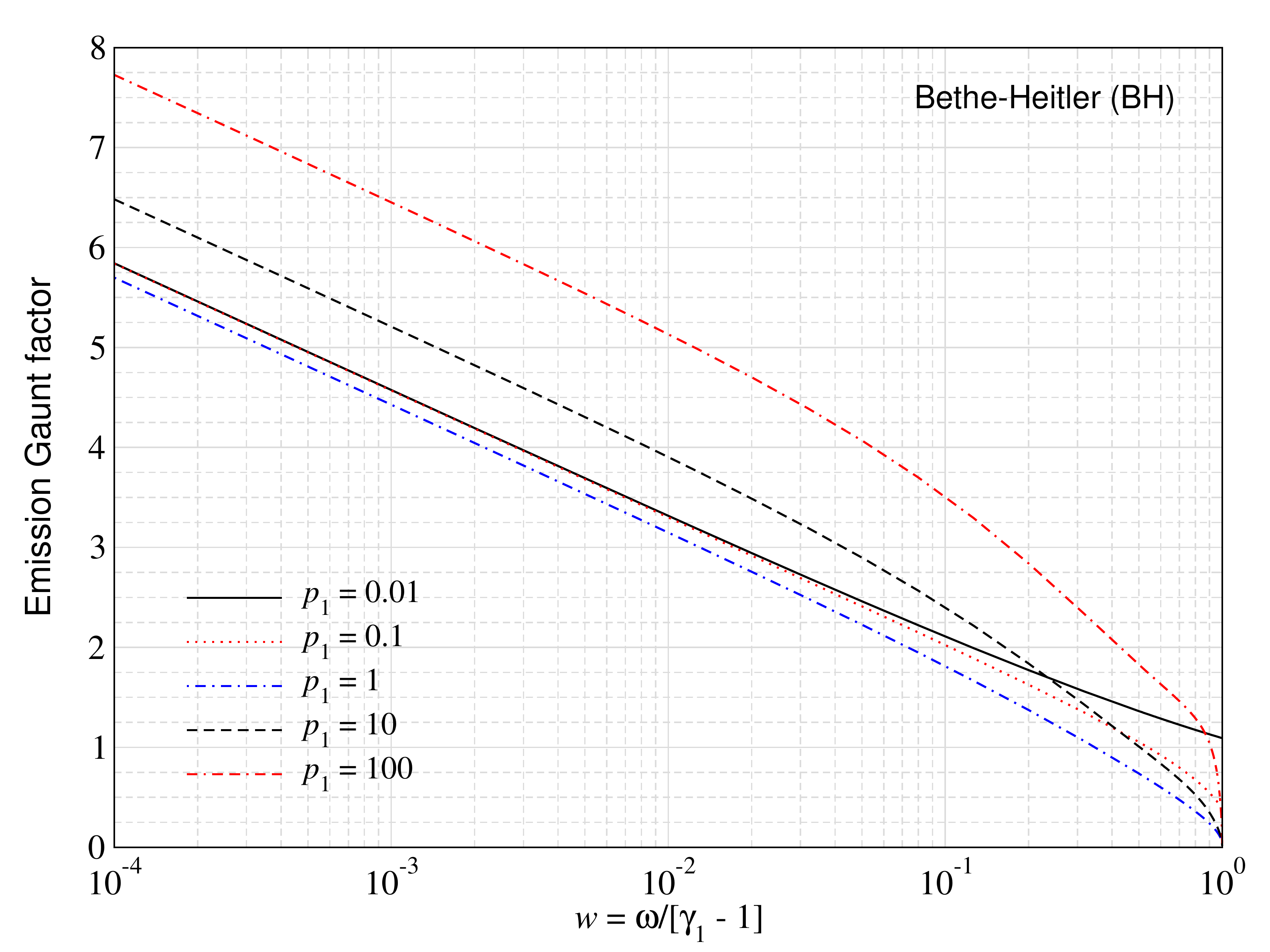}
\caption{Relativistic Gaunt factor using the Bethe-Heitler approximation with the Elwert factor, Eq.~\eqref{eq:Gaunt_BH_EW}. In addition to scaling the total cross section by the standard Kramers formula, we also scaled out a factor of $\gamma_1^2$ (see Sect.~\ref{sec:BH_sigma} for discussion). The Gaunt factor does not change significantly for electron momenta $p_1\lesssim 10^{-2}$ and indeed is inapplicable in that regime.}
\label{fig:gaunt_BH}
\end{figure}

\subsection{Relativistic Bethe-Heitler approximation}
In Fig.~\ref{fig:gaunt_BH}, we illustrate the Gaunt factor in the relativistic regime using the Bethe-Heitler approximation with the Elwert factor, Eq.~\eqref{eq:Gaunt_BH_EW}. We scaled out a factor of $\gamma_1^2$ (see discussion in Sect.~\ref{sec:BH_sigma}) to moderate the Gaunt factor variations. As we will see below, this modification also reduces the dynamic range for the thermally-averaged Gaunt factor at high frequencies (compare Fig.~\ref{fig:Gaunt_BH_therm_Z1} and \ref{fig:Gaunt_BH_therm_Z1_mod}). At low electron momenta ($p_1\lesssim 10^{-2}$) the Bethe-Heitler formula overestimates the Gaunt factor significantly. The BH formula also does not explicitly depend on the charge $Z$ and thus is unable to capture Coulomb corrections that become important for larger values of $Z$ and for low values of $p_1$ \citep[e.g.,][]{ElwertHaug1969}.

\begin{figure}
\includegraphics[width=\columnwidth ]{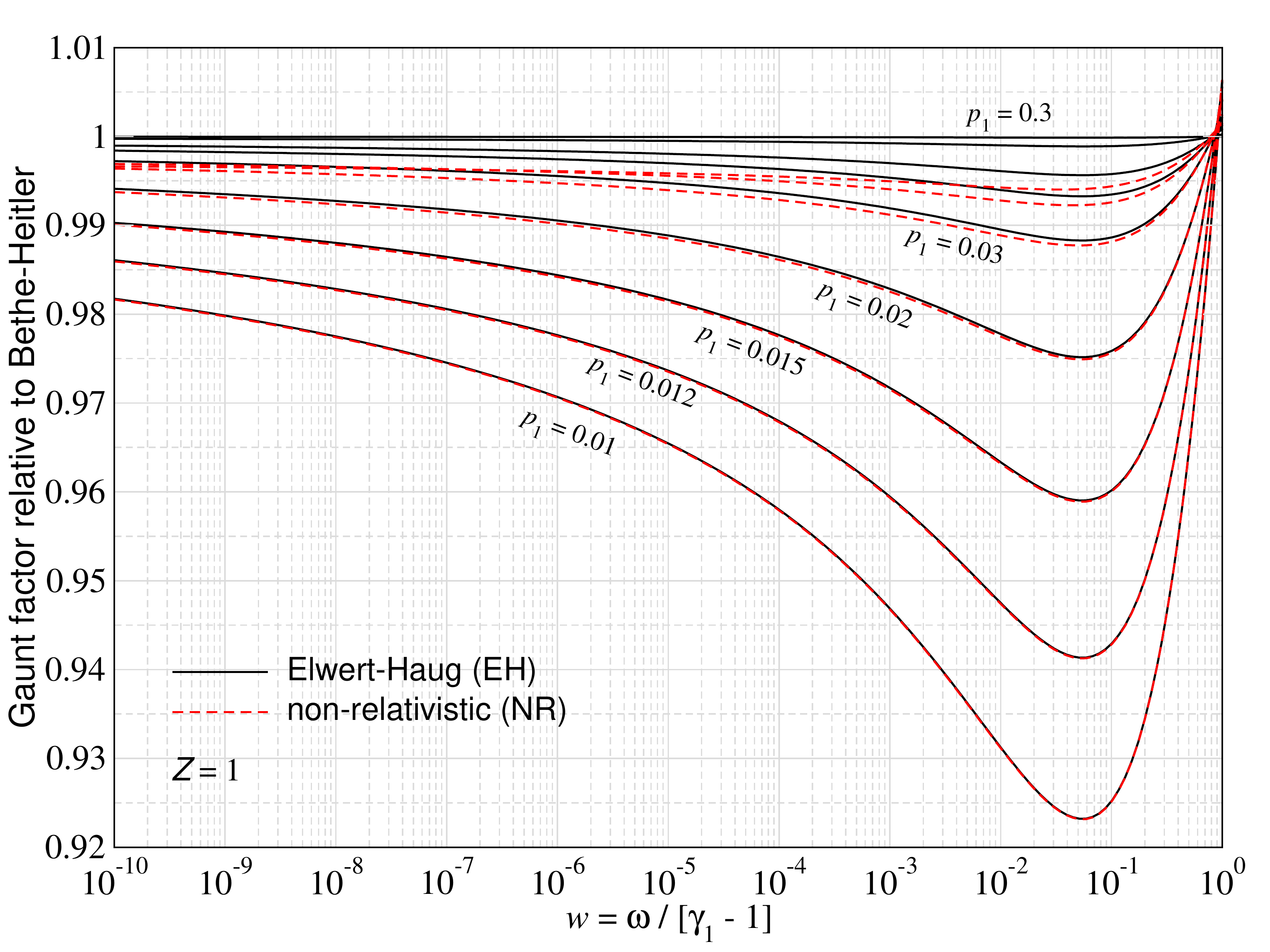}
\\[-0.5mm]
\includegraphics[width=\columnwidth ]{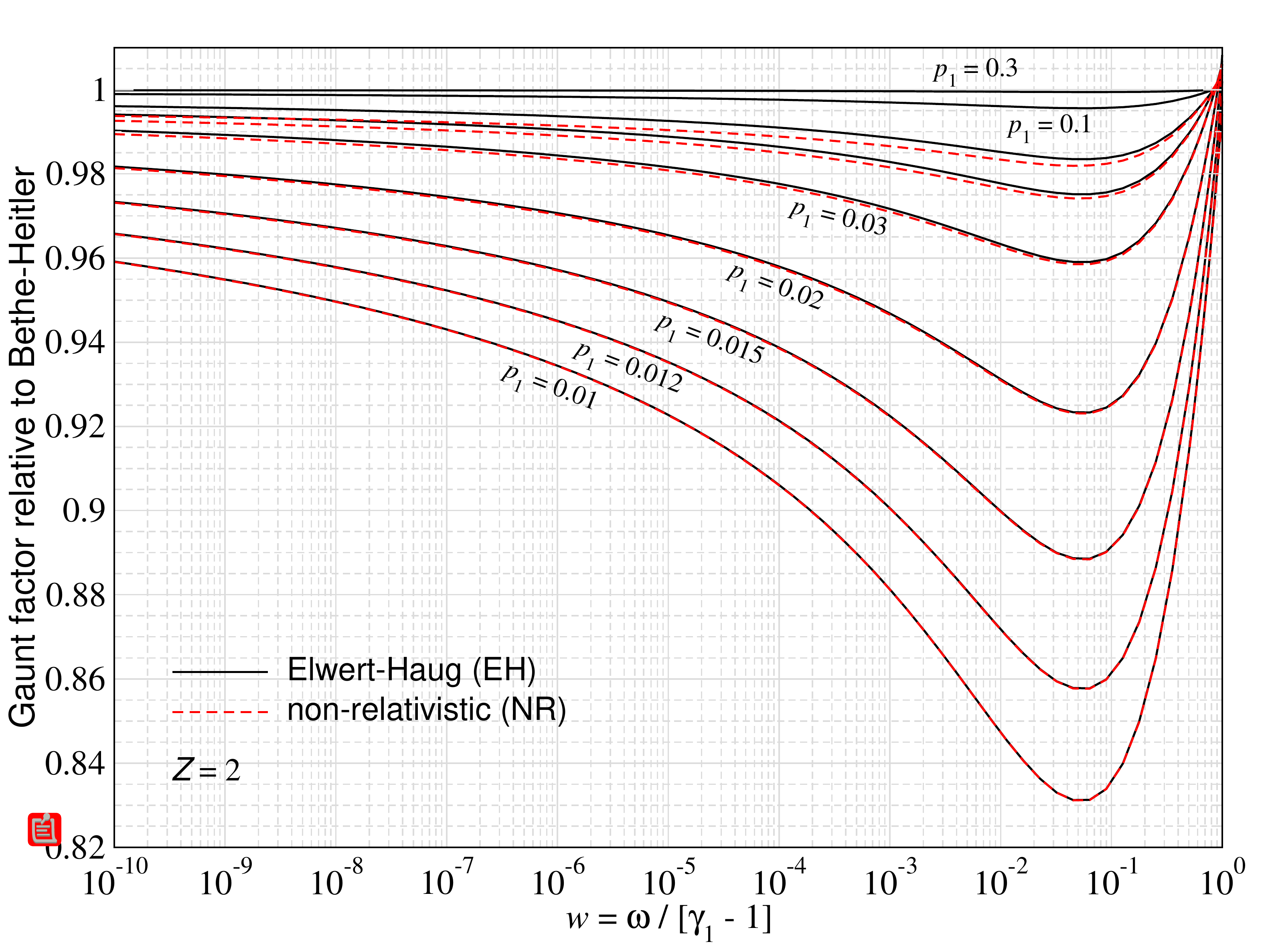}
\caption{Comparison of the Elwert-Haug and non-relativistic Gaunt factors with the Bethe-Heitler formula for $Z=1$ (upper panel) and $2$ (lower panel). At $p_1\simeq 0.01$ the NR and EH Gaunt factors agree extremely well while for $p_1\gtrsim 0.03$ the NR Gaunt factor underestimates the EH result. The EH formula converges towards the BH approximation at $p_1\gtrsim 0.2-0.3$.}
\label{fig:gaunt_EH}
\end{figure}

\vspace{-0mm}
\subsection{Intermediate regime and domains of validity of the various approximations}
\label{sec:EH}
For large electron momenta, the EH cross section asymptotes towards the BH formula as long as $Z$ is not too large (i.e., $Z\lesssim 10$). To illustrate the Gaunt factor based on the expressions given by EH, it is thus convenient to compare the values directly to the BH formula.
 In Fig.~\ref{fig:gaunt_EH}, we present the results for various electron momenta $p_1=\{0.01,\, 0.012,\, 0.015,\,0.02, \,0.03, \,0.04, \,0.05, \,0.1, \,0.3\}$. Note that some of the curves are not labeled explicitly and that for the NR case only those for $p_1\leq 0.05$ are presented as the others significantly underestimate the EH result. 
 
 For $Z=1$, the departures of the EH Gaunt factor from the BH approximation are smaller than $8\%$ for the chosen $p_1$ values . Even the NR formula works very well up to\footnote{This conclusion is not changed significantly when using the original version for the non-relativistic Gaunt factor with $\eta_i=\alpha Z/p_i$.} $p_1 \simeq 0.05$. As expected, the EH formula approaches the BH cross section at $p_1\gtrsim 0.2-0.3$, corresponding to kinetic energies in excess of $E_1\simeq 10\,{\rm keV}$. For charge $Z=2$ (lower panel of Fig.~\ref{fig:gaunt_EH}), similar trends can be observed, however, the departures from the BH formula generally are bigger for fixed $p_1$ since $p^*_1\approx p_1/Z$ reduces. This means that for larger values of $Z$, the Gaunt factor remains closer to the NR case up to larger values of $p_1$, i.e., $g_{\rm EH}(p_1, \omega, Z)\approx g_{\rm EH}(p_1/Z, \omega/Z^2, Z=1)$. 
 
\begin{figure}
\includegraphics[width=1.\columnwidth]{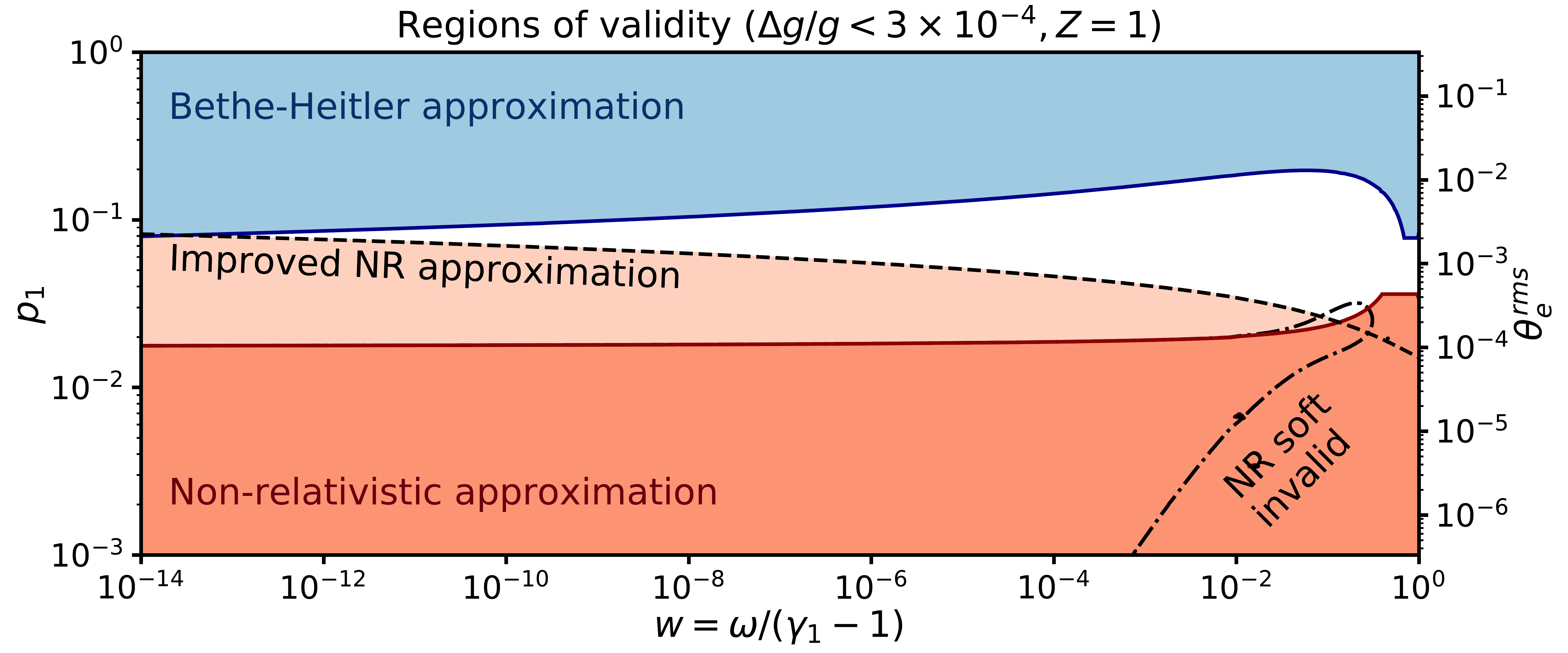}
\\[2mm]
\includegraphics[width=1.\columnwidth]{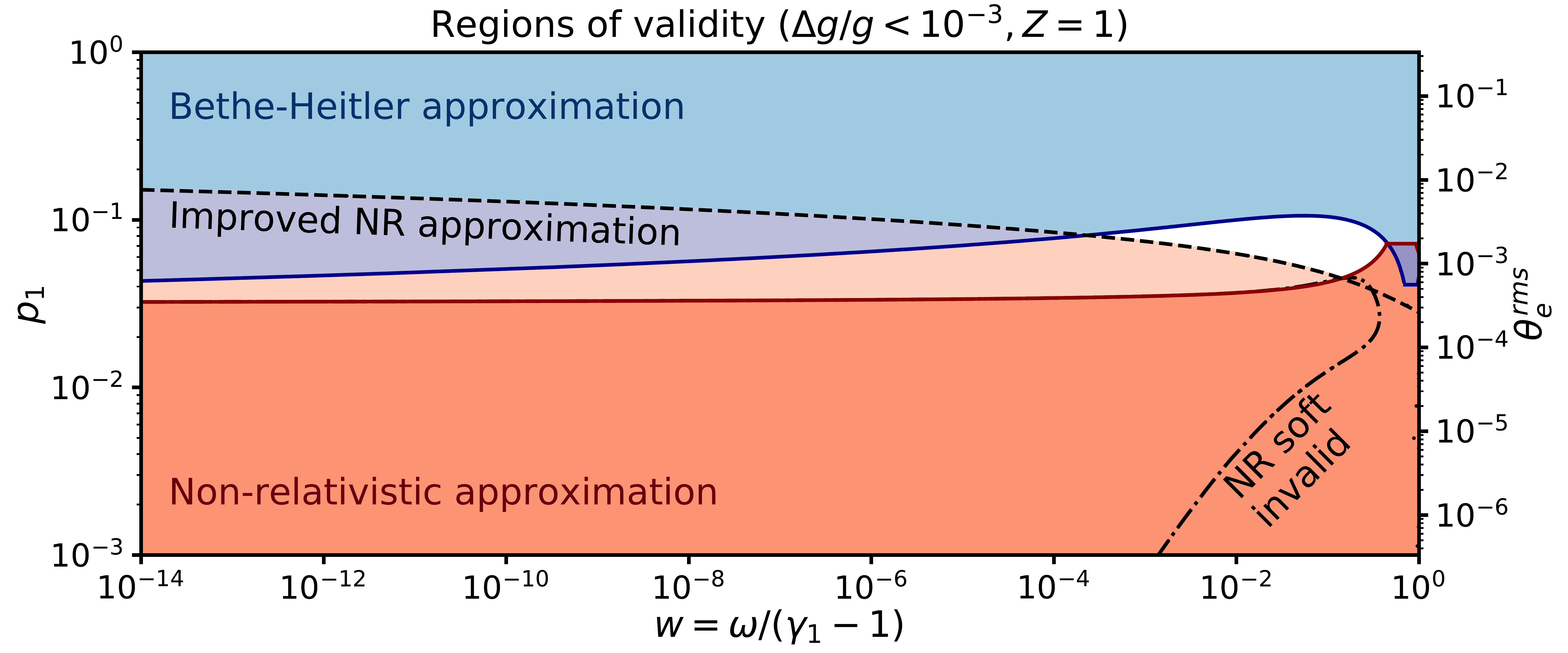}
\\[2mm]
\includegraphics[width=1.\columnwidth]{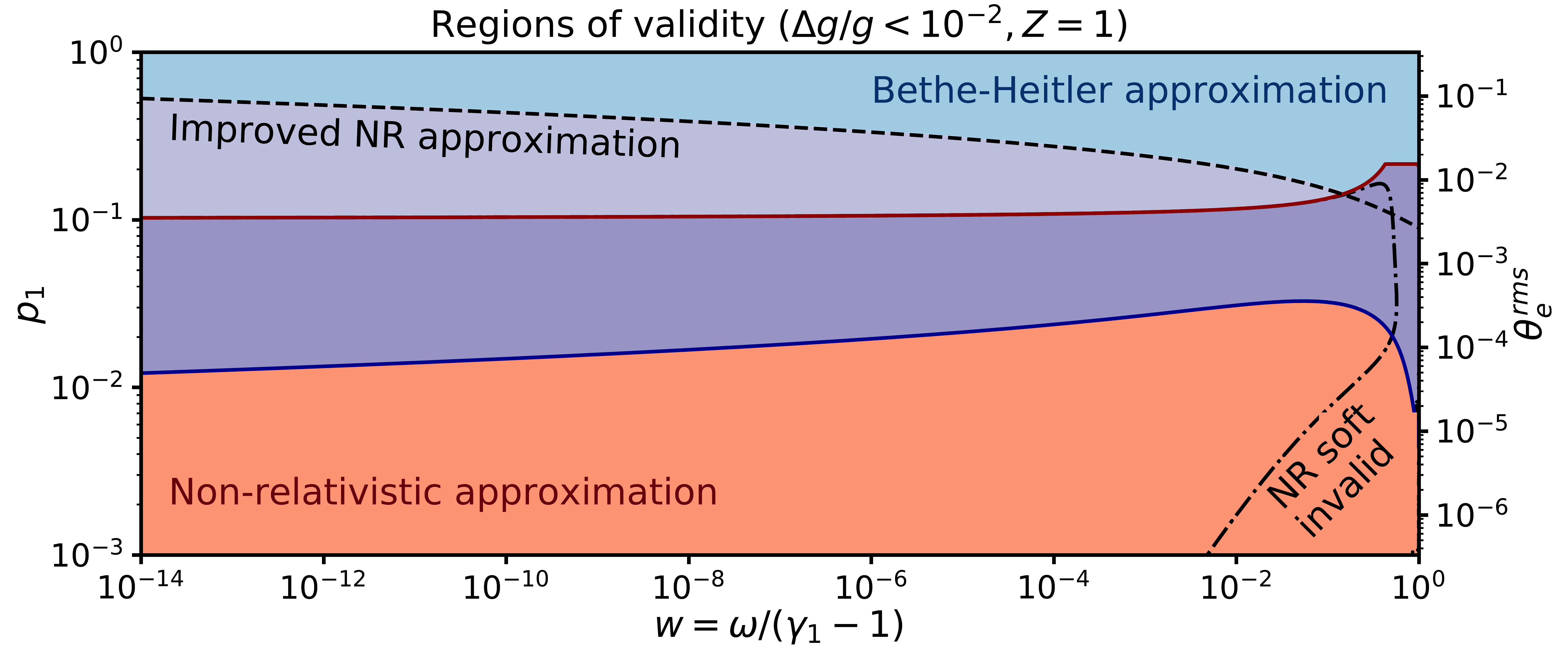}
\caption{Regions in which the NR approximations and BH formula are valid. 
The colored areas are the $(p_1,w)$ sub-spaces where the relative difference between the EH Gaunt factor and the labeled formula is less than 0.3\textperthousand~(Top panel), 1\textperthousand~(Middle panel), or 1\%~(Bottom panel).
The white area is where the calculation using the EH Gaunt factor is required to achieve the required precision. The improved NR approximation significantly extends the reach of the NR expression. The root-mean-square temperature $\The^{\rm rms} =k\Te^{\rm rms}/\me c^2= p_1^2/3$ is shows for comparison.}
\label{fig:comp_formulas_Z1}
\end{figure}
To more quantitatively assess the validity of various approximations we ran comparisons for the Gaunt factors asking when they depart by more than a fixed relative precision from the EH calculation. The results of this comparison for $Z=1$ are summarized in Fig.~\ref{fig:comp_formulas_Z1}. 
As expected, the Bethe-Heitler formula correctly approximates the cross section in the relativistic regime (blue region), namely above $p_1 \gtrsim 0.05-0.15$ if we require a maximum 1{\textperthousand} deviation (middle panel).
Relaxing this requirement (bottom panel), the region in which BH is valid overlaps with the region, in red, where the non-relativistic (NR) approximation is applicable, i.e., the two approximation depart from each other by less than 2\%.
For clarity we mention that in the purple areas mark the overlap of the red (NR) and blue (BH) regions.
Our improved NR approximation, i.e., Eq.~\eqref{eq:g_NR_rel}, which takes into account the leading order  relativistic correction, significantly enlarges the applicability of the NR formula (orange area). We also highlight that the NR soft photon approximation, Eq.~\eqref{eq:NR_approx_more}, works extremely well below and to the left of the dot dashed line; for higher photon energies the respective full expression has to be evaluated.

The presence of regimes in which neither the NR limit nor the BH approximations are valid (white areas in Fig.~\ref{fig:comp_formulas_Z1}) makes it clear that any precise calculation involving bremsstrahlung processes needs to carefully assess when the simplified expressions can be used, and eventually resort to the EH cross section evaluation in the intermediate regime. It is however impressive that for $Z=1$ at $1\%$ precision only the BH and NR expressions are needed and a simple switch at $p_1\simeq 0.05$ should suffice when combining these two. With {\tt BRpack} all cases can be efficiently modeled using one function evaluation with appropriate arguments.

\begin{figure}
\includegraphics[width=1.\columnwidth]{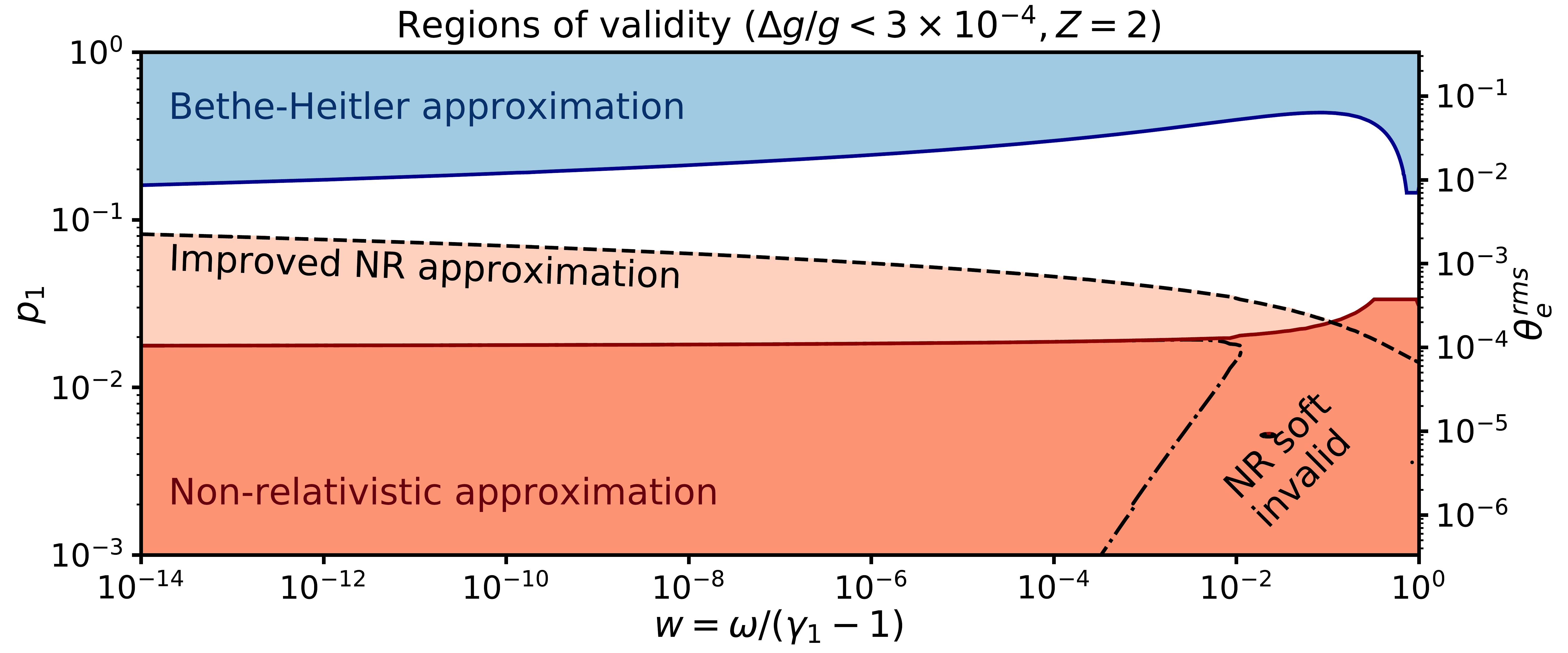}
\\[2mm]
\includegraphics[width=1.\columnwidth]{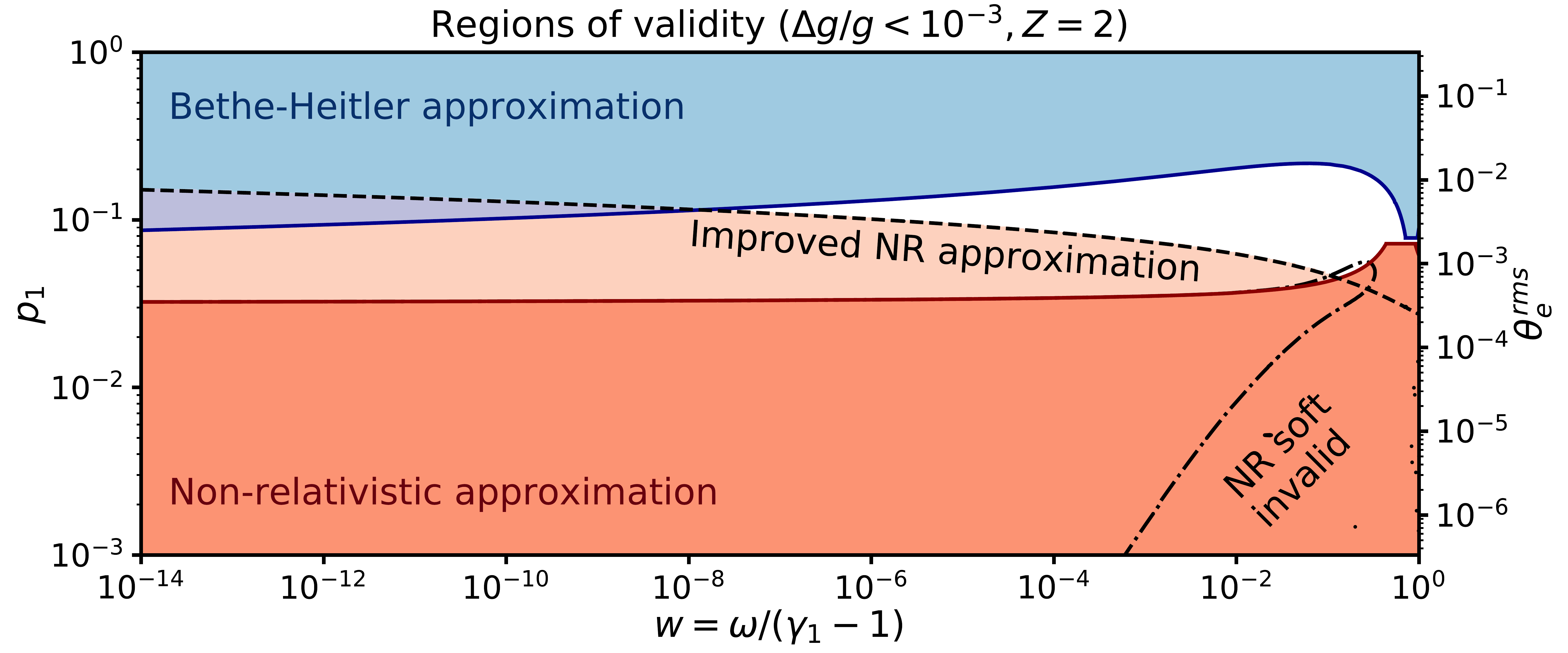}
\\[2mm]
\includegraphics[width=1.\columnwidth ]{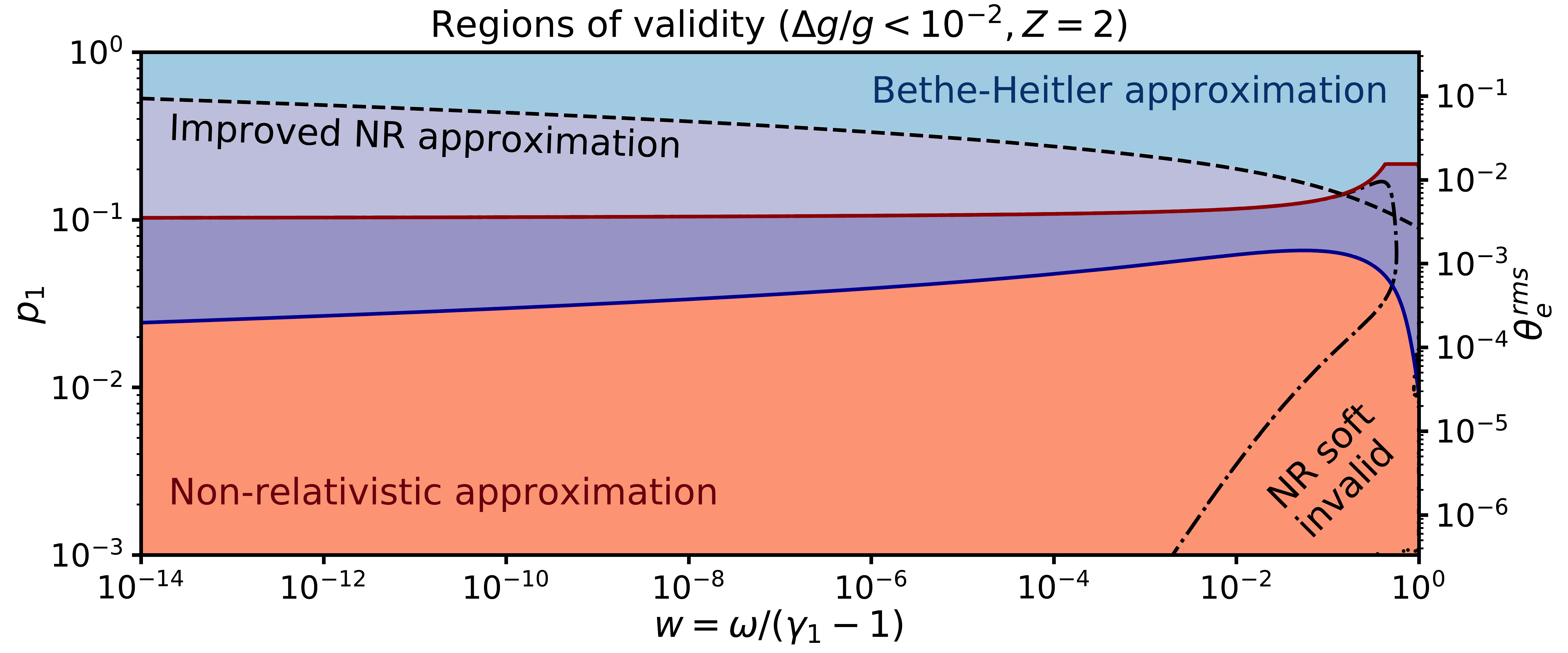}
\caption{Same as in Fig.~\ref{fig:comp_formulas_Z1} but for $Z=2$. The areas requiring the full EH evaluation slightly increased. Also, in comparion to $Z=1$, the boundary of the NR formula is shifted downward by roughly a factor of 2.}
\label{fig:comp_formulas_Z2}
\end{figure}
For $Z=2$, we reach similar conclusions as for $Z=1$ (Fig.~\ref{fig:comp_formulas_Z2}). The regions requiring the full EH evaluation slightly increase given the importance of terms $\propto \alpha Z$. Overall, the boundary of the NR approximation shifts roughly by a factor of $2$, which is expected from $g_{\rm EH}(p_1, \omega, Z)\approx g_{\rm EH}(p_1/Z, \omega/Z^2, Z=1)$. Again at $1\%$ precision the BH and NR formulae are sufficient for representing the intermediate Gaunt factor, while at $\lesssim 0.1\%$ the EH result is needed.
With {\tt BRpack} all cases can be considered and compared.

\subsection{Gaunt factors for $2<Z\leq 10$}
We have seen that for $Z=1$ and $2$, the departures from the EH calculation only become visible at the $\lesssim 0.1\%$ level (Fig.~\ref{fig:comp_formulas_Z1} and \ref{fig:comp_formulas_Z2}). For larger ion charge, corrections become increasingly important ultimately exceeding the $1\%$ level. Here we restrict our discussion to cases with $Z\leq 10$ as higher order Coulomb corrections are expected to become relevant for larger ion charge \citep{Roche1972, Haug2008, Haug2010}, a problem that we leave to future work.
 
\begin{figure}
\includegraphics[width=1.\columnwidth]{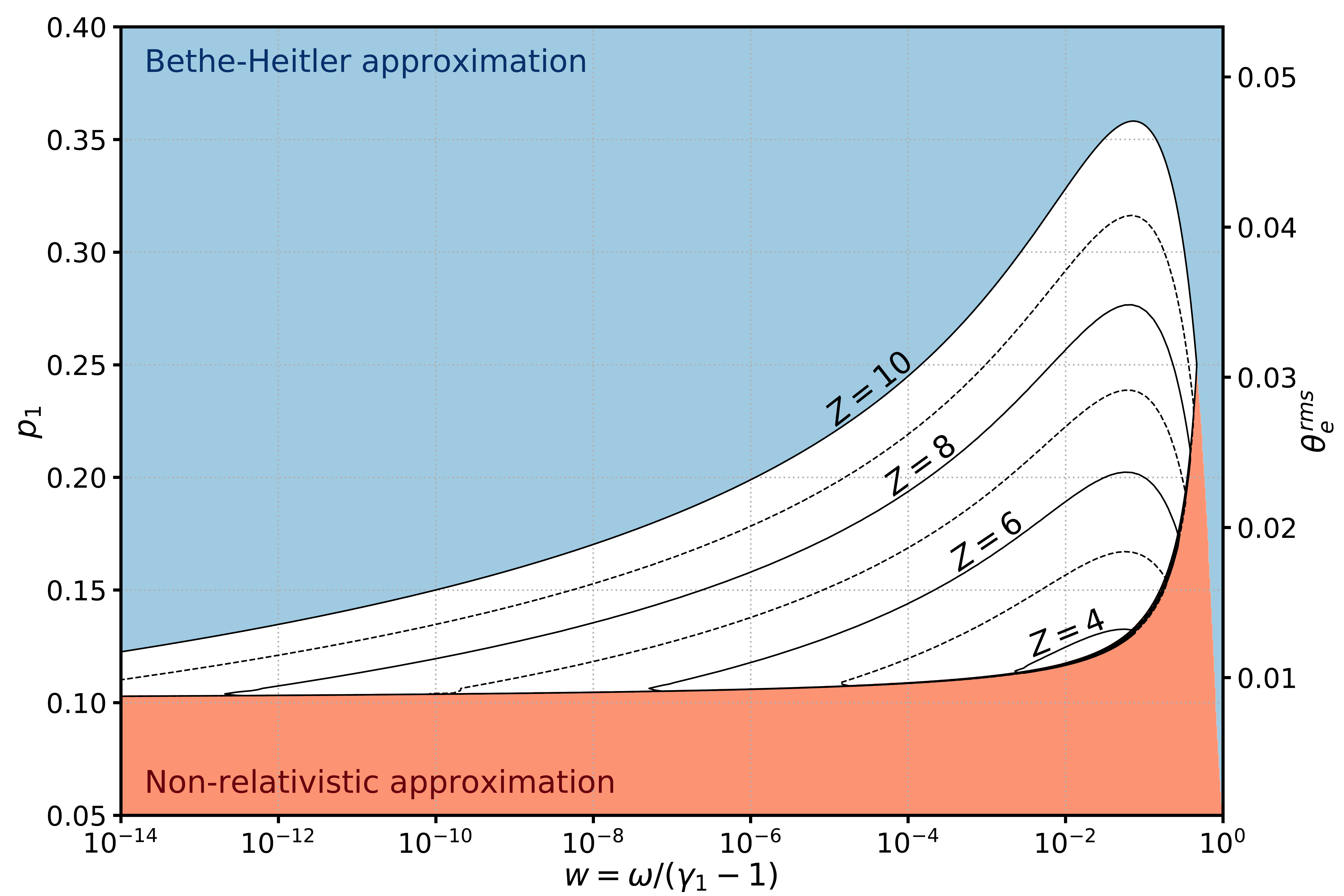}
\caption{Regions requiring EH evaluation to achieve 1\% accuracy for ion charges $Z=4-10$. {\tt BRpack} allows representing the Gaunt factor over the whole domain. The colored regions, where approximations can be safely used, refer to the case for $Z=10$.}
\label{fig:comp_formulas_Z1_10}
\end{figure}
In Fig.~\ref{fig:comp_formulas_Z1_10}, we show the domains in which the full EH Gaunt factor evaluation is required to achieve $1\%$ precision. For $Z=3$ we find that a combination of the BH and NR Gaunt factors remains sufficient at this precision, but for $Z=4$ a small domain requiring the EH evaluation appears.
As expected, this domain grows with increasing charge $Z$. For $Z=10$, one expects $\simeq 1\%$ corrections over a significant range of photon energies at $p_1\simeq 0.2$, corresponding to rms temperature $\Te \simeq 10^8\,{\rm K} \,(\The\simeq 0.02)$.

\begin{figure}
\includegraphics[width=1.\columnwidth]{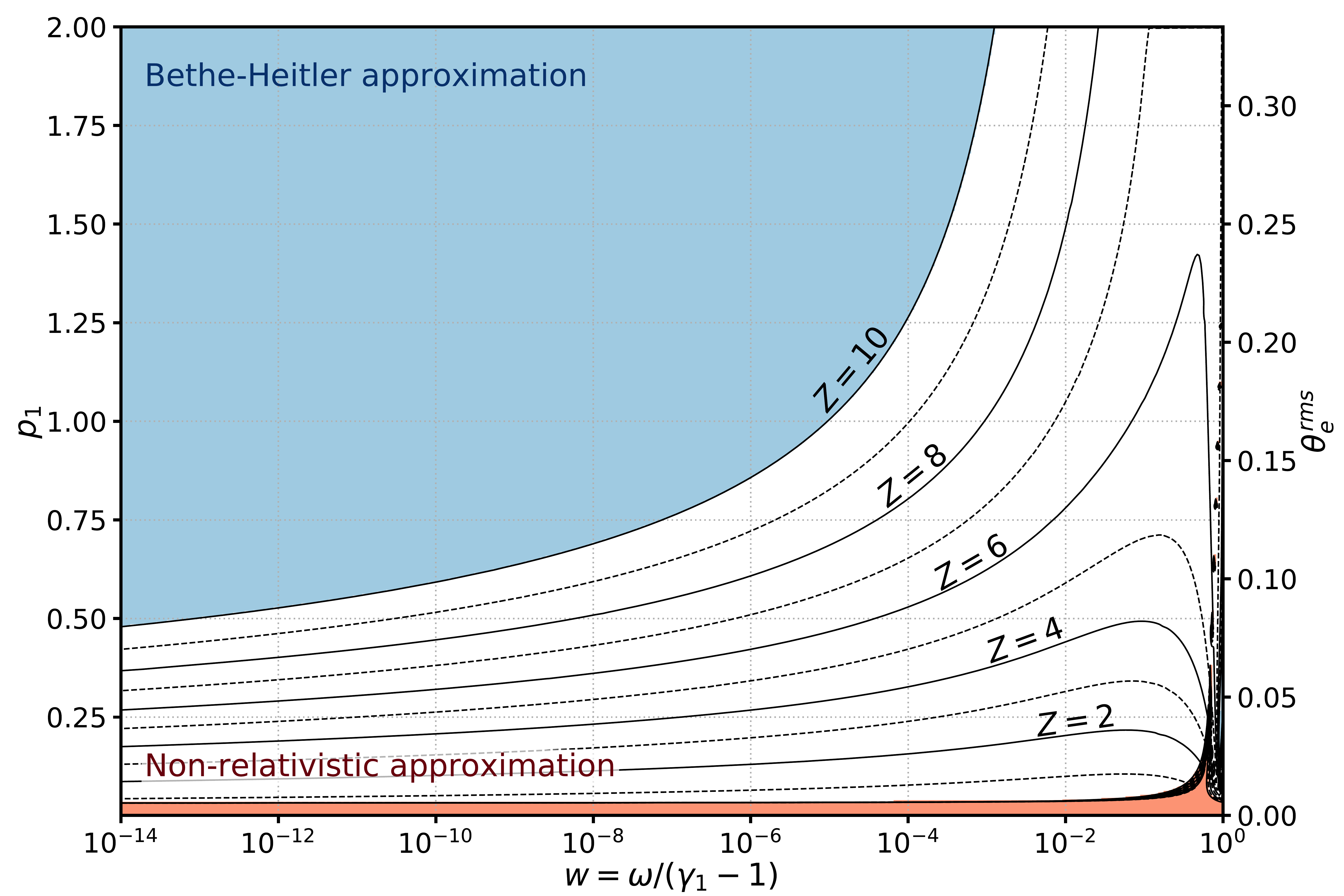}
\caption{Same as Fig.~\ref{fig:comp_formulas_Z1_10} but for precision $0.1\%$. At this precision, the BH formula is inaccurate for $Z\geq 7$ and $10^{-3}\lesssim w \leq 1$. The non-relativistic region is quite narrow (orange region) for the considered temperatures.}
\label{fig:comp_formulas_Z1_10_0.001}
\end{figure}
\begin{figure}
\includegraphics[width=1.\columnwidth]{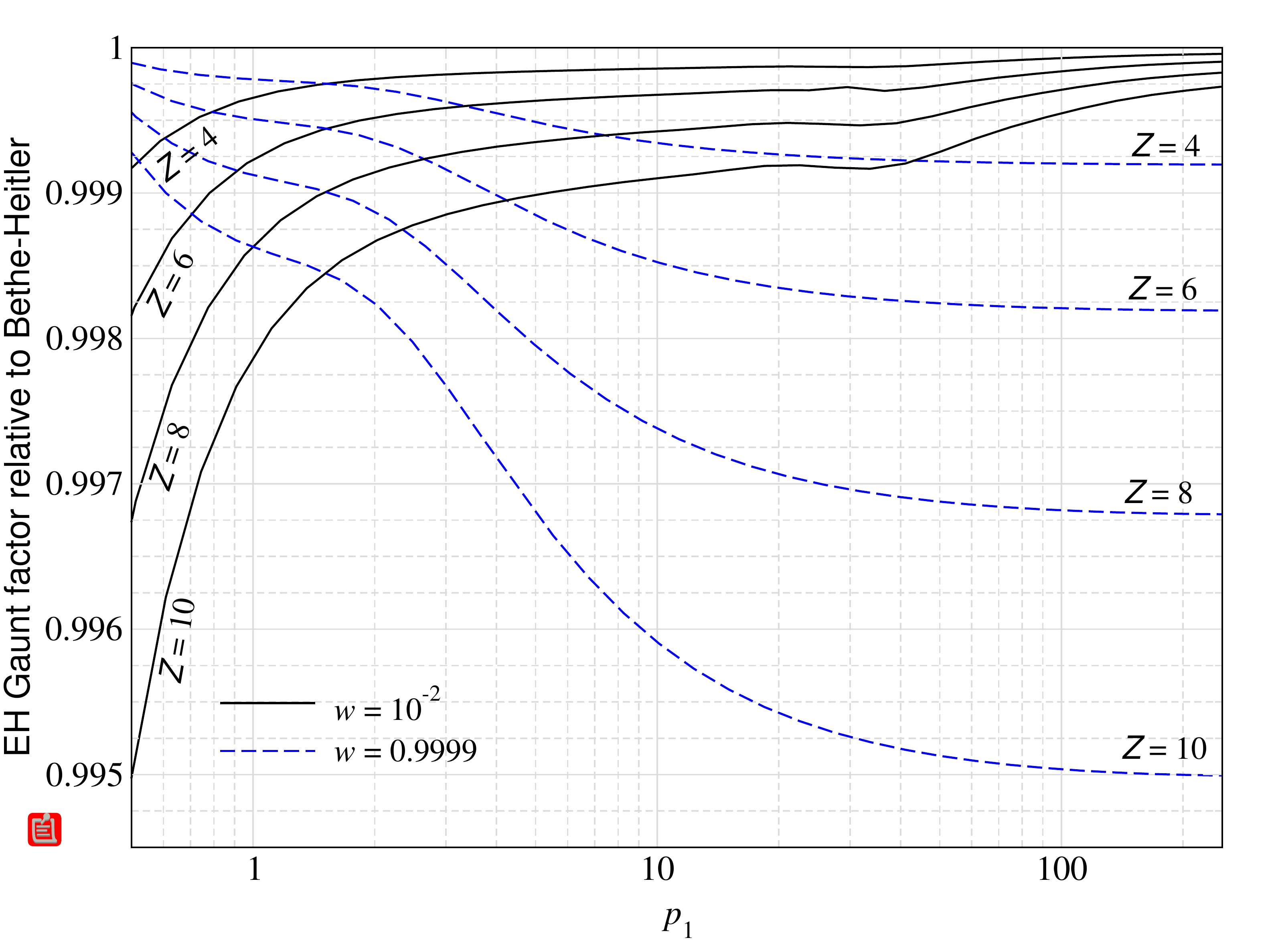}
\caption{EH Gaunt factor relative to BH formula for $w=10^{-2}$ and $w=1$ as a function of $p_1$ and varying values of $Z$. At low photon energies, the EH expression clearly approached the BH formula when increasing $p_1$, while in the short-wavelength limit departures from the BH formula remain visible in the shown range of $p_1$.}
\label{fig:High_p1}
\end{figure}

When tightening the precision requirement to $0.1\%$, we obtain the domains shown in Fig.~\ref{fig:comp_formulas_Z1_10_0.001}. We only computed the EH Gaunt factor up to $p_1=2$ (kinetic energy $\simeq 600\,{\rm keV}$), finding that for $Z\geq 7$ and $10^{-3}\lesssim w \leq 1$ the BH formula is inaccurate. Since for higher kinetic energies, additional corrections become important \citep[e.g.,][]{Haug2010}, we limited our tables to $p_1\leq 2$.
For accurate and efficient representation of the EH Gaunt factor, {\tt BRpack} can be used at $\simeq 0.01\%$ numerical accuracy up to $p_1=2$. Above this value of $p_1$, we resort to the BH formula. This causes inaccuracies in the high frequency tail of the thermally-averaged Gaunt factor, as we explain below. However, the differences are limited to $\lesssim 0.1\%$ for $Z\leq 4$, and remain smaller than $0.5\%$ even for $Z\leq 10$ (see Fig.~\ref{fig:High_p1}).

\subsection{High electron momenta}
\label{sec:large_e}
In our discussion, we only considered cases up to electron momenta $p_1=2$. For $Z=10$, this already revealed that the BH formula is inaccurate at the level $\simeq 0.1\%$ in the short-wavelength limit. Using the EH cross section, we can explore this aspect a little further. In Fig.~\ref{fig:High_p1}, we illustrate the departures of the EH Gaunt factor from the BH formula for $w=10^{-2}$ and $w=1$ and several values of $Z$. For $Z=4$, this shows that even up to very high electron momentum, the EH Gaunt factor does not depart by more than $0.1\%$ from the BH formula. This statement extends to the cases $Z<4$. {\tt BRpack}, which only contains tables up to $p_1=2$, thus represents the EH Gaunt factor to better than $\lesssim 0.1\%$ for $Z\leq 4$. For $Z\leq 2$, even a precision $\lesssim 0.03\%$ can be guaranteed.

At $Z>4$, the departures of the EH Gaunt factor from the BH formula exceed the level of $0.1\%$ in the short-wavelength limit ($\omega \simeq \gamma_1-1$). 
For $Z>4$ and $w\simeq 1$, {\tt BRpack} thus does not reproduce the EH Gaunt factor at $p_1>2$ beyond the $\simeq 0.5\%$ level. This causes inaccuracies in the thermally-average EH Gaunt factor at very high photon energies (see next Section). 
At lower values of $w$, the BH limit is again approached, with departures $\lesssim 0.15\%$ at $p_1>2$, $w\leq 10^{-2}$ and $Z\leq 10$.
Therefore the low frequency tail of the EH Gaunt factor should be reproduced to high precision.

We emphasize again that with {\tt BRpack} we did not attempt to represent the EH Gaunt factor at $p_1>2$ more rigorously as it is clear that other corrections will also become relevant there. 
However, at those energies, the total number of emitted photon is exponentially small, such that this should not cause any major limitations for most applications.

\vspace{-0mm}
\section{Thermally-averaged Gaunt Factors}
\label{sec:therm_results}
Describing the interactions of photons and electrons in the general case is quite complicated. However, for many astrophysical applications, one can neglect anisotropies in the medium (at least locally) and simply describe the evolution of the average electron and photon distribution functions. Coulomb interactions further drive the electron distribution quickly towards a relativistic Maxwell-Boltzmann distribution function [see Eq.~\eqref{eq:f_rMB} below]. Electron-ion degeneracy effects can furthermore be neglected (but can be easily added) unless temperatures in excess of the pair-production threshold are being considered. In particular for the evolution of CMB spectral distortions, the above conditions are the most relevant \citep[e.g.,][]{Chluba2011therm, Chluba2014, Lucca2019}.

\subsection{Average BR emissivity in the Kramers limit}
\label{sec:Kramers_therm}
To define the thermally-averaged Gaunt factors we first introduce the averaged BR photon production term of the plasma:
\begin{align}
\label{eq:dN_dt}
\left.\frac{\id N_\gamma}{\id t \id \omega} \right|_{\rm em}
&=N_{\rm i}  \Ne\, \int p_1^2 f(p_1) \, |\varv_{\rm rel}| \frac{\id \sigma(\omega, p_1)}{\id\omega}\id p_1.
\end{align}
Here, $N_{\rm i}$ the ion number density of change $Z$; $\Ne$ the electron number density corresponding to the electron momentum distribution function, $f(p_1)$, which we normalized as $\int p_1^2 f(p_1)\id p_1=1$. The relative speed of the colliding particles is further given by $|\varv_{\rm rel}|=c p_1/\gamma_1$, which becomes $|\varv_{\rm rel}|\approx c p_1$ in the non-relativistic limit. Equation~\eqref{eq:dN_dt} assumed that the ions are at rest (i.e., recoil effects due to the finite mass of the nucleus can be neglected) and that the momentum distribution of the electrons is described in this frame. Inserting the Kramers cross section, Eq.~\eqref{eq:dsig_domega_K}, into Eq.~\eqref{eq:dN_dt} the BR emissivity in the classical limit then reads
\begin{align}
\label{eq:dN_dt_K}
\left.\frac{\id N_\gamma}{\id t \id \omega} \right|^{\rm K}_{\rm em}
&\approx\frac{2\alpha Z^2\,}{\sqrt{3}} \frac{\Ne N_{\rm i} \sigT c}{\omega} 
\int_{p_{\rm min}}^\infty p_1 f(p_1) \id p_1 
\nonumber \\[1.5mm]
&\approx \frac{2\sqrt{2}\alpha Z^2\,}{\sqrt{3\pi \The}} \frac{\Ne N_{\rm i} \sigT c}{\omega} \,\expf{-\omega/\The},
\end{align}
where $p_{\rm min}$ is the minimal electron momentum that is required to produce a photon of energy $\omega=h\nu/\me c^2$. This is determined by $\omega = \gamma_{\rm min}-1$, which yields $p_{\rm min}=\sqrt{\omega(2+\omega)}\approx \sqrt{2\omega}$. In the last step of Eq.~\eqref{eq:dN_dt_K} we used a non-relativistic Maxwell-Boltzmann distribution function, $f_{\rm nr}(p)=\sqrt{2/\pi}\,\The^{-3/2}\,\expf{-p^2/2\The}$ with dimensionless electron temperature $\The=k\Te/\me c^2$ to carry out the integral.

The expressions above explicitly assume $p_1,\omega \ll 1$. Even without quantum corrections, to generalize the Kramers approximation to higher temperatures / energies, we shall use the relativistic Maxwell-Boltzmann distribution function
\begin{align}
\label{eq:f_rMB}
f_{\rm rMB}(p)
&=\frac{\expf{-\gamma(p)/\The}}{K_2(1/\The)\, \The},
\end{align}
where $K_2(x)$ is the modified Bessel function of second kind\footnote{The relativistic Maxwell-Boltzmann distribution is defined with the normalization $\int p^2 f_{\rm rMB}(p)\id p =1$.}.
We also keep the full relativistic expression for $|\varv_{\rm rel}|=c p_1/\gamma_1$ and furthermore realize that at low frequencies the overall BR cross section scales as $\simeq \gamma_1^2/[p_1^2\omega]$ towards higher electron energies (see discussion about Bethe-Heitler limit). Thus, after multiplying the Kramers approximation, Eq.~\eqref{eq:dsig_domega_K}, by $\gamma_1^2$ and carrying out the thermal average with the above modification, we have the relativistically-improved Kramers approximation for the BR emissivity:
\bsub
\label{eq:dN_dt_K_rel}
\begin{align}
\left.\frac{\id N_\gamma}{\id t \id \omega} \right|^{\rm K, rel}_{\rm em}
&=\frac{2\sqrt{2}\alpha Z^2\,}{\sqrt{3\pi \The}} \frac{\Ne N_{\rm i} \sigT c}{\omega} \,\expf{-\omega/\The} \,\mathcal{I}(\omega, \The)
 \\[2mm]
\label{eq:dN_dt_K_rel_b}
\mathcal{I}(\omega, \The) &=\sqrt{\frac{\pi\The}{2}} \,\expf{\omega/\The} \,\int_{p_{\rm min}}^\infty p_1 \gamma_1 f(p_1)\id p_1 
\nonumber\\[1.5mm]
&=\sqrt{\frac{\pi\The}{2}}\,\frac{\expf{-1/\The}}{K_2(1/\The)} \, \left[(1+\omega)^2 + 2\The (1+\omega) + 2\The^2\right]
\nonumber\\[1.5mm]
&\approx (1+\omega)^2 \left[1+\left(\frac{2}{1+\omega}-\frac{15}{8}\right)\The \right].
\end{align}
\esub
This shows that in the classical treatment the improved asymptotic scales as $\propto (1+\omega)^2\,\expf{-\omega/\The}$ for low temperatures. The origin of this correction is not quantum-mechanical but simply due to special relativistic effects. This modification absorbs the leading order corrections towards the BH limit, as we discuss next.
We also mention that in Eq.~\eqref{eq:dN_dt_K_rel} the temperature-dependent factor,
\begin{align}
\label{eq:norm_ratio}
\mathcal{R}(\The)
&=\sqrt{\frac{\pi\The}{2}}\,\frac{\expf{-1/\The}}{K_2(1/\The)}
\equiv \frac{\int_0^\infty p^2 \expf{-p^2/2\The}\id p }{\int_0^\infty p^2 \expf{-(\gamma-1)/\The} \id p }
\nonumber\\[1mm]
&\approx 1-\frac{15}{8}\The +\frac{345}{128}\The^2-\frac{3285}{1024}\The^3,
\end{align}
is directly related to the differences in the normalization of the non-relativistic and relativistic Maxwell-Boltzmann distribution. At high temperatures, the corrections can become sizable, giving $\mathcal{R}(\The)\simeq 0.98$ at $k\Te = 5\,{\rm keV}$ and $\mathcal{R}(\The)\simeq 0.84$ at $k\Te \approx 50\,{\rm keV}$, and thus should be taken into account for accurate calculations.

\subsection{Definition of thermally-averaged Gaunt factors}
\label{sec:Gen_therm}
The main quantity that enters the BR emission term in the photon Boltzmann equation as well as the electron temperature evolution equation, is the thermally-averaged Gaunt factor. It can be simply obtained by comparing the total plasma emissivity with the emissivity in the Kramers limit and is usually computed as
\begin{align}
\label{gaunt_def_usual}
  \bar{g}(\omega, \The)
  &=\frac{\int_{p_{{\rm min}}}^{\infty}\frac{p_{1}^{3}}{\gamma1}f(p_{1})
  \left.\frac{\id\sigma(\omega, p_{1})}{\id\omega}\right|_{\rm K} g(\omega,p_1)\id p_{1}}
  {\int_{p_{{\rm min}}}^{\infty}\frac{p_{1}^{3}}{\gamma1}f(p_{1})\left.\frac{\mathrm{d}\sigma(\omega,p_{1})}{\id\omega}\right|_{\rm K}\id p_{1}}
  \nonumber \\[1mm]
  &=\frac{\int_{p_{{\rm min}}}^{\infty}\frac{p_1}{\gamma1}f(p_{1}) g(\omega,p_1)\id p_{1}}
  {\int_{p_{{\rm min}}}^{\infty}\frac{p_1}{\gamma1} f(p_{1}) \id p_{1}}
 \nonumber \\[1mm]
 &\equiv
\int_{0}^{\infty}\expf{-\xi} g\left(\omega, p_1=\sqrt{(\omega+\The\xi)(2+\omega+\The\xi)}\right)\id \xi.
\end{align}
In the last step we explicitly assumed that the electrons follow a non-degenerate, relativistic Maxwell-Boltzmann distribution\footnote{We have $\int_{p_{{\rm min}}}^{\infty}\frac{p_1}{\gamma1} f(p_{1}) \id p_{1}=\int_{p_{{\rm min}}}^{\infty}\frac{p_1}{\gamma1} f(p_{1}) \id p_{1}=\expf{-(1+\omega)/\The}/K_2(1/\The)$ in this case, which cancels a corresponding factor from the numerator.}.

In the non-relativistic limit ($\gamma_1\simeq 1$), Eq.~\eqref{gaunt_def_usual} is a very good choice. However, for $p_1 \gtrsim 1$, the Gaunt factor scales $\propto \gamma_1^2$ at low frequencies (see Sect.~\ref{sec:BH_sigma}). It is thus useful to multiply the Kramers cross section by $\gamma_1^2$, which then yields a slightly modified definition for the Bremsstrahlung Gaunt factor:
\begin{align}
\label{gaunt_def_rel_mod}
  \bar{g}^{\rm rel}(\omega, \The) &=\frac{\int_{p_{{\rm min}}}^{\infty} p_1 \gamma_1  f(p_{1}) \, g^{\rm rel}(\omega,p_1)\id p_{1}}
  {\int_{p_{{\rm min}}}^{\infty} p_1 \gamma_1 f(p_{1}) \id p_{1}}
 \nonumber \\
 &=
 \frac{\int_{p_{{\rm min}}}^{\infty} \frac{p_1}{\gamma_1} f(p_{1}) \id p_{1}}
 {\int_{p_{{\rm min}}}^{\infty} p_1 \gamma_1 f(p_{1}) \id p_{1}}
 \,\bar{g}(\omega, \The)
 \nonumber \\
 &\equiv
 \frac{\bar{g}(\omega, \The)}{(1+\omega)^2+2(1+\omega)\The + 2\The^2}
 \nonumber \\[1mm]
  g^{\rm rel}(\omega, p_1) &=\gamma_1^{-2}\,g(\omega,p_1).
  \end{align}
This redefinition reduces the dynamic range of the Gaunt factor and is thus very useful for compressing the data in tabulations. 
The final Bremsstrahlung emission term then takes the form
\bsub
\begin{align}
\label{eq:dN_dt_stim}
\left.\frac{\id N_\gamma}{\id t \id \omega} \right|_{\rm em}
&=\frac{2\sqrt{2}\alpha Z^2\,}{\sqrt{3\pi \The}} \frac{\Ne N_{\rm i} \sigT c}{\omega} \,\expf{-\omega/\The} 
\,\mathcal{I}(\omega, \The)\,
\bar{g}^{\rm rel}(\omega, \The)
 \\[1mm]
&=\frac{2\sqrt{2}\alpha Z^2\,}{\sqrt{3\pi \The}} \frac{\Ne N_{\rm i} \sigT c}{\omega} \,\expf{-\omega/\The} 
\,\mathcal{R}(\The)\,
\bar{g}(\omega, \The),
\end{align}
\esub
where $\mathcal{I}(\omega, \The)$ given by Eq.~\eqref{eq:dN_dt_K_rel_b} and $\mathcal{R}(\The)$ by Eq.~\eqref{eq:norm_ratio}.
Both definitions of course give exactly the same answer for the overall Bremsstrahlung emission term. Nevertheless, in applications $\bar{g}^{\rm rel}(\omega, \The)$ is beneficial since it does not scale as strongly with temperature and can also be extrapolated towards high photon energies without further computation (see discussion below).

\subsection{Thermally-averaged NR and BH Gaunt factors}
\label{sec:NR-thermal}
In this section we illustrate the effects of thermal averaging on the non-relativistic and Bethe-Heitler Gaunt factors. We also consider the improvements by adding a factor of $\gamma_1^2$ to the Kramers' and NR formulae to capture the main relativistic correction. This leads to a more moderate scaling of the Gaunt factor at high frequencies and also improves the agreement with the EH result.

\subsubsection{Karzas-Latter case}
\label{sec:KL-thermal}
We start our discussion by reproducing the results from KL for the non-relativistic Gaunt factor. Similar figures can also be found in \citet{vanHoof2014}. To obtain this result we need to use $\eta_i^{\rm KL}=\alpha Z /p_i$ instead of $\eta_i=\alpha Z \gamma_i/p_i$ in Eq.~\eqref{eq:dsig_domega_K}. We furthermore approximate the relative speed by $|\varv_{\rm rel}|\approx p_1$ and assume a non-relativistic Maxwellian, $f_{\rm nr}(p)=\sqrt{2/\pi}\,\The^{-3/2}\,\expf{-p^2/2\The}$. The minimal momentum is furthermore set to $p_{{\rm min}}\approx\sqrt{2\omega}$. With this the Gaunt factor's thermal average, Eq.~\eqref{gaunt_def_usual}, reduces to
\begin{align}
\label{gaunt_def_KL}
  \bar{g}^{\rm KL}(\omega, \The)
   &=\frac{\int_{\sqrt{2\omega}}^{\infty}p_1 \expf{-p_1^2/2\The} g^{\rm KL}(\omega,p_1)\id p_{1}}
  {\int_{\sqrt{2\omega}}^{\infty} p_1 \expf{-p_1^2/2\The} \id p_{1}}
  \nonumber \\
  &=
  \frac{\int_{\omega/\The}^{\infty}\expf{-\xi} g^{\rm KL}\left(\omega, p_1=\sqrt{2 \The\xi}\right)\id \xi}
  {\int_{\omega/\The}^{\infty} \expf{-\xi} \id \xi}
  \nonumber \\
&\equiv
\int_{0}^{\infty}\expf{-\xi} g^{\rm KL}\left(\omega, p_1=\sqrt{2(\omega+\The\xi)}\right)\id \xi,
\end{align}
which is equivalent to Eq.~(21) of KL after switching to the absorption Gaunt factor (exchange of the roles of the incoming and outgoing electrons and use of energy conservation). It also directly follows from Eq.~\eqref{gaunt_def_usual} for $\The, \omega\ll 1$. 

\begin{figure}
\includegraphics[width=1. \linewidth ]{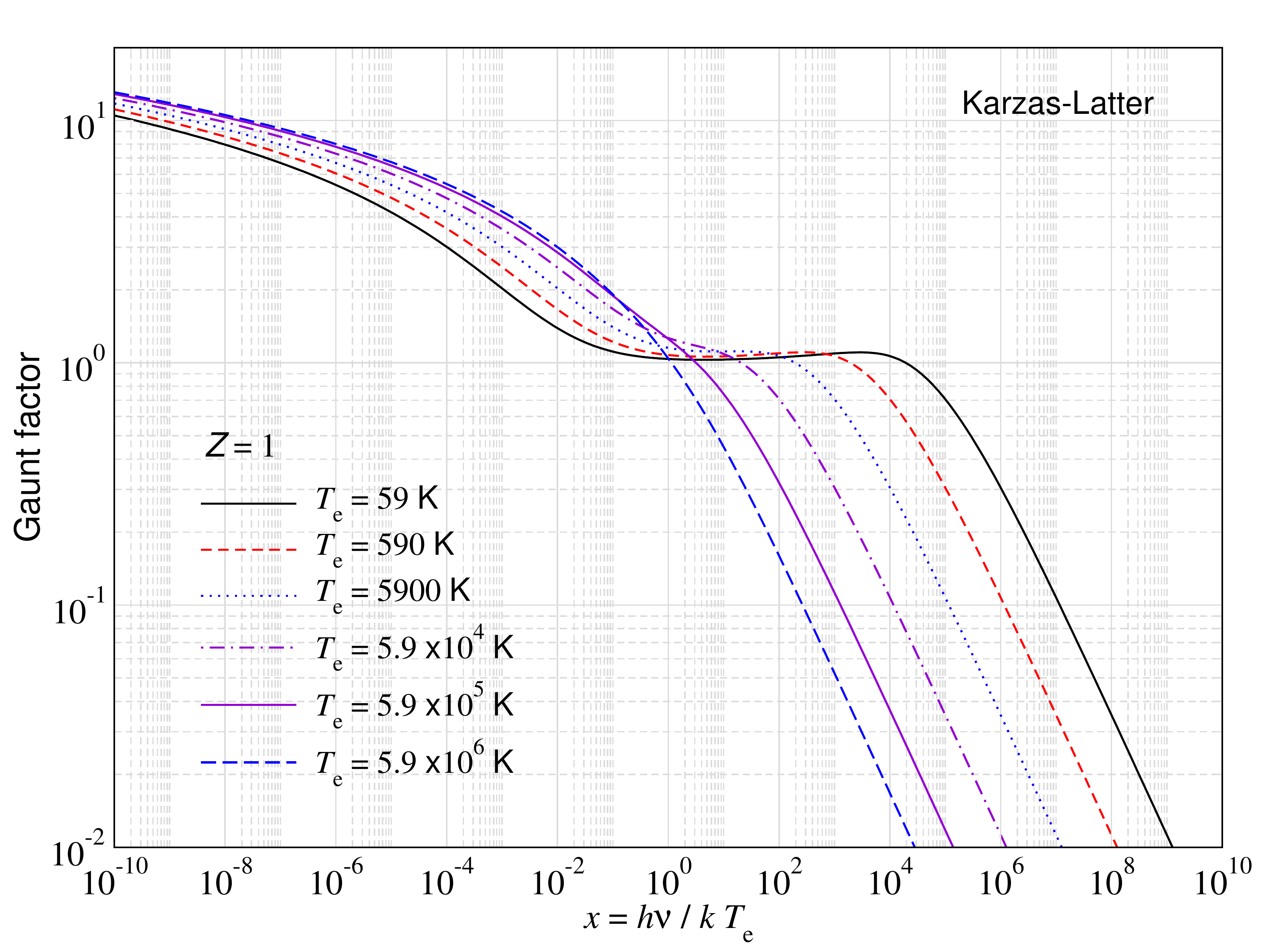}
\caption{Thermally-averaged Gaunt factor with definitions as in KL for $Z=1$ (see Sect.~\ref{sec:KL-thermal} for details). The steep drop at high photon energies is because relativistic boosting is not accounted for, an effect that is cured by our modified non-relativistic expression (see Fig.~\ref{fig:Gaunt_NR_therm_Z1} and \ref{fig:Gaunt_NR_therm_Z1-high}).}
\label{fig:Gaunt_NR_KL_therm_Z1}
\end{figure}

\begin{figure}
\includegraphics[width=1. \linewidth ]{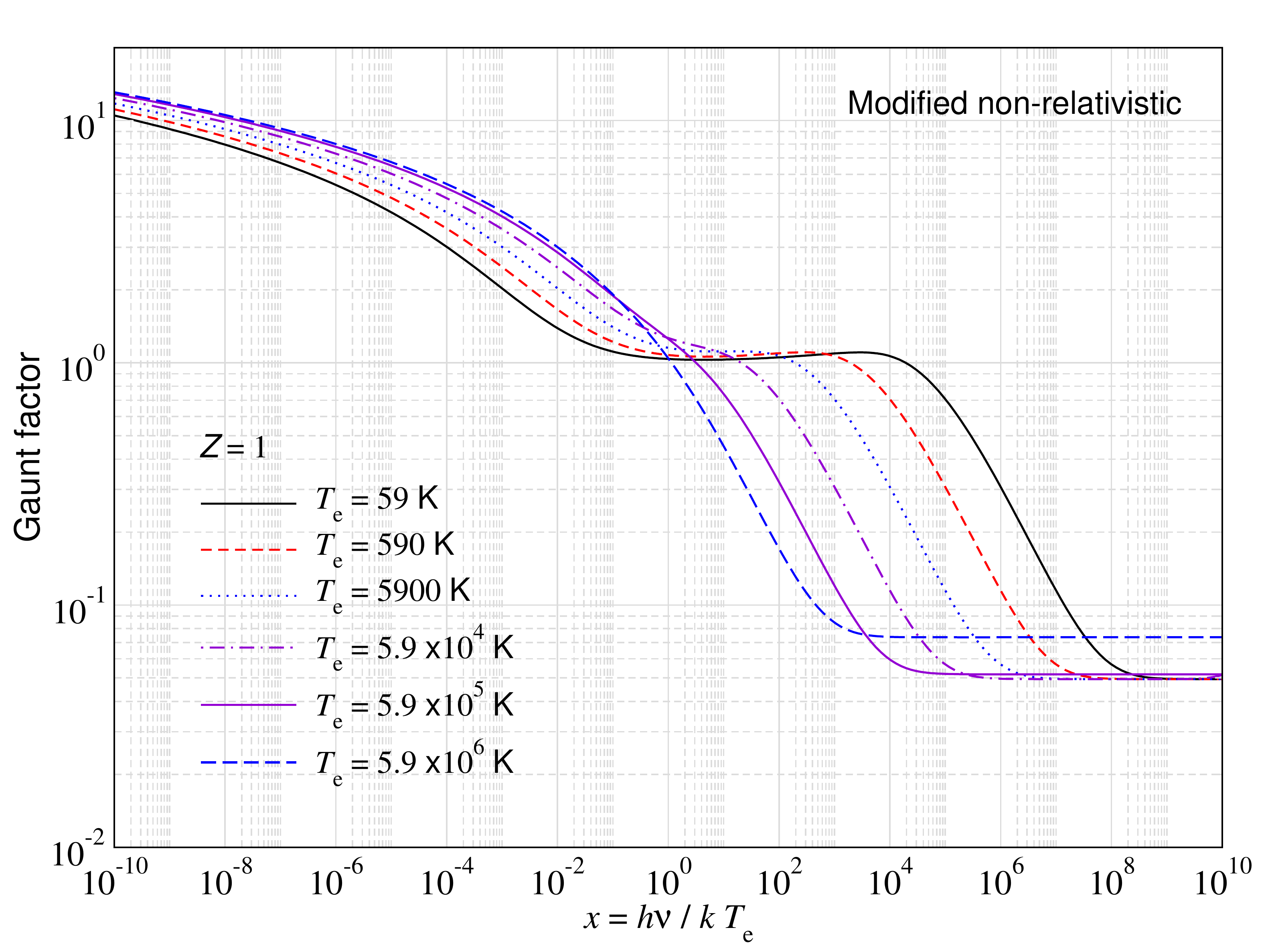}
\caption{Thermally-averaged Gaunt factor in the non-relativistic limit [Eq.~\eqref{eq:dsig_domega_K}] for $Z=1$ when using the standard definition, Eq.~\eqref{gaunt_def_usual}, for the thermal average. The replacement of $\eta_i^{\rm KL}=\alpha Z /p_i$ by $\eta_i=\alpha Z \gamma_i/p_i$ removes the unphysical drop of the KL Gaunt factor at high energies (cp. Fig.~\ref{fig:Gaunt_NR_KL_therm_Z1}).}
\label{fig:Gaunt_NR_therm_Z1}
\end{figure}

Figure~\ref{fig:Gaunt_NR_KL_therm_Z1} illustrates the thermally-averaged Gaunt factor for varying temperature and $Z=1$ using the approximations of KL. At high photon energies, a steep drop of the Gaunt factor is observed. This is not found for the EH result even at these relatively low temperatures and is simply caused by the fact that in the tail of the electron distribution function relativistic correction cannot be neglected. By switching back to $\eta_i=\alpha Z \gamma_i/p_i$ instead of $\eta_i^{\rm KL}=\alpha Z /p_i$ and inserting this into Eq.~\eqref{gaunt_def_usual} [i.e., not setting factors of $\gamma_i$ to unity], we obtain the results in Fig.~\ref{fig:Gaunt_NR_therm_Z1}. The unphysical drop of the Gaunt factor at high energies is removed by this transformation and the Gaunt factor become constant at $x=\omega/\The \gtrsim 3/\The$ or $h\nu\gtrsim 3\me c^2$. Although this already is an improvement of the non-relativistic expression, it still underestimates the  result at high frequencies. However, the Gaunt factor can now be extrapolated to any higher frequency using a finite range in $x$.
Another improvement can be achieved by adding a factor of $\gamma_1^2$ to the NR cross section, as will be discussed below (see Fig.~\ref{fig:Gaunt_NR_therm_Z1-high}). 

\subsubsection{Bethe-Heitler case}
\label{sec:BH-thermal}
\begin{figure}
\includegraphics[width=1. \linewidth ]{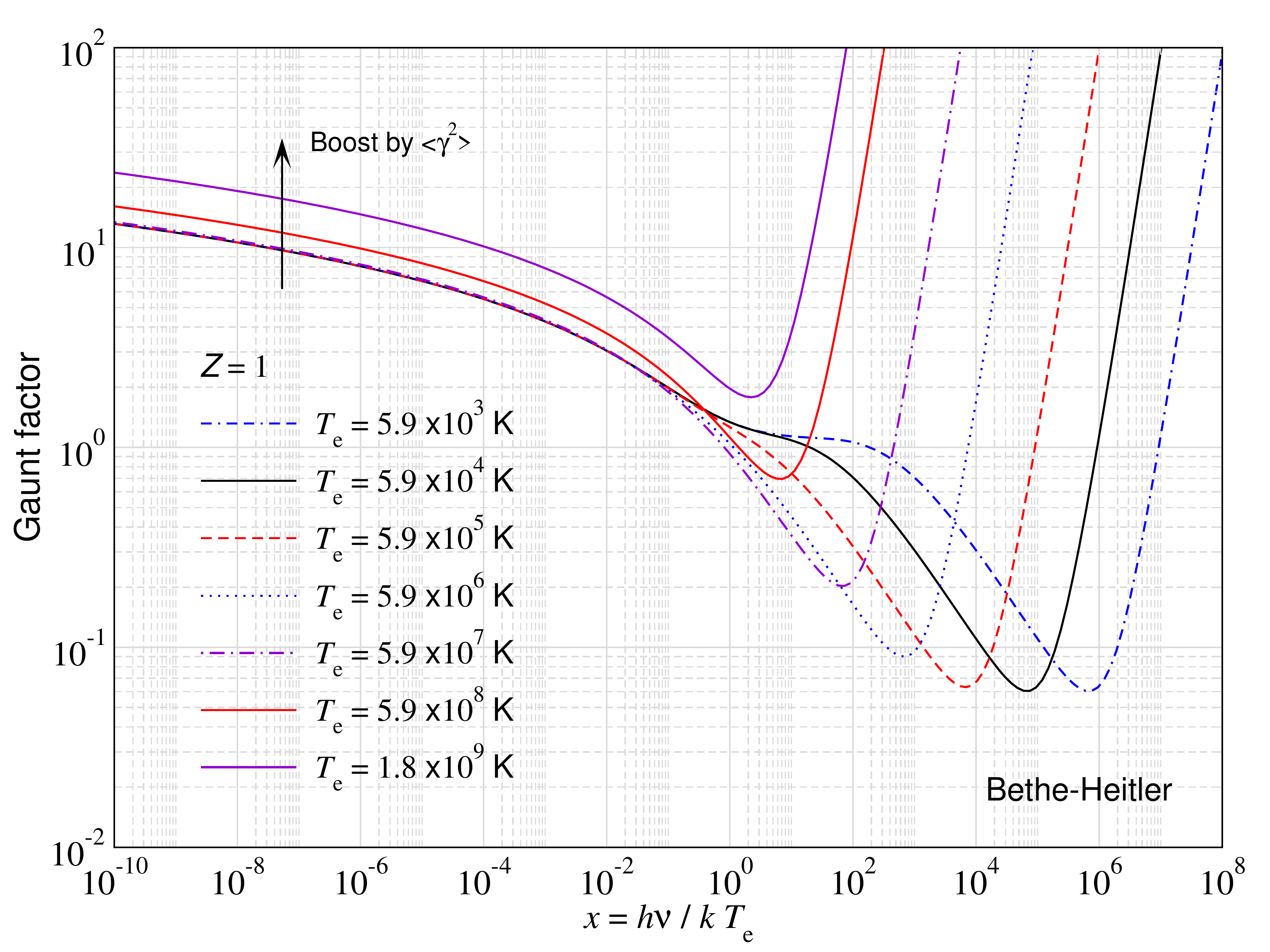}
\caption{Thermally-averaged relativistic BH Gaunt factor for $Z=1$ and varying temperature ($\The=\{10^{-6}, 10^{-5}, 10^{-4}, 10^{-3}, 10^{-2}, 0.1, 0.3\}$). The BH Gaunt factor exhibits a steep increase at high frequencies. At temperature $\Te\gtrsim \pot{5.9}{8}\,{\rm K}$ ($\The \gtrsim 0.1$), extra boosting by $\simeq \left<\gamma_1^2\right>$ becomes relevant. Both aspects can be captured by redefining the thermal average (see Fig.~\ref{fig:Gaunt_BH_therm_Z1_mod}), which reduces the dynamic range of the Gaunt factors.}
\label{fig:Gaunt_BH_therm_Z1}
\end{figure}

\begin{figure}
\includegraphics[width=1. \linewidth ]{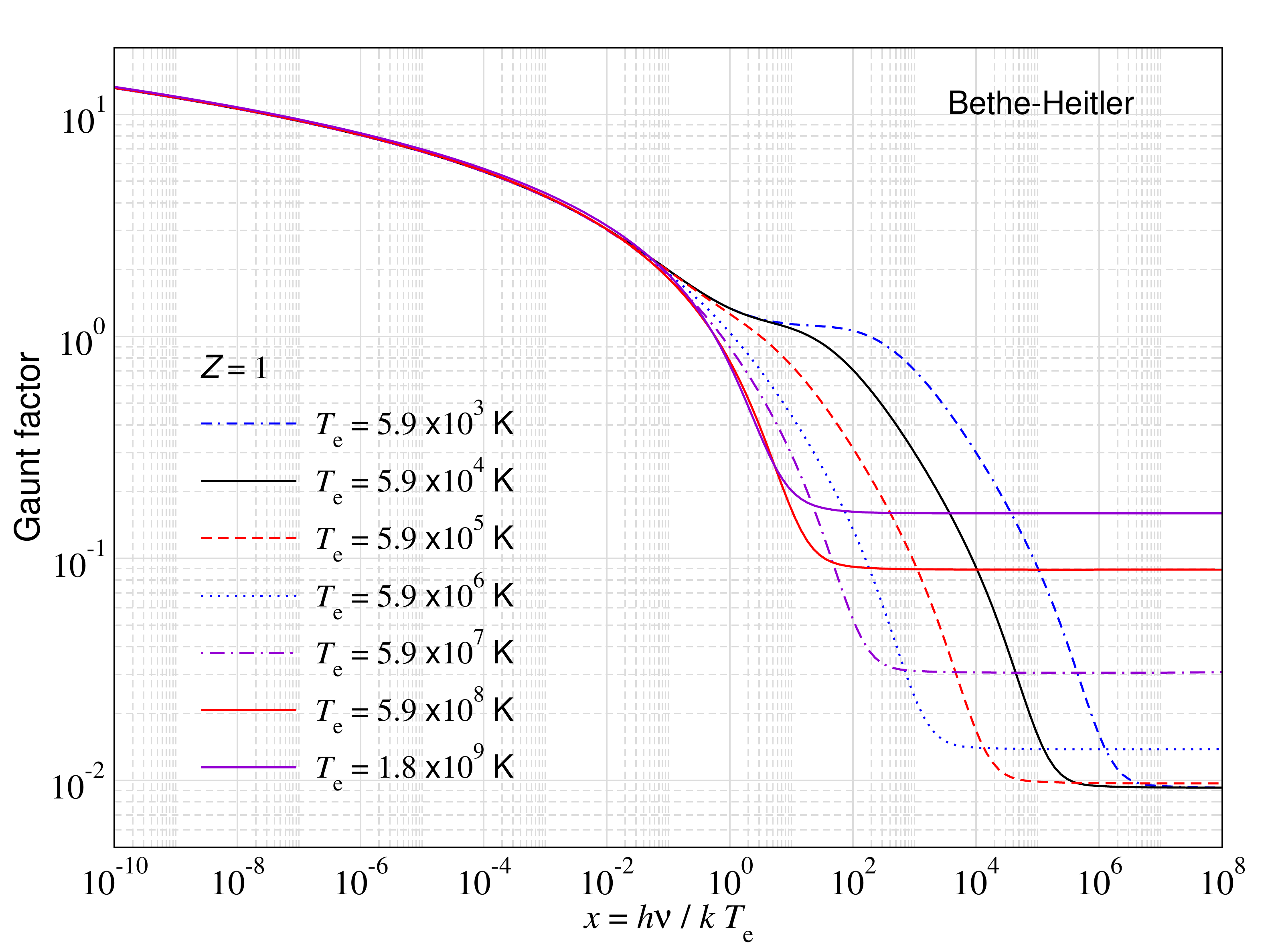}
\caption{Same as in Fig.~\ref{fig:Gaunt_BH_therm_Z1} but using our modified definition for the thermal average, Eq.~\eqref{gaunt_def_rel_mod}. The dynamic range is greatly reduced by the redefinition of the thermal average.}
\label{fig:Gaunt_BH_therm_Z1_mod}
\end{figure}

To illustrate the BH case, we first use Eq.~\eqref{eq:dsig_domega_BH} in Eq.~\eqref{gaunt_def_usual}, obtaining the results presented in Fig.~\ref{fig:Gaunt_BH_therm_Z1}. In this case, our results are in very good agreement with those obtained by \citet{Nozawa1998} and \citet{vanHoof2015}. The BH Gaunt factor shows a steep increase towards high frequencies. 
At temperature $\Te\gtrsim \pot{5.9}{8}\,{\rm K}$ ($\The \gtrsim 0.1$), an additional increase of the overall Gaunt factor amplitude by $\simeq \left<\gamma_1^2\right>\simeq 1+3\The+15 \The^2/2$ furthermore becomes noticeable. Both aspects can be avoided by using the relativistically-improved Kramers cross section for reference. This approach is taken for our modified thermal average, Eq.~\eqref{gaunt_def_rel_mod}, and illustrated in Fig.~\eqref{fig:Gaunt_BH_therm_Z1_mod}. The modification captures the main relativistic effects and greatly reduces the dynamic range of the Gaunt factors, which is beneficial for numerical applications. Again, extrapolation of the Gaunt factor to very high energies is possible using a finite range in $x$, since $\bar{g}^{\rm rel}(\omega, \The)$ becomes roughly constant at $x=\omega/\The \gtrsim 3/\The$ or $h\nu\gtrsim 3\me c^2$.
In {\tt BRpack}, we make use of this property.

\begin{figure}
\includegraphics[width=1. \linewidth ]{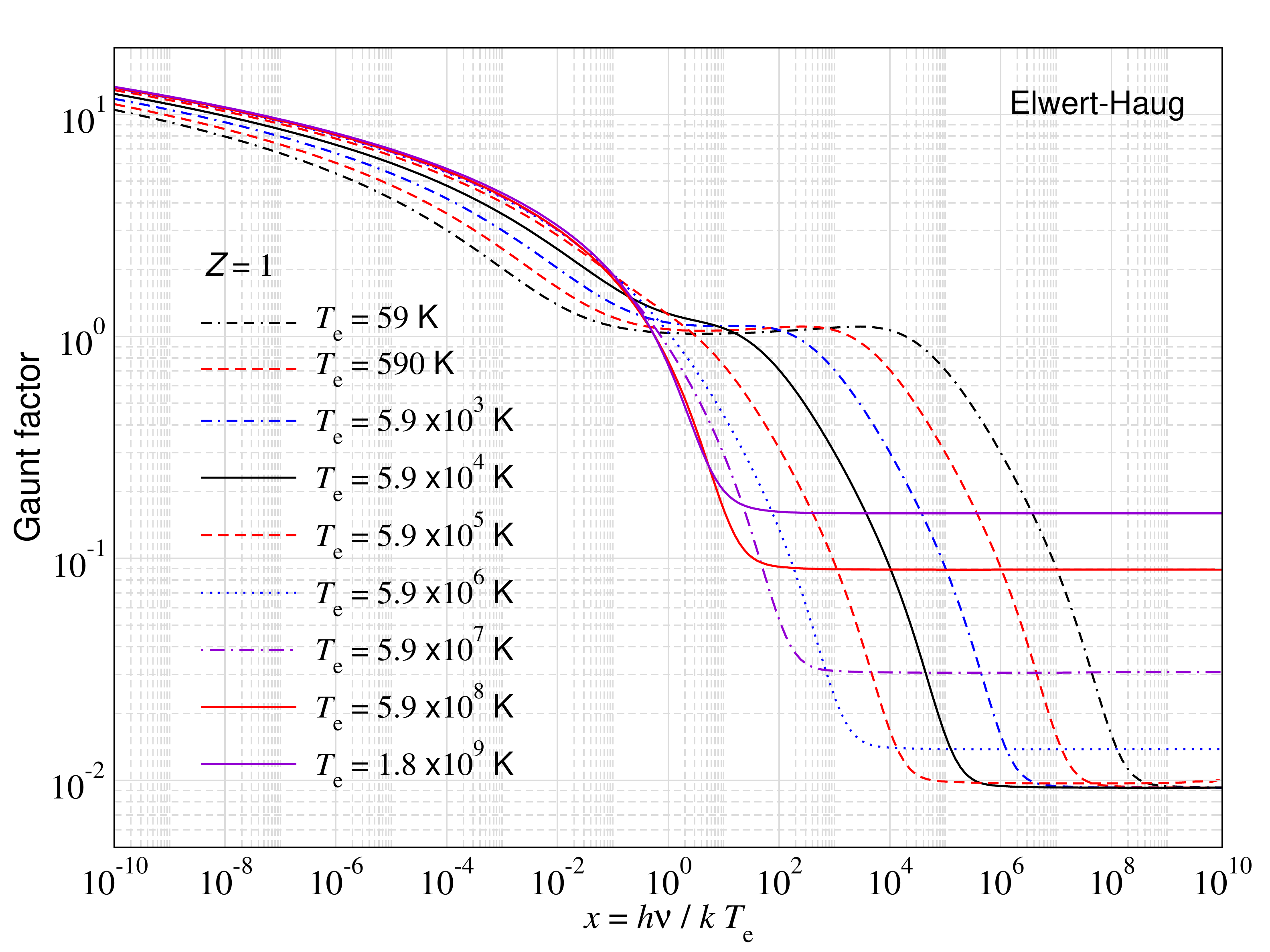}
\caption{Thermally-averaged Gaunt factor using the integrated expressions of EH in Eq.~\eqref{gaunt_def_rel_mod} for $Z=1$ and varying temperature (corresponding to $\The=\{10^{-8}, 10^{-7}, 10^{-6}, 10^{-5}, 10^{-4}, 10^{-3}, 10^{-2}, 0.1, 0.3\}$). }
\label{fig:Gaunt_EH_therm_Z1_mod}
\end{figure}

\subsection{Thermally-averaged Gaunt factor for the EH case}
\label{sec:EH-thermal}
We are now in the position to compute the thermally-averaged EH Gaunt factor. In Fig.~\ref{fig:Gaunt_EH_therm_Z1_mod} we illustrate the results over a wide range of temperatures and photon energies. We directly used our modified definition for the thermal average, Eq.~\eqref{gaunt_def_rel_mod}, which greatly reduces the dynamic range. This definition is ideal for tabulation of the result, and is used in {\tt BRpack}. At high photon energies the result can be obtained by extrapolation, however, the net emission vanishes in this limit for any practical purposes. To our knowledge, this is the first precise representation of the thermally-averaged EH Gaunt factor for hydrogen over an as vast range of energies. Cases for $Z\leq10$ can also be quickly computed using {\tt BRpack}. 

\begin{figure}
\includegraphics[width=\columnwidth]{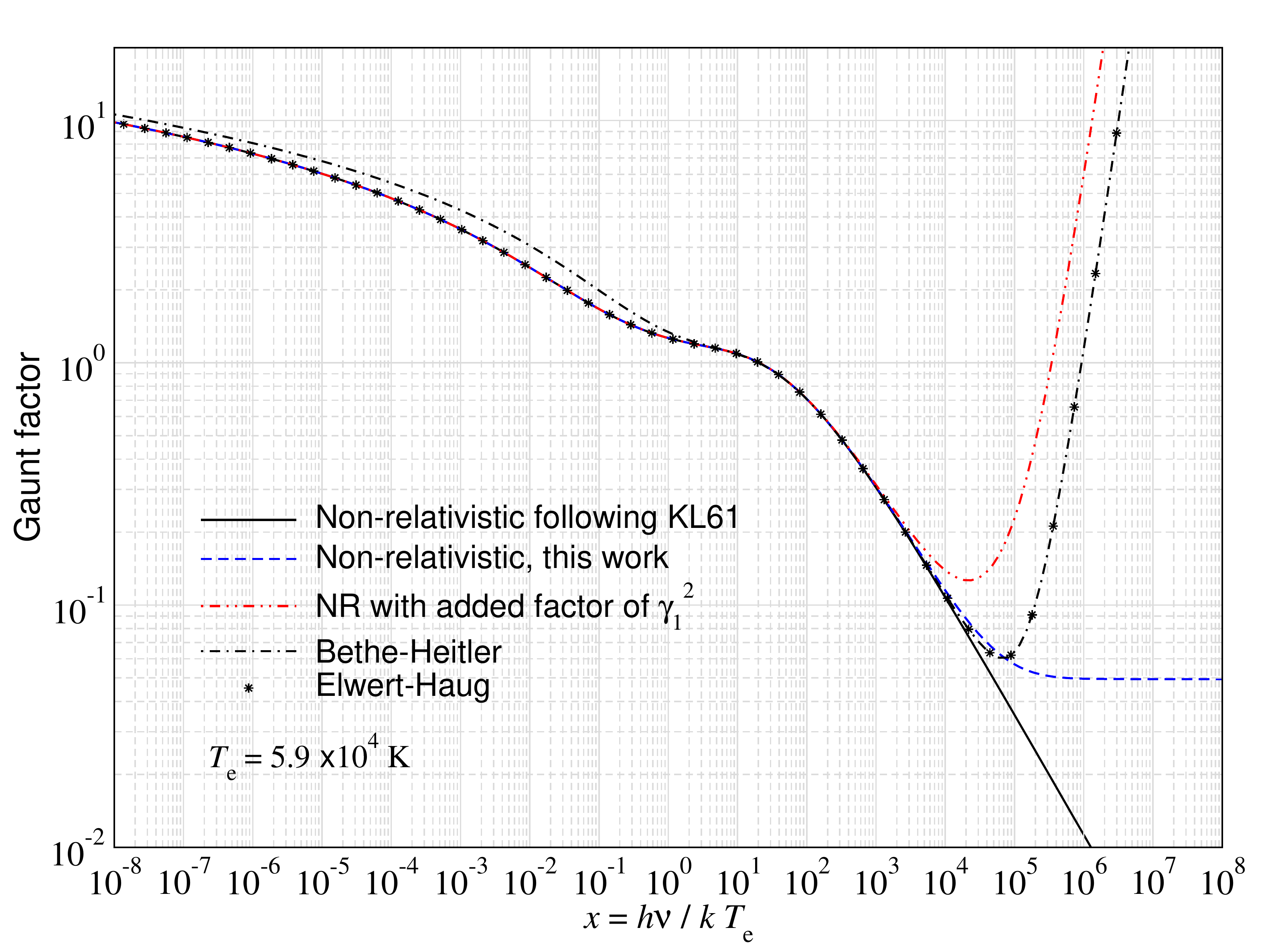}
\caption{Thermally-averaged Gaunt factors for $Z=1$ when using the standard definition, Eq.~\eqref{gaunt_def_usual}, for the thermal average and various limits for the cross section. The EH result is shown for comparison. BH works very well at high photon energies, while the non-relativistic expressions capture the behavior at low energies. An improvement of the non-relativistic expression is obtained when adding a factor of $\gamma_1^2$.}
\label{fig:Gaunt_NR_therm_Z1-high}
\end{figure}
\subsubsection{Comparison with simple approximations}
\label{sec:rel-NR-thermal}
We now compare the various approximations for the thermally-averaged Gaunt factor with the those obtained from the EH expressions. For electron temperature $\Te=\pot{5.9}{4}\,{\rm K}$ ($\The=10^{-5}$) and $Z=1$ the results are shown in Fig.~\ref{fig:Gaunt_NR_therm_Z1-high}, using the standard definition for the Gaunt factor thermal average, Eq.~\eqref{gaunt_def_usual}. 
As expected, the Bethe-Heitler approximation works extremely well at high frequencies, where all contributions indeed arise from relativistic electrons of the Maxwellian. 
In contrast, the NR expressions work very well at low frequencies. As already shown in Sect.~\ref{sec:EH}, the agreement with the EH result can be further improved by multiplying the cross section by a factor of $\gamma_1^2$, which captures the main relativistic boosting effect. The overall scaling of the EH result is well-represented by our improved non-relativistic expression given in Eq.~\eqref{eq:g_NR_rel}.

\begin{figure}
\includegraphics[width=1.0\columnwidth]{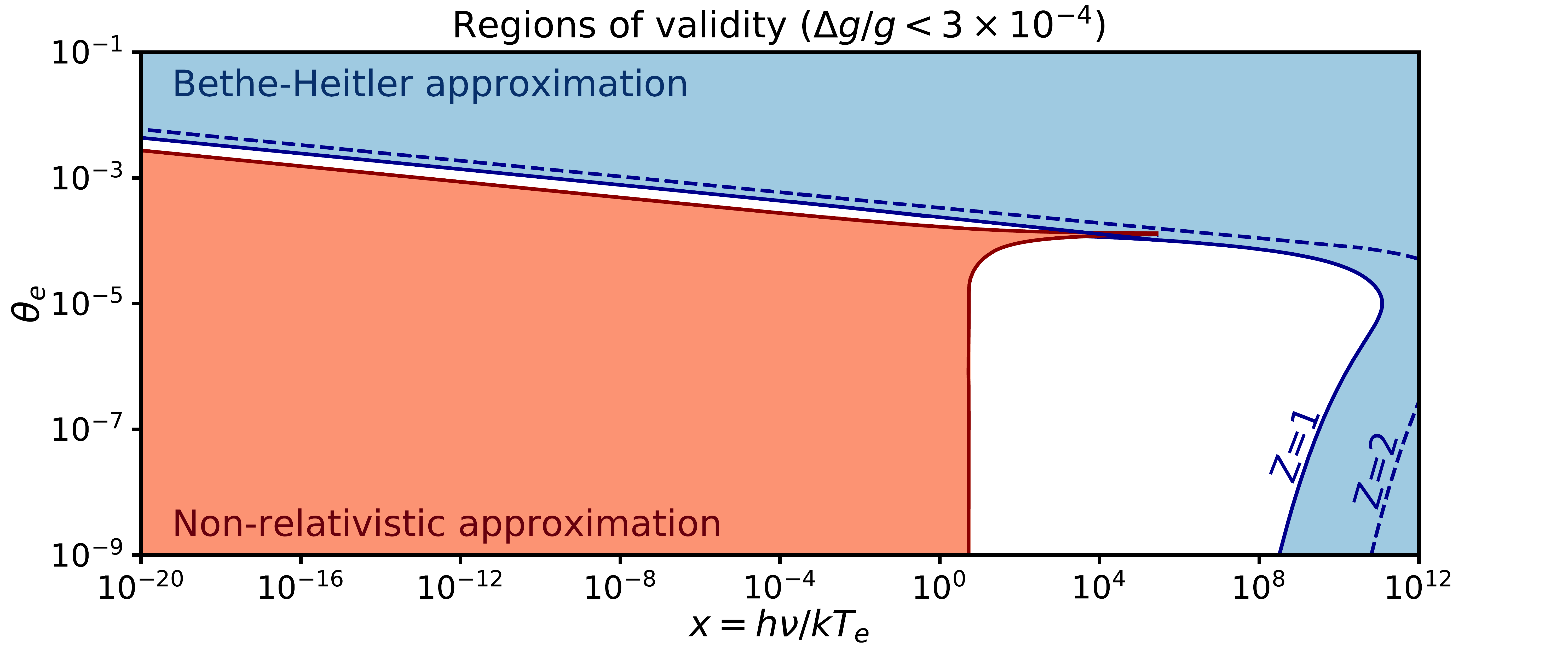}
\\[1.5mm]
\includegraphics[width=1.0\columnwidth]{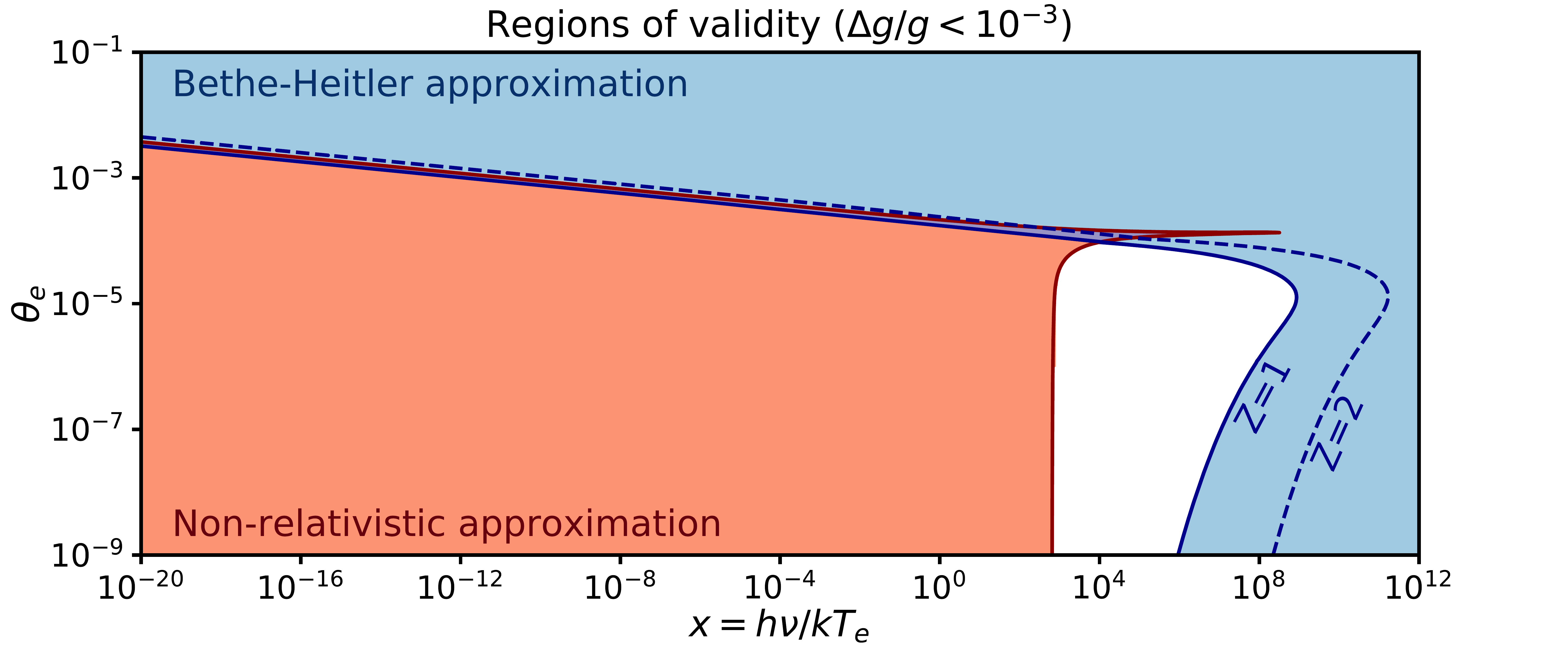}
\\[1.5mm]
\includegraphics[width=1.0\columnwidth ]{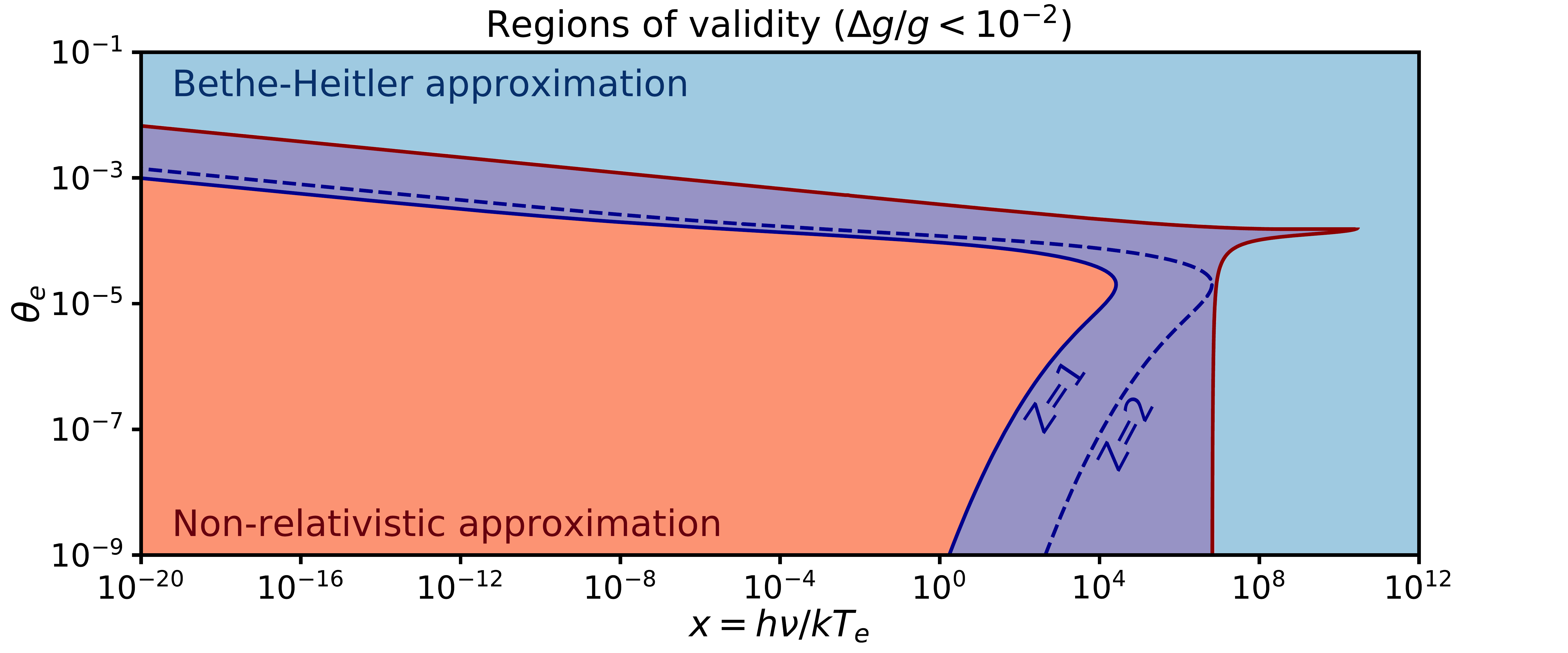}
\caption{Regions in which the NR and BH approximations can be used to calculate the thermally-averaged Gaunt factor.
The colored areas show the ($x, \The$) domains where the relative difference with respect to the thermally-averaged EH Gaunt factor is 0.3\textperthousand~(Top panel), 1\textperthousand~(Middle panel), or 1\% (Bottom panel) for $Z=1$. The dashed line displays the boundary of the BH region for $Z=2$. The NR regions for $Z=1$ and $2$ coincide at the plot resolution. In the white areas, the EH formula is required.}
\label{fig:comp_formulas_th}
\end{figure}
\subsubsection{Domains of validity}
\label{sec:rNR_BH-thermal}
For a more quantitative accuracy assessment of the NR and BH formula, we again perform a comparison similar to the one discussed in Sect.~\ref{sec:EH}. We compute the thermally-averaged Gaunt factor using solely the BH formula or the NR expression and then ask for which pairs $(x, \The)$ it deviates from the one obtained using EH by less than a given threshold. The corresponding regions for $Z=1$ and $Z=2$ are displayed in Fig.~\ref{fig:comp_formulas_th}, in red for the NR approximation and in blue for the BH formula. 
From top to bottom the required agreement is at least 0.3\textperthousand, 1\textperthousand, and 1\%.
Some caveats about the BH validity region should be mentioned here. Due to the structured behavior of the averaged Gaunt factor (cf. Fig.~\ref{fig:Gaunt_NR_therm_Z1-high}), several disconnected regions are identified as valid when using the described thresholding procedure, especially for high-accuracy thresholds. We thus only highlight the points $(x,\The)$ such that $(\tilde{x},\The) \in $~BH validity region for all $\tilde{x}\geq x$.
Overall we find that a combination of the BH and NR expressions for $Z\lesssim 10$ leads to a good description of the full EH Gaunt-factor over a wide range of photon energies and electron temperatures unless precision $\lesssim 1\%$ is required. For $Z=1$ and $2$, we expect {\tt BRpack} to provide a better than $0.03\%$ level representation of the thermally-averaged EH Gaunt factor at all photon energies. In particular the low-frequency part of the EH emission spectrum should be represented very accurately up to mildly-relativistic temperatures $k\Te \simeq 50\,{\rm keV}$.

We also note that since for $Z\leq 10$ we only tabulated the EH Gaunt-factor up to $p_1=2$, at $x\gtrsim 1.2 / \The$, we always switch to the BH result. The error with respect to the full EH evaluation should be limited to $\lesssim 0.5\%$ (see Fig.~\ref{fig:High_p1}), which again should not cause any severe limitations for astrophysical applications at $\The \lesssim 0.1$.

\subsection{Comparison with previous works}
\label{sec:Itoh_and_vH}
The thermally-averaged Gaunt factor for the cross section expressions of KL and Bethe-Heitler formula were previously considered in detail \citep{Itoh1985, Nozawa1998, Itoh2000, vanHoof2014, vanHoof2015}.
For the non-relativistic regime, the KL definition for the thermally-average Gaunt factor, Eq.~\eqref{gaunt_def_KL}, was used.
For the BH limit, the Gaunt factor definition of the aforementioned works relates to ours, Eq.~\eqref{gaunt_def_usual}, by
\bsub
\begin{align}
\label{eq:Gaunt_Itoh}
\bar{g}^{\rm Itoh}(\omega, \The)
&=\mathcal{R}(\The)\, \bar{g}(\omega, \The).
\end{align}
\esub
In \citet{Itoh2000}, fits were given over a limited range of photon energies and temperatures, while \citet{vanHoof2014, vanHoof2015} provided extensive tables covering a wide range of temperatures, photon energies and ion charges $Z$. 
We were able to reproduce the results of \citet{vanHoof2014, vanHoof2015} for the KL and BH limits, finding excellent agreement. We also confirmed the results of \citet{Itoh2000} at $x=10^{-4}-20$ finding very good agreement.

In \citet{vanHoof2015}, the KL and BH limits were 'merged' to mimic the transition between the non-relativistic and relativistic regimes. However, no explicit assessment of the accuracy of this procedure was provided.  As we saw in Sec.~\ref{sec:EH}, for low ion charge $Z\leq 10$ we can expect departures to become visible at the level of $0.1-1\%$. 
For $Z=1$ and $Z=2$, we find the EH calculation to agree with \citet{vanHoof2015} at the $0.1\%$ level, while for higher charges the differences do exceed this level. Again, this outcome is expected given the discussion of Sect.~\ref{sec:EH}.

\begin{figure}
\includegraphics[width=\columnwidth]{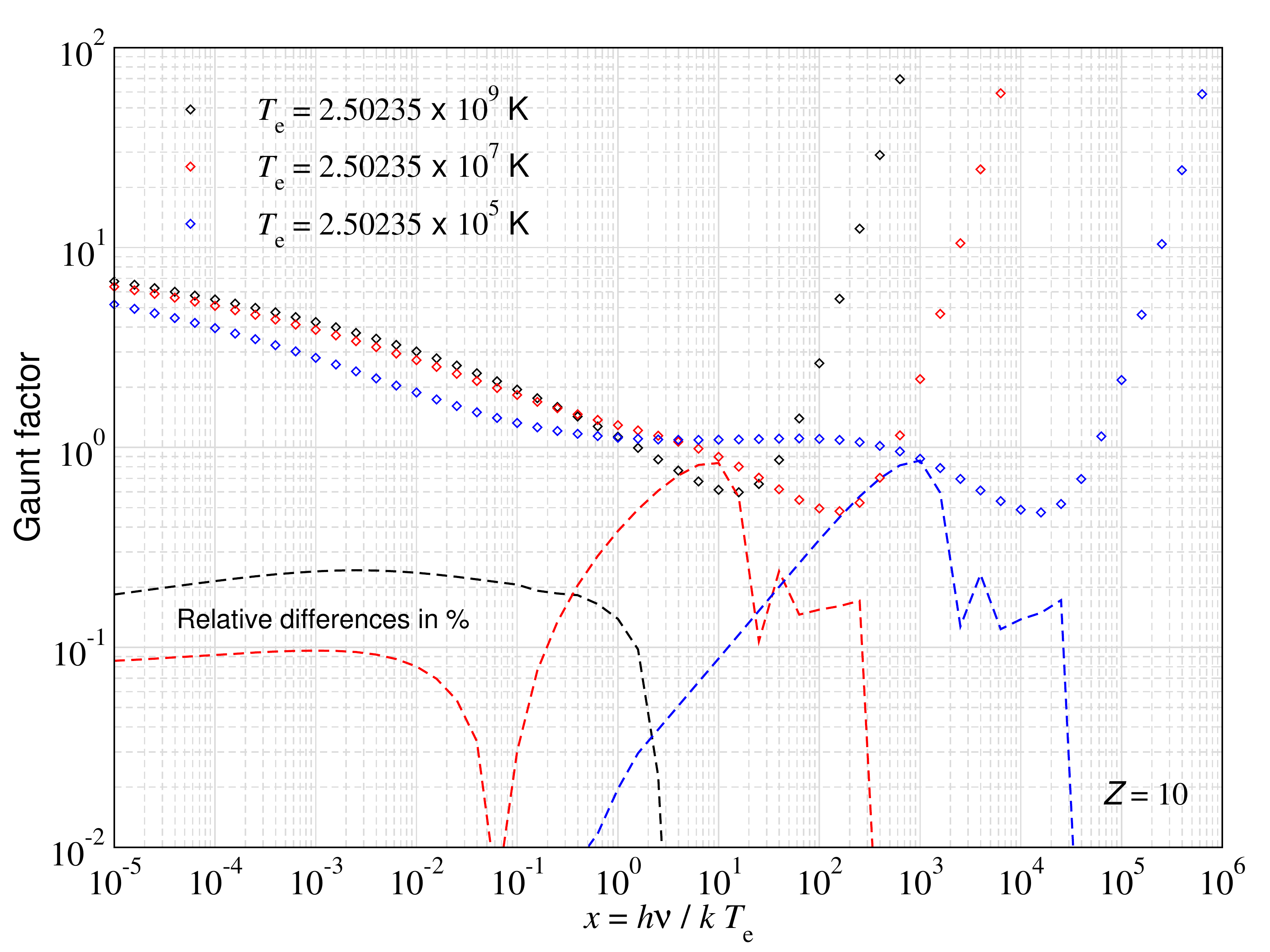}
\caption{Direct comparison of the EH Gaunt factor with the values given by \citet{vanHoof2015} for $Z=10$. The dashed lines show the relative differences in percent. At $x\simeq 1-10^3$, the departures can reach the percent level at low and intermediate temperatures. The abrupt drop of the relative differences (absolute value) at high frequencies are due to our limited tables of the total EH Gaunt factor (see text for explanation).}
\label{fig:Gaunt_vH_EH_therm}
\end{figure}

In Fig.~\eqref{fig:Gaunt_vH_EH_therm}, we show our result for the EH Gaunt factor and relative difference with respect to \citet{vanHoof2015} for several temperatures and ion charge $Z=10$. For the comparison, we took the exact values from the tables provided by \citet{vanHoof2015} without any interpolation. 
The departures are visible at the $\simeq 0.1-1\%$ level around $x\simeq 1-10^4$ and low temperatures, $\The \lesssim 0.01$. We also see an abrupt drop in the relative difference around $x\simeq 3$, $x\simeq 300$ and $\pot{3}{4}$ for the three shown cases. This is because our tables for the EH Gaunt factor only extend up to $p_1=2$, such that at very high photon energies we always converge to the BH result, and thus agree with \citet{vanHoof2015} to high precision. Note however, that at these high photon energies hardly any BR emission is expected such that errors should remain minor for astrophysical applications. The low-frequency region is much more crucial in this respect and we expect our thermally-averaged Gaunt factor to be highly accurate there ($\Delta g/g \lesssim 0.1\%$ at $x\lesssim 1$ for $Z\leq 10$).

For larger values of $Z$, the departures exceed the percent-level. We numerically evaluated the case $Z=20$ finding differences with respect to \citet{vanHoof2015} at the level of $\simeq 2-3\%$ around $x\simeq 1-10^4$ and temperatures $\The \lesssim 0.01$. However, for higher ionic charge also additional Coulomb corrections and shielding effects should also be accounted for, such that we leave a more quantitative comparison to future work. {\tt BRpack} should yield reliable results for $Z\leq 10$ and $\The \lesssim 0.1$ at $\lesssim 0.5\%$ precision. For $Z\leq 4$, the EH Gaunt factor should be reproduced at the level of $\lesssim 0.1\%$ precision. 

\section{Conclusion}
\label{sec:conclusion}
We presented a comprehensive study of the free-free Gaunt factor, $g(\omega, p_1)$, and its thermally-averaged version, which is relevant to many astrophysical applications. Our focus was on ions with low ionic charge ($Z\leq 10$), for which we computed the BR Gaunt factors using the differential cross section expressions given by EH. 
We compared our results with various approximations and previous Gaunt factor computations, illustrating the domains of validity and their precision (e.g., Fig.~\ref{fig:comp_formulas_Z1}). Our results for $g_{\rm EH}(\omega, p_1)$ should be accurate at the level of $\lesssim 0.03\%$ for $Z\leq 10$ and $p_1\leq 2$. 
For the thermally-averaged EH Gaunt factor we expect our computations to yield $\lesssim 0.1\%$ precision at $k\Te \lesssim 50\,{\rm keV}$ for $Z\leq 4$ and slightly better ($\Delta g/g \lesssim 0.03\%$) for $Z\leq 2$. For $Z\leq 10$ we expect an overall precision of $\lesssim 0.5\%$ for the thermally-averaged EH Gaunt factor at temperatures $k\Te \lesssim 50\,{\rm keV}$. 

We simplified the computations of the EH differential cross section, showing that the hypergeometric function evaluations can be reduced to an evaluation of one real function. This function can be computed using an ordinary differential equation and thus improves the computational precision and efficiency greatly. In a similar manner we showed that the non-relativistic Gaunt factor can also be related to the same real function [see Eq.~\eqref{eq:gaunt_NR_mod}]. Overall, our numerical procedure allow us to precisely compute the EH Gaunt factor over a wide range of energies, with extensions to low and high photon energies obtained using analytic expressions.
Coulomb corrections and shielding effects are expected to become important for $Z>10$ and at high electron energies. These can in principle be added using our computational method.

We developed new software package, {\tt BRpack}, which allows efficient and accurate representation of the NR, BH and EH Gaunt factors for $Z\leq 10$, both for individual values of the electron and photon momenta as well as for thermally-averaged cases. It should prove useful for computations of CMB spectral distortions and radiative transfer problems in the intergalactic medium at low redshifts. We can furthermore use the Gaunt factor for improved modeling of the free-free emission from our own galaxy, potentially even taking non-thermal contributions into account without mayor complications. 
Our procedure can also be applied to computations of the $e-e$ and $e^--e^+$ Bremsstrahlung processes \citep[e.g.,][]{Haug1985, Haug1975, Itoh2002}, which will be important at higher plasma temperatures ($k\Te/\me c^2 \gtrsim 1$). 

While with {\tt BRpack} a numerical precision of better than ${\simeq0.01\%}$ can be reached for any photon and electron energy, it is clear that this does not fully reflect the accuracy of the Gaunt factor. Higher order Coulomb corrections, shielding effects and radiative corrections are not accounted for by the EH expression. These invalidate the cross section at higher energies and for large ion charge \citep{Tseng1971, Roche1972, Haug2008}. 
However, even at the temperatures and photon energies of interest to us ($k\Te \lesssim {\rm few}\times {\rm keV}$), corrections may become relevant at $\lesssim 0.1\%$ accuracy. In this case, exact calculations using Dirac-wave functions \citep[e.g.,][]{Tseng1971, Andrius2018, Andrius2019} for the electron may be required. Given the many applications of the BR process in astrophysics, accurate calculations with the goal to provide comprehensive, user-friendly, quasi-exact representations of the process for a wide range of conditions should be undertaken. We look forward to further investigations of the problem.

\vspace{0mm}
{\small
{\noindent \it Acknowledgments:} This work was supported by the ERC Consolidator Grant {\it CMBSPEC} (No.~725456) as part of the European Union's Horizon 2020 research and innovation program.
JC was supported by the Royal Society as a Royal Society URF at the University of Manchester, UK.
}

{\small
\begin{appendix}

\section{Properties of $G_\ell$} 
\label{app:G_ell_Real}
We first prove that $G_\ell$ is real. Starting from Eq.~\eqref{eq:def_Gell}, this can be seen with
\begin{align}
G^*_\ell(\eta_1, \eta_2, x)
&=
\left(-x\right)^{\ell +1}
(1-x)^{\frac{- i(\eta_1+\eta_2)}{2}}\,\expf{-\pi\eta_1}
\nonumber\\
&\qquad\quad _2F_1\left(1+\ell-i\eta_1, 1+\ell-i\eta_2, 2\ell+2, x \right)
\nonumber\\
&
=\left(-x\right)^{\ell +1}
(1-x)^{\frac{- i(\eta_1+\eta_2)}{2}} \,(1-x)^{i(\eta_1+\eta_2)}  \,\expf{-\pi\eta_1}
\nonumber\\
&\qquad\quad {_2}F_1\left(1+\ell+i\eta_1, 1+\ell+i\eta_2, 2\ell+2, x \right)
\nonumber\\
&\equiv G_\ell(\eta_1, \eta_2, x),
\end{align}
where we used ${_2}F_1\left(a, b, c, x \right)=(1-x)^{c-a-b}\,{_2}F_1\left(c-a, c-b, c, x \right)$ for the hypergeometric function. More generally one can show that
\begin{align}
f(x)\,(1-x)^{\frac{\pm i(a+b)}{2}} {_2}F_1\left(c\pm ia, c \pm ib, 2c, x \right)
\\
f(x)\,(1-x)^{\frac{\pm i(a-b)}{2}} {_2}F_1\left(c\pm ia, c\mp ib, 2c, x \right)
\end{align}
are all real functions for real $a, b, c$ and $x$. These relations are very useful when studying recurrence relations for $G_\ell$ (Appendix~\ref{sec:recursion}). In particular it is beneficial to include $f(x)=\sqrt{1-x}$ in the definition of $G_\ell(x)$.

\vspace{-3.8mm}
\subsection{Relation between $G_0$ and $G_1$}
\label{app:G_ell_Real_rewrite}
To simplify the computation of the non-relativistic Gaunt factor it is useful to study the relation between $G_0$ and $G_1$. The hypergeometric function related to $G_\ell(x)$ are $F_\ell(x)={_2F_1}\left(1+\ell+i\eta_1,1+\ell+i\eta_2, 2(\ell+1), x \right)$. Taking the first and second derivatives of $F_0$ with respect to $x$, we find
\bsub
\begin{align}
F_0'&=\frac{1}{2} (1+i\eta_1) (1+i\eta_2) {_2F_1}\left(2+i\eta_1, 2+i\eta_2, 3, x \right)
\nonumber\\
&=\frac{1}{2} (1+i\eta_1) (1+i\eta_2) (1-x)^{-1-i\eta_+} {_2F_1}\left(1-i\eta_1, 1-i\eta_2, 3, x \right)
\\
\label{app:DF_0}
&=\left[G'_0 - \left(\frac{1}{x}-\frac{i\eta_+}{2(1-x)}\right)G_0\right] \expf{\pi\eta_1}\, \frac{(1-x)^{-i\frac{\eta_+}{2}}}{(-x)}
\\
\label{app:DDF_0}
F_0''&=\frac{1+i\eta_+}{1-x}\,F_0'+\frac{(1+\eta_1^2)(1+\eta_2^2)}{6(1-x)}\,F_1
\end{align}
\esub
with $\eta_\pm=\eta_1\pm \eta_2$. For the differential equation of $F_0$ we have
\begin{align}
\nonumber
x (1-x) F_0'' +  [2-(3+i\eta_+)x] F_0' - (1+i\eta_1) (1+i\eta_2) F_0 = 0.
\end{align}
Inserting Eq.~\eqref{app:DDF_0} then yields
\begin{align}
\nonumber
\frac{(1+\eta_1^2)(1+\eta_2^2)}{6}\,x F_1 +  2(1-x) F_0' - (1+i\eta_1) (1+i\eta_2) F_0 = 0.
\end{align}
Multiplying this equation by $G_0/F_0=\left(-x\right)(1-x)^{\frac{i\eta_+}{2}}\,\expf{-\pi\eta_1}$ and combining with Eq.~\eqref{app:DF_0}, we then obtain
\begin{align}
&-\frac{(1+\eta_1^2)(1+\eta_2^2)}{6}\,G_1 +  2(1-x) G_0'  - \left[\frac{2}{x}-2-i\eta_+ +(1+i\eta_1) (1+i\eta_2) \right]G_0 
\nonumber \\
&\quad = -\frac{(1+\eta_1^2)(1+\eta_2^2)}{6}\,G_1 + 2(1-x) G_0' 
+ \left[\eta_1 \, \eta_2+\frac{1}{2}\left(\frac{\eta_1}{\eta_2}+\frac{\eta_2}{\eta_1}\right)\,\right] G_0 = 0.
\nonumber
\\[0.1mm]
&\quad\longleftrightarrow \quad 
G'_0=\left[\eta_1 \, \eta_2+\frac{1}{2}\left(\frac{\eta_1}{\eta_2}+\frac{\eta_2}{\eta_1}\right)\,\right] G_0
-\frac{(1+\eta_1^2)(1+\eta_2^2)}{6}\,G_1
\end{align}
By comparing with Eq.~\eqref{eq:dsig_domega_K}, we can thus obtain Eq.~\eqref{eq:gaunt_NR_mod}.

\vspace{-4.5mm}
\subsection{Differential equation for $G_0$}
\label{app:G_0_ODE}
Since $G_0$ and $G'_0$ are both real functions it is useful to study the associated differential equation directly. From the differential equation for the hypergeometric function $F_0$ we find
\begin{align}
x (1-x)^2 G_0'' - x(1-x)G_0' +\left[\eta_1\eta_2 +\frac{\eta_-^2}{4} x \right] G_0= 0.
\end{align}
This can be converted into a set of first order equations
\bsub
\begin{align}
G'_0 &= H_0
\\
H'_0 &= \frac{H_0}{1-x}-\left[\frac{\eta_1\eta_2}{x} +\frac{\eta_-^2}{4} \right] \frac{G_0}{(1-x)^2}.
\end{align}
\esub
This system has regular singular points at $x=0, 1, \infty$. For our problems we need the solution at $x<0$. Choosing a starting point very close to the origin, convenient initial conditions are 
\bsub
\begin{align}
G_0(x) &\approx - \expf{-\pi\eta_1}x\left[1+\frac{x}{2} (1-\eta_1\eta_2)\right]
\\[1mm]
H_0(0) &= G_0'(0) \approx - \expf{-\pi\eta_1}\left[1-x (1-\eta_1\eta_2)\right].
\end{align}
\esub
These allow solving the problem for various values of $\eta_1$ and $\eta_2$ of interest using a solver based on the Gear's method \citep{Chluba2010}. Due to the factor $\expf{-\pi\eta_1}$, this procedure is limited to $p_1\gtrsim \pot{8}{-5}\,Z$. 
At lower values of $p_1$, we can start with rescaled initial conditions and then reinitialize the solver after appropriate intervals multiplying portions of $\expf{-\pi\eta_1}$. For required values of $x$, this leads to numerically stable results.

\subsection{Recursion relation for $G_\ell$}
\label{sec:recursion}
Here we briefly rederive recurrence relations for $G_\ell$ following a procedure that is similar to that of \citet{Karzas1961, Hummer1988}. The same relations are useful for the EH cross section computation, as we show below. 
The starting point is 
\begin{align}
\label{app:initial_G_ell}
G_\ell(x)&=\left(-x\right)^{\ell +1}(1-x)^{\frac{i\eta_+}{2}}\,\expf{-\pi\eta_1}
\nonumber\\
&\qquad {_2}F_1\left(1+\ell+i\eta_1,1+\ell+i\eta_2, 2(\ell+1), x \right),
\end{align}
with $\eta_\pm=\eta_1\pm \eta_2$. To obtain the recurrence relations one expresses $_2F_1$ in terms of $G_\ell$. We first define\footnote{This reduces the number of terms.}  $\tilde{G}_\ell=\sqrt{1-x}\,G_\ell$ and then write
\begin{align}
\label{app:initial_F_ell}
F_\ell(x)=\tilde{G}_\ell(x)\left(-x\right)^{-(\ell +1)}(1-x)^{- \frac{i\eta_+}{2}-\frac{1}{2}}\,\expf{\pi\eta_1}.
\end{align}
This can then be inserted into the differential equation for the hypergeometric functions (which $F_\ell(x)$ fulfills), yielding 
\begin{align}
x^2(1-x)^2\tilde{G}''_\ell(x)=\left\{\ell(\ell+1)-\left[\eta_1\eta_2+\ell(\ell+1)\right]x - \frac{1+\eta_-^2 }{4}\,x^2\right\}\tilde{G}_\ell(x).
\nonumber
\end{align}
In the evaluation of the non-relativistic cross section, we always have $x<0$. Assuming $|x|< 1/2$, one can use the Ansatz $\tilde{G}_\ell(x)=(-x)^{\ell +1}\,\expf{-\pi\eta_1}\sum_n a_n x^n$, which yields 
\begin{align}
\label{eq:rec_I}
a_0&=1, \qquad a_1=\frac{\ell(\ell+1)-\eta_1\eta_2}{2\ell+2}
\nonumber\\[1mm]
a_n
&=(\kappa_{n\ell}-\lambda_{n\ell})\,a_{n-1} -(\mu_{n\ell}+\nu_{n\ell})\,a_{n-2}
\nonumber
\\[1mm]
\kappa_{n\ell}&=\frac{\ell(\ell+1)+2(n-1)(2\ell+n)}{n(2\ell+1+n)}, \qquad 
\lambda_{n\ell}=\frac{\eta_1\eta_2}{n(2\ell+1+n)}, 
\nonumber\\[1mm] 
\mu_{n\ell}&=\frac{[2(\ell+n)-3]^2}{4n(2\ell+1+n)}, \qquad 
\nu_{n\ell}=\frac{(\eta_1-\eta_2)^2}{4n(2\ell+1+n)}
\end{align}
for the coefficients, and thus $G_\ell(x)=(-x)^{\ell +1}\,\expf{-\pi\eta_1}\sum_n a_n\, x^n /\sqrt{1-x}$. 

\citet{Hummer1988} directly evaluated $\alpha_n=a_n\,x^n$, however, in our applications we also evaluate $G_\ell$ for varying $x$ at fixed values of $\eta_1$ and $\eta_2$. In this case it is better to store the required values of $a_n$ instead. The real gains are marginal in any case, in particular since for the evaluation of the sum one can compute $x^n=x \,(x^{n-1})$ in each step at hardly any extra cost. We also did not find the stability of the expressions to improve by changing the procedure. Stability issues could be solved using arbitrary number precision or resorting to the differential equation for $G_0$.

For $-2<x<-1/2$, to accelerate convergence one should rewrite the expressions in terms of $y=x/(x-1)$, which maps the interval into $1/3 < y < 2/3$. Applying hypergeometric function relations, we find
\bsub
\begin{align}
G_\ell(y)
&=
y^{\ell +1}
(1-y)^{\frac{i\eta_-}{2}}\,\expf{-\pi\eta_1}
\nonumber\\
&\qquad\quad _2F_1\left(1+\ell+i\eta_1, 1+\ell-i\eta_2, 2\ell+2, y \right)
\\[1mm]
F_\ell(y)&=\tilde{\tilde{G}}_\ell(y)\left(-y\right)^{-(\ell +1)}(1-y)^{-\frac{i\eta_-}{2} -\frac{1}{2}} \,\expf{\pi\eta_1}.
\end{align}
\esub
with $\tilde{\tilde{G}}_\ell(y)=(-1)^{\ell+1}\,\sqrt{1-y}\,G_\ell(y)$. Comparing with Eq.~\eqref{app:initial_F_ell}, we thus can again apply the recurrence relations, Eq.~\eqref{eq:rec_I}, for $\tilde{\tilde{G}}_\ell(y)$ after replacing $x\rightarrow y = x/(x-1)$ and $\eta_2\rightarrow -\eta_2$.

To treat the problem at $x<-2$, we use $z=1/x$, which maps the interval into $-1/2<z<0$. In this case, {\it two} new recursions are needed. Applying the hypergeometric function relations, we can write
\begin{align}
G_\ell(z)
&=
\left(-z\right)^{-(\ell +1)}
[-(1-z)/z]^{\frac{i\eta_+}{2}}\,(-z)^{\ell +1}\,\expf{-\pi\eta_1}\,
\nonumber\\
&\quad \times \Bigg[ (-z)^{i\eta_1}\Lambda_\ell(\eta_1, \eta_2) \, {_2}F_1\left(1+\ell+i\eta_1, i\eta_1-\ell, 1+i\eta_-, z \right)
\nonumber\\
&\qquad +(-z)^{i\eta_2}\Lambda_\ell^*(\eta_1, \eta_2) \, {_2}F_1\left(1+\ell+i\eta_2, i\eta_2-\ell, 1-i\eta_-, z \right) \Bigg]
\nonumber\\
&= (-z)^{\frac{i\eta_-}{2}} \Lambda_\ell(\eta_1, \eta_2) \, H_\ell(\eta_1, \eta_2, z) +(-z)^{-\frac{i\eta_-}{2}}\Lambda_\ell^*(\eta_1, \eta_2) \, H^*_\ell(\eta_1, \eta_2,z)
\nonumber\\
&= 2{\rm Re}\left[(-z)^{\frac{i\eta_-}{2}}\,\Lambda_\ell(\eta_1, \eta_2)\, H_\ell(\eta_1, \eta_2, z)\right]
\nonumber\\
&={\rm Re}\left[J_\ell(z)\right] \cos\left[ \frac{\eta_-}{2}\ln(-z)\right] -{\rm Im}\left[J_\ell(z)\right] \sin\left[ \frac{\eta_-}{2}\ln(-z)\right]
\nonumber
\\[1mm]
&\!\!\!\!\!\!H_\ell(\eta_1, \eta_2,z)=(1-z)^{\frac{i\eta_+}{2}}\,{_2}F_1\left(1+\ell+i\eta_1, i\eta_1-\ell, 1+ i\eta_-, z \right)
\nonumber
\\[1mm]
&\!\!\!\!\!\!J_\ell(\eta_1, \eta_2,z)=2\Lambda_\ell(\eta_1, \eta_2)\, H_\ell(\eta_1, \eta_2,z)
\nonumber
\\[1mm]
&\!\!\!\!\!\!\Lambda_\ell(\eta_1, \eta_2)=\frac{\Gamma(2\ell+2)\,\Gamma(-i\eta_-)\,\expf{-\pi\eta_1}}{\Gamma(1+\ell-i\eta_1)\,\Gamma(1+\ell+i\eta_2)}.
\end{align}
%
This means that, similar to KL and H88, one can make the Ansatz
\begin{align}
\label{eq:Re_II}
G_\ell(z)&=A_\ell(z)  \cos\left[ \frac{\eta_-}{2}\ln(-z)\right] +B_\ell(z)  \sin\left[ \frac{\eta_-}{2}\ln(-z)\right],
\end{align}
where $A_\ell(\eta_1, \eta_2, z)={\rm Re}\left[J_\ell(\eta_1, \eta_2, z)\right]$ and $B_\ell(z)=-{\rm Im}\left[J_\ell(\eta_1, \eta_2, z)\right]$ can both be written as a real power series in $z$. We also have to rewrite the differential equation for $\tilde{G}_\ell(x)$ using $x=1/z$, which gives
\begin{align}
&z^2(1-z)^2\tilde{G}''_\ell(z)+2z(1-z)^2\tilde{G}'_\ell(z)
\nonumber\\
&\qquad\qquad=\left\{\ell(\ell+1)z^2-\left[\eta_1\eta_2+\ell(\ell+1)\right]z - \frac{1+\eta_-^2}{4} \right\}\tilde{G}_\ell(z).
\nonumber
\end{align}
Multiplying $G_\ell(z)$ this by $\sqrt{(z-1)/z}$ and inserting it into the differential equation for $\tilde{G}_\ell(z)$, we find the two equations
\bsub
\begin{align}
\label{eq:Re_AB_II}
2\vek{a}^{(n)}_\ell \cdot \vek{A}^{(n)}_\ell  + \eta_- \vek{b}^{(n)} \cdot \vek{B}^{(n)}_\ell&=0,
\\
\eta_- \vek{b}^{(n)} \cdot \vek{A}^{(n)}_\ell  - 2\vek{a}^{(n)}_\ell \cdot \vek{B}^{(n)}_\ell&=0,
\\
\vek{a}^{(n)}_\ell=
    \begin{pmatrix}
    n^2  
    \\
    \ell(\ell+1)-n(2n-3)-1+\frac{\eta_+^2+\eta_-^2}{4}
    \\
    -\ell(\ell+1)+n(n-3)+2-\frac{\eta_-^2}{4}
    \end{pmatrix}, 
  \;
&
\vek{b}^{(n)}=
    \begin{pmatrix}
    2 n 
    \\
    -4n+3
    \\
    2n-3
    \\
    \end{pmatrix}.
 \end{align}
 \esub
for $\vek{A}^{(n)}_\ell=\left(A^{(n)}_\ell,A^{(n-1)}_\ell, A^{(n-2)}_\ell \right)^{T}$ and similar for $\vek{B}_\ell$. The symmetries of the coefficients imply $B^{(n)}_\ell(\eta_1,\eta_2)=A^{(n)}_\ell(\eta_2,\eta_1)$. This also means that the recursion relations for $B^{(n)}_\ell(\eta_1,\eta_2)$ are identical to those for $A^{(n)}_\ell(\eta_1,\eta_2)$ when switching $\eta_1 \leftrightarrow \eta_2$ and $A^{(n)}_\ell\rightarrow B^{(n)}_\ell$. Solving for $A^{(n)}_\ell$ we find
\begin{align}
\label{eq:Re_A_sol}
A^{(n)}_\ell&=\frac{1+n(2n-3)-\ell(\ell+1)-\eta_1\eta_2+\frac{3}{2}\eta_-^2(1-1/n)}{\eta_-^2+n^2}\,A^{(n-1)}_\ell
\nonumber \\
&\quad -\frac{(n+\ell-1)(n-\ell-2)+\frac{3}{4}\eta_-^2(1-2/n)}{\eta_-^2+n^2}A^{(n-2)}_\ell
\nonumber \\
&\qquad +\eta_-\,\frac{2\ell(\ell+1)+3n-2+\frac{1}{2} (\eta_+^2+\eta_-^2)}{n(\eta_-^2+n^2)}B^{(n-1)}_\ell
\nonumber \\
&\quad \qquad -\eta_-\,\frac{2\ell(\ell+1)+3n-4+\frac{1}{2} \eta_-^2}{n(\eta_-^2+n^2)}B^{(n-2)}_\ell
\end{align}
The equations for $B^{(n)}_\ell$ are obtained by switching variables, as mentioned above. The initial conditions are
\begin{align}
\label{eq:Re_A_sol_initial}
A^{(0)}_\ell&=2{\rm Re}\left[\frac{\Gamma(2\ell+2)\,\Gamma(-i\eta_-)\,\expf{-\pi\eta_1}}{\Gamma(1+\ell-i\eta_1)\,\Gamma(1+\ell+i\eta_2)}\right]
\nonumber\\
B^{(0)}_\ell&=-2{\rm Im}\left[\frac{\Gamma(2\ell+2)\,\Gamma(-i\eta_-)\,\expf{-\pi\eta_1}}{\Gamma(1+\ell-i\eta_1)\,\Gamma(1+\ell+i\eta_2)}\right],
\end{align}
which both can be easily computed using standard libraries. This then gives the solutions $A_\ell(\eta_1, \eta_2, z)=\sum_n A^{(n)}_\ell\,z^n$ and similar for $B_\ell(\eta_1, \eta_2, z)$. 

\subsubsection{Dealing with $\Lambda_\ell(\eta_1, \eta_2)$}
\label{sec:Lambda_deal}
To evaluate the initial conditions for the recurrence relations we need 
\begin{align}
\Lambda_\ell(\eta_1, \eta_2)&=\frac{\Gamma(2\ell+2)\,\Gamma(-i\eta_-)\,\expf{-\pi\eta_1}}{\Gamma(1+\ell-i\eta_1)\,\Gamma(1+\ell+i\eta_2)}
\nonumber\\
&=\frac{\Gamma(2\ell+2)\,\Gamma(-i\eta_-)\,\Gamma(1+\ell+i\eta_1)\,\Gamma(1+\ell-i\eta_2)\,\expf{-\pi\eta_1}}{|\Gamma(1+\ell-i\eta_1)\,\Gamma(1+\ell+i\eta_2)|}
\nonumber\\
&=\Gamma(-i\eta_-)\,\Gamma(1+\ell+i\eta_1)\,\Gamma(1+\ell-i\eta_2)\,\Xi_\ell.
\end{align}
Here we have 
\bsub
\begin{align}
\Xi_\ell&=\frac{\Gamma(2\ell+2)\,\expf{-\pi\eta_1}}{|\Gamma(1+\ell-i\eta_1)\,\Gamma(1+\ell+i\eta_2)|}
\\
\Xi_0&=\frac{{\rm Sinh}(\pi\eta_1)\,{\rm Sinh}(\pi\eta_2)\,\expf{-\pi\eta_1}}{\pi^2 \eta_1\,\eta_2}
\\
\Xi_1&=\frac{{\rm Sinh}(\pi\eta_1)\,{\rm Sinh}(\pi\eta_2)\,\expf{-\pi\eta_1}}{\pi^2 \eta_1\,\eta_2}\,\frac{6}{(1+\eta_1^2)(1+\eta_2^2)}.
\end{align}
\esub
This also shows that 
\begin{align}
\mathcal{F}(\eta_1,\eta_2) &=\frac{4\pi^2\eta_1\eta_2}{(1-\expf{-2\pi\eta_1})(1-\expf{-2\pi\eta_2})}\equiv \frac{\expf{\pi\eta_2}}{\Xi_0},
\end{align}
which appears in the normalization of the non-relativistic Gaunt factor.

\vspace{-3mm}
\section{Low-frequency approximation for the Gaunt factor}
\label{sec:NR_low_approx}
At low frequencies, the calculation of the Gaunt factor for the non-relativistic limit become highly unstable. To obtain an approximation at low frequencies we take the limits of $G_\ell$ for $\eta_2\rightarrow \eta_1$. With $z=-1/x=\eta_-^2/4\eta_1\eta_2$, this yields
\begin{align}
\label{eq:NR_approx_G0}
G_0&\approx 2\Big( {\rm Re}\left[\Lambda_0\right] \cos\left[ \frac{\eta_-}{2}\ln(z)\right] -{\rm Im}\left[\Lambda_0\right] \sin\left[ \frac{\eta_-}{2}\ln(z)\right]\Big)
\nonumber\\
\Delta G&=
\left[\eta_1 \, \eta_2+\frac{1}{2}\left(\frac{\eta_1}{\eta_2}+\frac{\eta_2}{\eta_1}\right)\,\right]\,G_0 - \frac{(1+\eta_1^2)(1+\eta_2^2)}{6} G_1
\nonumber\\
&\approx  4 \eta_- \Big( {\rm Im}\left[\Lambda_0\right]  \cos\left[ \frac{\eta_-}{2}\ln(z)\right] 
- {\rm Re}\left[\Lambda_0\right]  \sin\left[ \frac{\eta_-}{2}\ln(z)\right] \Big)
\nonumber \\ 
{\rm Re}\left[\Lambda_0\right]&\approx - \frac{1-\expf{-2\pi\eta_1}}{2\pi \eta_1}\, {\rm Re}\left[H(i \eta_1)\right]
\nonumber\\
{\rm Im}\left[\Lambda_0\right]&\approx  - \frac{1-\expf{-2\pi\eta_1}}{2\pi \eta_1}\, {\rm Im}\left[H(i \eta_1)\right]+\frac{1}{\eta_-},
\end{align}
where $H(z)$ denotes the harmonic number of $z$. From this one finds 
\begin{align}
\label{eq:NR_approx}
g_{\rm NR}(\omega, p_1)
&\approx \frac{\sqrt{3}}{\pi} \,\mathcal{F}_{\rm E}(\eta_1,\eta_2)\,\mathcal{C}_0(\eta_1,\eta_2)\,\Delta \mathcal{C}(\eta_1,\eta_2)
\nonumber\\
\mathcal{F}_{\rm E}(\eta_1,\eta_2)&=\frac{\eta_2}{\eta_1}\frac{1-\expf{-2\pi\eta_1}}{1-\expf{-2\pi\eta_2}}, 
\nonumber\\
\Sigma_1&={\rm Re}\left[H(i \eta_1)\right],
\quad \Sigma_2={\rm Im}\left[H(i \eta_1)\right]-\frac{1}{\Delta \eta},
\\
\nonumber
\mathcal{C}_0(\eta_1,\eta_2)
&=\Sigma_1 \cos\left[ \frac{\Delta \eta}{2}\ln(-x)\right] +\Sigma_2 \sin\left[ \frac{\Delta \eta}{2}\ln(-x)\right]
\\
\nonumber
\Delta\mathcal{C}(\eta_1,\eta_2)
&=\eta_- \Bigg(\Sigma_2\cos\left[ \frac{\Delta \eta}{2}\ln(-x)\right] +\Sigma_1 \sin\left[ \frac{\Delta \eta}{2}\ln(-x)\right]\Bigg),
\end{align}
with $\Delta \eta=\eta_-=\eta_1-\eta_2$. Since at low frequencies $\Delta \eta\ll 1$, one can set $\cos\left[ \frac{\Delta \eta}{2}\ln(-x)\right]\simeq 1$ and $\sin\left[ \frac{\Delta \eta}{2}\ln(-x)\right]\simeq \frac{\Delta \eta}{2}\ln(-x)$. Equation~\eqref{eq:NR_approx} therefore further simplifies to the expression given in Eq.~\eqref{eq:NR_approx_more}.

To evaluate the real and imaginary parts of the harmonic number we use the explicit series when $0<\eta_1<10$:
\begin{align}
\label{eq:Harmon}
{\rm Re}\left[H(i a)\right]&= \sum_{m=1} \frac{a^2}{m(a^2+m^2)}, \qquad {\rm Im}\left[H(i a)\right]= \sum_{m=1} \frac{a}{(a^2+m^2)},
\end{align}
while for large argument we have 
\begin{align}
\label{eq:Harmon_approx}
{\rm Re}\left[H(i a)\right]&\approx \gamma_{\rm E} + \ln a +\frac{1}{12 a}\left(1+\frac{1}{10 a^2}+\frac{1}{21 a^4}
+\frac{1}{20 a^6}+\frac{1}{11 a^8}
\right)
\nonumber\\
{\rm Im}\left[H(i a)\right]&\approx \frac{\pi}{2}-\frac{1}{2a}+2\pi\left[{\rm coth}\left(\frac{\pi}{a}\right)-1\right],
\end{align}
where $\gamma_{\rm E}$ is the Euler-constant. These approximation are extremely useful at very low frequencies, $\omega/\omega_{\rm max} \lesssim 10^{-6}$, and large $\eta_1$, i.e., $p_1\lesssim  10^{-3}$. 

\section{EH cross section}
\label{sec:EH_rewrite}
The starting point for our computations is the EH cross section \citep{ElwertHaug1969}. 
We shall use all definitions as in Sect.~\ref{sec:BR_definitions} and add the auxiliary variables
\bsub
\begin{align}
\mu_i&=\frac{\vek{p}_i\cdot\vek{k}}{p_1\omega},\quad \kappa_i=2(\gamma_i-p_i\mu_i)=2(\gamma_i-\pi_i)
\\
\mu_{12}&=\frac{\vek{p}_1\cdot\vek{p}_2}{p_1 p_2}=\mu_1\mu_2+\cos(\phi_2)\sqrt{1-\mu_1^2}\sqrt{1-\mu_2^2}
\\
\pi_{i}&=p_i\mu_1, \quad \pi_{12}=p_1p_2\mu_{12}, 
\quad \eta_\infty=\alpha Z,
\quad\eta_\pm=\eta_1\pm\eta_2
\\
\chi_i&=p_i\sqrt{1-\mu_i^2}, \quad \chi_{12}=\chi_1 \chi_2 \cos(\phi_2)
\\
\tau_i&=4\gamma_i^2-q^2, \quad \tau_{12}=4\gamma_1\gamma_2-q^2, \quad \zeta_i=\chi_i^2-\chi_{12},
\\[2mm]
q^2&=|\vek{p}_1-\vek{p}_2-\vek{k}|^2
=p_1^2+p_2^2+\omega^2+2\left[\omega(\pi_2-\pi_1)-\pi_{12}\right]
\\[1mm]
\xi&=\frac{\tilde{\mu} q^2}{\kappa_1\kappa_2},
\quad \tilde{\mu}=\frac{\mu}{\omega^2}=\left(\frac{p_1+p_2}{\omega}\right)^2-1\equiv \frac{2(\gamma_1\gamma_2+p_1p_2-1)}{\omega^2}, 
\\[1mm]
\rho&=\frac{1}{p_1}+\frac{1}{p_2}, \quad
\kappa=\frac{\gamma_1}{p_1}+\frac{\gamma_2}{p_2},
\end{align}
\esub
for further convenience. The Gaunt factor (differential in three angles) can then be written as \citep{ElwertHaug1969}
\bsub
\begin{align}
\label{eq:EH_start}
\frac{\id^3 g_{\rm EH}}{\id\mu_1\!\id\mu_2\!\id\phi_2}
&=\frac{3\sqrt{3}}{8\pi^2}\,p_1 p_2\,\mathcal{F}(\eta_1,\eta_2)\,\mathcal{M}^2(\omega, p_1, \mu_1, \mu_2, \phi_2)
\\
\mathcal{M}^2&=\frac{\expf{-2\pi\eta_1}}{q^4}\Bigg\{\frac{E_1}{\kappa_1^2} \,|A_1|^2+\frac{E_2}{\kappa_2^2} \,|A_2|^2 - \frac{2 E_3}{\kappa_1\kappa_2} \,{\rm Re}[A_1^*A_2]
\\ \nonumber
&\quad + \frac{F_3\,q^4}{\kappa^2_1\kappa^2_2}\,|B|^2 
-\frac{2q^2\left(
F_1\,\kappa_2\,{\rm Re}[A_1^*B]+
F_2\,\kappa_1\,{\rm Re}[A_2^*B]
\right)}{\kappa^2_1\kappa^2_2}
\Bigg\} .
\end{align}
Here $\mathcal{M}^2$ depends on the following functions:
\begin{align}
V(\eta_1, \eta_2, \xi)&={_2}F_1\left(i\eta_1, i\eta_2, 1, 1-\xi \right)
\\
W(\eta_1, \eta_2, \xi)&={_2}F_1\left(1+i\eta_1, 1+i\eta_2, 2, 1-\xi \right)=\frac{1}{\eta_1\eta_2}\partial_\xi V(\eta_1, \eta_2, \xi)
\\
A_i&=V- i\xi \, \eta_i W, \quad B=i \eta_\infty W
\\
E_1&=(4\gamma_2^2-q^2)\,\chi_1^2	+ \left[\chi_1^2-\chi_{12}+2(1+\chi_2^2)\frac{\omega}{\kappa_2}\right]\,\kappa_1\omega
\\
E_2&=(4\gamma_1^2-q^2)\,\chi_2^2	- \left[\chi_2^2-\chi_{12}-2(1+\chi_1^2)\frac{\omega}{\kappa_1}\right]\,\kappa_2\omega
\\
E_3&=(4\gamma_1\gamma_2-q^2)\,\chi_{12}	+ 2(1+\chi_{12})\omega^2
\nonumber \\ 
&\qquad + \left[\left(\chi_1^2-\chi_{12}\right)\kappa_2-\left(\chi_2^2-\chi_{12}\right)\kappa_1\right]\frac{\omega}{2}
\\[2mm]
F_1&=\rho(\chi_1^2-\chi_{12})+\kappa\left[\pi_1(\pi_{12}+p_2^2)+(2-\pi_1\pi_2)\omega\right]
\\
&\qquad-\left(\kappa p_1p_2-2\frac{\omega}{p_1}\right)\left[\pi_1+\pi_2-\omega\right]
\\
F_2&=\rho(\chi_2^2-\chi_{12})+\kappa\left[\pi_2(\pi_{12}+p_1^2)-(2-\pi_1\pi_2)\omega\right]
\\
&\qquad-\left(\kappa p_1p_2+2\frac{\omega}{p_2}\right)\left[\pi_1+\pi_2+\omega\right]
\\
F_3&=\mu\left[1-\frac{\pi_1\pi_2}{p_1p_2}+\frac{\gamma_1+\gamma_2}{p_1p_2}\,\frac{\gamma_1+\gamma_2+\pi_1+\pi_2}{p_1p_2}\right] - 2\rho^2.
\end{align}
\esub
We followed the original definitions as closely as possible but already performed a few trivial simplifications. A few symmetries are worth noting. The function $E_2$ can be obtained from $E_1$ by switching the roles of $\vek{p}_1\leftrightarrow \vek{p}_2$ and negating $\omega\rightarrow -\omega$ or equivalently $\vek{k}\rightarrow -\vek{k}$. The function $E_3$ remains invariant under this transformation. Similar statements apply to the $F_i$'s.

It is difficult to work with the matrix element in the above form, which furthermore suffers from severe numerical cancelation issues. To simplify matters, we collect all coefficients of the main functions that are appearing. These are $\propto |V|^2$, $\propto {\rm Im}[V^*W]$ and $\propto |W|^2$ and after gathering terms we find
\bsub
\begin{align}
|A_i|^2&=|V|^2+2\eta_i \xi \, {\rm Im}[V^*W]+\eta^2_i \xi^2 |W|^2
\\
|B|^2&=\eta_\infty^2\, |W|^2
\\
{\rm Re}[A_1^*A_2]&=|V|^2+\eta_+ \xi \, {\rm Im}[V^*W]+\eta_1\eta_2 \xi^2 |W|^2
\\
{\rm Re}[A_i^*B]&=-\eta_\infty {\rm Im}[V^*W] - \eta_i \eta_\infty \xi \, |W|^2.
\end{align}
\esub
We also note that in comparison to EH a factor of $\expf{-2\pi \eta_1}$ can be absorbed into the definitions of $V$ and $W$, since it is canceled directly by a corresponding factor from the hypergeometric function, and thus avoids spurious numerical instabilities for $p_1\ll 1$. By regrouping terms we then find
\begin{align}
\mathcal{M}^2&=\frac{\expf{-2\pi\eta_1}}{q^4}
\Bigg\{\left( \frac{E_1}{\kappa_1^2}+\frac{E_2}{\kappa_2^2}-\frac{2E_3}{\kappa_1\kappa_2}\right) |V|^2 
+ 2\xi \left[ \frac{\eta_1 E_1}{\kappa_1^2}+\frac{\eta_2 E_2}{\kappa_2^2} -\frac{E_3 \eta_+}{\kappa_1\kappa_2} \right.
\nonumber\\
&\quad
\left.
+\left(\frac{F_1}{\kappa_1}+\frac{F_2}{\kappa_2}\right)\,\frac{\eta_\infty}{\tilde{\mu}}
\right] {\rm Im}[V^*W]
+ \xi^2 \left[ \frac{\eta^2_1 E_1}{\kappa_1^2}+\frac{\eta^2_2 E_2}{\kappa_2^2} -\frac{2E_3 \eta_1\eta_2}{\kappa_1\kappa_2} 
\right.
\nonumber\\
&\qquad
\left.
+\left(\frac{\eta_1 F_1}{\kappa_1}+\frac{\eta_2 F_2}{\kappa_2}\right)\,\frac{2\eta_\infty}{\tilde{\mu}}
+F_3\,\frac{\eta_\infty^2}{\tilde{\mu}^2}
\right] |W|^2
\Bigg\},
\end{align}
where we used $q^2/\kappa_1\kappa_2 = \xi / \tilde{\mu}$. After simplifying the expressions and defining
$\mathcal{V}=\expf{-\pi\eta_1}V$ and $\mathcal{W}=-\xi \expf{-\pi\eta_1} W$ this then yields
\bsub
\label{eq:EH_G_final}
\begin{align}
\mathcal{M}^2&=\frac{1}{q^4}\Bigg\{ J_1 |\mathcal{V}|^2  - 2J_2 \,{\rm Im}[\mathcal{V}^* \mathcal{W}] + J_3 |\mathcal{W}|^2 \Bigg\}
\\
J_1&\equiv J_{\rm BH}=\tau_1 \frac{\chi_2^2}{\kappa_2^2}+\tau_2 \frac{\chi_1^2}{\kappa_1^2} -\tau_{12} \frac{2\chi_{12}}{\kappa_1\kappa_2}
+ \left(\zeta_1+\zeta_2\right)\,\frac{2\omega^2}{\kappa_1\kappa_2}
\\
J_2&=\tau_1\frac{\eta_2\chi_2^2}{\kappa_2^2}+\tau_2\frac{\eta_1\chi_1^2}{\kappa_1^2}
-\tau_{12}\frac{\eta_+\,\chi_{12}}{\kappa_1\kappa_2}
+\frac{\eta_-}{2}\left(\frac{\zeta_1}{\kappa_1}+\frac{\zeta_2}{\kappa_2}\right)\omega
\nonumber \\ 
&\qquad
+ \left(\eta_1\zeta_2+\eta_2\zeta_1\right) \frac{2\omega^2}{\kappa_1\kappa_2}
+ \left(\frac{F_1}{\kappa_1}+\frac{F_2}{\kappa_2}\right)\frac{\eta_\infty}{\tilde{\mu}}
\\
J_3&=\tau_1\frac{\eta_2^2\chi_2^2}{\kappa_2^2}+\tau_2\frac{\eta_1^2\chi_1^2}{\kappa_1^2}
-\tau_{12}\frac{2\eta_1\eta_2\chi_{12}}{\kappa_1\kappa_2}
+\eta_-\left(\frac{\eta_1 \zeta_1}{\kappa_1}+\frac{\eta_2 \zeta_2}{\kappa_2}\right)\omega
\nonumber \\ 
&\qquad
+ \left[\eta^2_1\zeta_2 +\eta_2^2\zeta_1+ \eta_-^2 (1+\chi_{12})\right]\frac{2\omega^2}{\kappa_1\kappa_2}
\nonumber \\ 
&\qquad\qquad
+\left(\frac{\eta_1 F_1}{\kappa_1}+\frac{\eta_2 F_2}{\kappa_2}\right)\frac{2\eta_\infty}{\tilde{\mu}}
+F_3\,\frac{\eta_\infty^2}{\tilde{\mu}^2}.
\end{align}
\esub
The function $J_1$ contains all terms relevant to the Bethe-Heitler approximation.
Even if written in this way the matrix element already becomes more transparent, further simplifications are possible. Firstly, we have the identity
\begin{align}
\frac{\eta_-}{2}\omega + \rho \frac{\eta_\infty}{\tilde{\mu}} \equiv 0 \qquad \longleftrightarrow \qquad\frac{\eta_\infty}{\tilde{\mu}}=-\frac{\eta_-\omega}{2\rho}
\end{align}
This eliminates the $\rho$-terms from $F_1$ and $F_2$ while canceling those directly $\propto \omega$ in $J_2$ and $J_3$:
\bsub
\label{eq:J12_rewrite}
\begin{align}
J_2&=\tau_1\frac{\eta_2\chi_2^2}{\kappa_2^2}+\tau_2\frac{\eta_1\chi_1^2}{\kappa_1^2}
-\tau_{12}\frac{\eta_+\,\chi_{12}}{\kappa_1\kappa_2}
\nonumber \\ 
&\qquad
+ \left(\eta_1\zeta_2+\eta_2\zeta_1\right) \frac{2\omega^2}{\kappa_1\kappa_2}
+ \left(\frac{\tilde{F}_1}{\kappa_1}+\frac{\tilde{F}_2}{\kappa_2}\right)\frac{\eta_\infty}{\tilde{\mu}}
\\
J_3&=\tau_1\frac{\eta_2^2\chi_2^2}{\kappa_2^2}+\tau_2\frac{\eta_1^2\chi_1^2}{\kappa_1^2}
-\tau_{12}\frac{2\eta_1\eta_2\chi_{12}}{\kappa_1\kappa_2}+\frac{\eta^2_-\omega^2}{2}
\nonumber \\ 
&\qquad
+ \left[\eta^2_1\zeta_2 +\eta_2^2\zeta_1+ \eta_-^2 (1+\chi_{12})\right]\frac{2\omega^2}{\kappa_1\kappa_2}
\nonumber \\ 
&\qquad\qquad
+\left(\frac{\eta_1 \tilde{F}_1}{\kappa_1}+\frac{\eta_2 \tilde{F}_2}{\kappa_2}\right)\frac{2\eta_\infty}{\tilde{\mu}}
+F_3\,\frac{\eta_\infty^2}{\tilde{\mu}^2}
\end{align}
\esub
with $\tilde{F}_i=F_i(\rho=0)$. 
The $J_1$ terms are most relevant in the Bethe-Heitler regime, and we thus recast $J_2$ and $J_3$ as
\bsub
\label{eq:J12_rewrite}
\begin{align}
J_2&=\eta_1 J_1-\eta_-\Delta_1, \qquad J_3=\eta_1^2 J_1-2 \eta_1 \eta_- \Delta_1+\eta_-^2 \Delta_2
\\[1mm]
\Delta_1&=
\frac{\tau_{1}\chi^2_{2}}{\kappa^2_2}-\frac{\tau_{12}\chi_{12}}{\kappa_1\kappa_2}
+\zeta_1\frac{2 \omega^2}{\kappa_1\kappa_2}
+ \left(\frac{L_1}{\kappa_1}+\frac{L_2}{\kappa_2}\right)\frac{\omega}{2\rho}
\\[0mm]
\Delta_2&=
\frac{\tau_{1}\chi^2_{2}}{\kappa^2_2}
+(1+\chi^2_1)\frac{2 \omega^2}{\kappa_1\kappa_2}
+\frac{L_2}{\kappa_2}\frac{\omega}{\rho}
+F_3\frac{\omega^2}{4\rho^2}
\\[1mm]
L_1&=\kappa\left[\pi_1(\pi_{12}+p_2^2)-\left(\pi_1+\pi_2-\omega\right)p_1p_2+(2-\pi_1\pi_2)\omega\right]
\nonumber\\
&\qquad\qquad+2\frac{\omega}{p_1}\left(\pi_1+\pi_2-\omega\right)
\\[0mm]
L_2&=\kappa\left[\pi_2(\pi_{12}+p_1^2)-\left(\pi_1+\pi_2+\omega\right)p_1p_2-(2-\pi_1\pi_2)\omega\right]
\nonumber\\
&\qquad\qquad-2\frac{\omega}{p_2}\left(\pi_1+\pi_2+\omega\right)
\label{eq:J_rewrite_I}
\end{align}
\esub
Inserting these expression back into Eq.~\eqref{eq:EH_G_final} and collecting terms we find
\bsub
\begin{align}
\mathcal{M}^2&=\frac{1}{q^4}\Bigg\{ J_1 |\mathcal{A}|^2  - 2\eta_- \Delta_1 |\mathcal{C}|^2 + \eta_-^2 \Delta_2 |\mathcal{W}|^2 \Bigg\}
\\
|\mathcal{A}|^2&=|\mathcal{V}|^2- 2\eta_1 {\rm Im}[\mathcal{V}^* \mathcal{W}]+ \eta_1^2|\mathcal{W}|^2
\\
|\mathcal{C}|^2&= \eta_1|\mathcal{W}|^2-{\rm Im}[\mathcal{V}^* \mathcal{W}]
\end{align}
\esub
Expressed in this way indeed simplifies the evaluation of the cross section significantly and also allows one to more easily read off limiting cases.

To explicitly evaluate the functions $|\mathcal{A}|^2$ and $|\mathcal{W}|^2$ we used precomputed tables obtained with {\tt Mathematica} to improve the precision as well as the Gauss series for the hypergeometric function. As shown below, this also determines $|\mathcal{C}|^2$ at no additional cost [see Eq.~\eqref{eq:relations_AWIM}]. At low frequencies ($w\lesssim 10^{-6}$), one can furthermore apply
\begin{align}
|\mathcal{A}|^2
&\approx \left(\frac{1-\expf{-2\pi \eta_1}}{2\pi \eta_1}\right)^2
\left[1-\frac{2\eta_1^2}{\xi}\left(1+\phi\right)\right]
\nonumber\\[1mm]
|\mathcal{W}|^2
&\approx \left(\frac{1-\expf{-2\pi \eta_1}}{2\pi \eta_1}\right)^2
\left[\phi^2\left(1+\frac{2}{\xi}\right)+\frac{2\eta_1^2}{\xi}\phi (2+\phi) \right]
\nonumber\\[1mm]
\phi&=\ln\xi - 2\,{\rm Re}[H(i\eta_1)]
\label{eq:Matrix_rewrite}
\end{align}
to ease the computations. However, the methods relying on $G_0$ and $G'_0$ (next section) was found to be more efficient.

\subsection{Relating $\mathcal{V}$ and $\mathcal{W}$ to $G_0$}
\label{sec:EH_G0}
While in principle we can already compute the EH cross section using the expression from the preceding section, one additional obstacle can be overcome by directly relating the function $\mathcal{V}$ and $\mathcal{W}$ to $G_0$ and $G_0'$. One of the important benefits is that $G_0$ and $G_0'$ are a real valued function and thus greatly simply matters. Progress can be made by starting from the differential equation for $V$, which with $z=1-\xi$ reads
\begin{align}
\nonumber
0 &= z (1-z) V'' +  [1-(1+i\eta_+)z] V' + \eta_1 \eta_2 V 
\nonumber \\
&= -\eta_1 \eta_2 \left\{z (1-z) {W}' + [1-(1+i\eta_+)z] {W} - {V} \right\}
\nonumber\\
\Rightarrow\quad {V}&=[(1-z)-i\eta_+z] {W}+z (1-z) {W}'.
\end{align}
Here primes denote derivatives with respect to $z$. By comparing $W$ with $G_0$ we find
\begin{align}
\label{eq:Re_EH_G0}
{W}&=\frac{\expf{\pi\eta_1}(1-z)^{-\frac{i \eta_+}{2}}}{-z} G_0(z)
\quad \leftrightarrow \quad
\mathcal{W}=\frac{(1-z)^{1-\frac{i \eta_+}{2}}}{z} G_0(z),
\end{align}
which directly implies
\begin{align}
\label{eq:Re_W2_G0}
|\mathcal{W}|^2&=\frac{\xi^2G^2_0(1-\xi)}{(1-\xi)^2} 
\equiv \xi^2\,\expf{-2\pi\eta_1} |{_2}F_1\left(1+i\eta_1,1+i\eta_2, 2, 1-\xi \right)|^2
\end{align}
We then also have 
\begin{align}
\label{eq:Re_W_prime}
W'&=\frac{\expf{\pi\eta_1}(1-z)^{-\frac{i \eta_+}{2}}}{-z^2(1-z)} \left[z(1-z) G'_0(z)-(1-z)G_0(z)+i \frac{\eta_+}{2} z G_0(z)\right]
\nonumber
\\
\nonumber
{V}&=\expf{\pi\eta_1}(1-z)^{-\frac{i \eta_+}{2}}  \left\{ i \frac{\eta_+}{2} G_0(z)- (1-z) G'_0(z)\right\},
\end{align}
which related both $W'$ and $V$ to $G_0$ and $G'_0$ only.

\noindent
Since $|\mathcal{V}|^2=\expf{-2\pi \eta_1}|{V}|^2$, this then yields
\begin{align}
\nonumber
|\mathcal{V}|^2&=\frac{\eta^2_+}{4} G^2_0(z) + (1-z)^2[G'_0(z)]^2
\\
{\rm Im}[\mathcal{V}^* \mathcal{W}]&=-\frac{\eta_+}{2} \frac{(1-z)}{z} G^2_0(z) =-\frac{\eta_+}{2} \frac{z}{1-z} |\mathcal{W}|^2
\nonumber
\end{align}
Putting everything together we finally obtain
\bsub
\label{eq:relations_AWIM}
\begin{align}
|\mathcal{A}|^2&=\frac{(\eta_+ + \eta_- \xi)^2}{4\xi^2} \, |\mathcal{W}|^2 + \xi^2 [G'_0(1-\xi)]^2
\\
|\mathcal{C}|^2&= \frac{\left(\eta_+ + \eta_- \xi\right) }{2\xi}\,|\mathcal{W}|^2
\\
|\mathcal{W}|^2&=\frac{\xi^2G^2_0(1-\xi)}{(1-\xi)^2},
\end{align}
\esub
which eliminates the need to compute $|\mathcal{C}|^2$ explicitly.
Collecting terms then results in
\begin{align}
\label{eq:EH_G_final_M2}
\mathcal{M}^2&
=\frac{1}{q^4}\Bigg\{ 
\left[J_1 - 2\delta \, D_1 +\delta^2\, D_2 \right] \frac{\eta_+^2 G_0^2(1-\xi)}{4(1-\xi)^2} + J_1 [\xi G'_0(1-\xi)]^2
\Bigg\}
\nonumber
\\[1.5mm]
\delta&=\frac{\eta_-}{\eta_+}\xi, \qquad
D_1=2\Delta_1-J_1, \qquad D_2=J_1-4\Delta_1+4\Delta_2.
\end{align}
This definition groups terms of similar order of magnitude in $\eta_\pm$ and $\xi$. Inserting the definitions of $\Delta_i$ and simplifying the expression then gives Eq.~\eqref{eq:EH_G_final_G0} for the EH matrix element.

\end{appendix}
}

{\small
\vspace{-3mm}
\bibliographystyle{mn2e}
\bibliography{Lit}
}

\end{document}